\documentclass[11pt,a4paper]{article}
\usepackage{jheppub_kim}
\usepackage{rotating}
\usepackage{graphicx,epsfig}
\usepackage{amsmath}
\usepackage {amssymb}
\usepackage{subfigure}

\usepackage{relsize}

\def\beq{\begin{equation}}
\def\eeq{\end{equation}}
\def\bea{\begin{eqnarray}}
\def\eea{\end{eqnarray}}

\begin{document}

\title{  Viscous Cosmology for  Early- and   Late-Time Universe}

\author[a]{Iver Brevik}

\author[b]{{\O}yvind  Gr{\o}n}

\author[c]{Jaume de Haro}

\author[d,e,f,g]{Sergei D. Odintsov}

\author[h,i,j]{Emmanuel N. Saridakis}

 \affiliation[a]{Department of Energy and Process Engineering, Norwegian University of
Science and Technology, N-7491 Trondheim, Norway}

\affiliation[b]{Oslo and Akershus University College of Applied Sciences, Faculty of
Technology, Art and Design, St. Olavs Plass, N-0130 Oslo, Norway}

\affiliation[c]{ Departament de Matem\`atica Aplicada, Universitat Polit\`ecnica de
Catalunya, Diagonal
647, 08028 Barcelona, Spain}

\affiliation[d] {ICREA, Passeig Luis Companys, 23, 08010 Barcelona, Spain}

\affiliation[e] {Institute of Space Sciences (IEEC-CSIC) C. Can Magrans s/n, 08193
Barcelona, Spain}

\affiliation[f]{Tomsk State Pedagogical University, 634061 Tomsk and Int. Lab. Theor.
Cosmology,
Tomsk State Univ. of Control Systems and Radioelectronics (TUSUR), 634050 Tomsk, Russia}

\affiliation[g]{Inst. of Physics, Kazan Federal University, Kazan 420008, Russia}

\affiliation[h]{Department of Physics, National Technical University of Athens, Zografou
Campus GR 157 73, Athens, Greece}

\affiliation[i]{National Center for Theoretical Sciences, Hsinchu,
Taiwan 300}

\affiliation[j]{CASPER, Physics Department, Baylor University, Waco, TX 76798-7310, USA}

\emailAdd{oyvind.gron@hioa.no}
\emailAdd{iver.h.brevik@ntnu.no}
\emailAdd{jaime.haro@upc.edu}
\emailAdd{ odintsov@ieec.uab.es }
\emailAdd{Emmanuel$_-$Saridakis@baylor.edu}

\abstract
{From a hydrodynamicist's point of view the inclusion of viscosity concepts in the
macroscopic theory of the cosmic fluid would appear  most natural, as an ideal fluid is
after all an abstraction (excluding special cases such as superconductivity). Making use 
of
modern observational results for the Hubble parameter plus standard Friedmann formalism,
we may extrapolate the description of the universe back in time up  to the inflationary
era, or we may go to the  opposite extreme and analyze the probable  ultimate fate of the
universe. In this review we discuss a variety of topics in cosmology when it is enlarged
in order to contain a bulk viscosity.  Various forms of this viscosity, when expressed in
terms of the fluid density or the Hubble parameter, are discussed. Furthermore, we
consider homogeneous as well as inhomogeneous equations of state. We investigate
viscous cosmology in the early universe, examining the viscosity effects on the various
inflationary observables. Additionally, we study viscous cosmology in the late universe,
containing current acceleration and the possible future singularities, and we investigate
how one may even unify inflationary and late-time acceleration.
Finally, we analyze the viscosity-induced crossing through the quintessence-phantom
divide, we examine the realization of viscosity-driven cosmological bounces, and we
briefly discuss  how the Cardy-Verlinde formula is affected by viscosity.}


\maketitle

 \newpage

\section{Introduction}

The introduction of viscosity coefficients in cosmology has itself a long history,
although the physical importance of these phenomenological parameters has traditionally
been assumed to be weak or at least subdominant. In connection with the very early
universe, the influence from viscosity is assumed to be the  largest at the time of
neutrino
decoupling (end of the lepton era), when the temperature was about $10^{10}~$K. Misner
\cite{Misner:1967uu} was probably the first to introduce the viscosity from the standpoint
of particle physics; see also Zel'dovich and Novikov \cite{Zeldovich:1983cr}.
Nevertheless, on a phenomenological level, the viscosity concept was actually introduced
much earlier, with the first such work being that of Eckart \cite{eckart40}.

When considering deviations from thermal equilibrium to the first order  in the
cosmic fluid, one should recognize that there are  in principle  two different  viscosity
coefficients, namely the bulk viscosity $\zeta$ and the shear viscosity $\eta$. In view
of the commonly accepted spatial isotropy of the universe, one usually omits the shear
viscosity. This is motivated by the WMAP \cite{Komatsu:2010fb} and Planck observations
\cite{Ade:2015xua}, and is moreover supported by theoretical calculations which show that
in a large class of homogeneous and anisotropic universes isotropization is quickly
established.
Eckart's theory, as most other theories, is maintained at first-order level. In
principle, a difficulty with this kind of theory is that one becomes confronted with a
non-causal behavior. In order to prevent this one has to go to the second order
approximation, away from thermal equilibrium.

The interest in viscosity theories in cosmology has increased in recent years, for various
reasons, perhaps especially from a
fundamental viewpoint. It is well  known among hydrodynamicists that the ideal
(nonviscous) theory is after all only an approximation to the real world.
For reviews on both causal and non-causal theories, the reader may  consult  Gr{\o}n
\cite{Gron:1990ew}  (surveying the literature up to 1990), and later treatises by
Maartens \cite{Maartens:1995wt,Maartens:1996vi}, and Brevik and Gr{\o}n
\cite{Brevik:2014cxa}.

The purpose of the present review is to explore how several parts of cosmological theory
become affected when a bulk viscosity is brought into the formalism. After highlighting
the basic formalism in the remaining of the present section,  in Section \ref{Section2}
we consider the very early (inflationary) universe. We briefly present the conventional
inflation theory, covering ``cold", ``warm" and ``intermediate"
inflation, and we extract various inflationary observables.  Thereafter we investigate
 the viscous counterparts in different models, depending on the form of   bulk
viscosity as well as on the equation of state.

In Section \ref{Section3} we turn to the late universe, including the characteristic
singularities in the far future, related also to the  phantom region in which the
equation-of-state parameter is less than $-1$. The different types of future
singularities are classified, and we explore the consequences of
letting the equation of state to be inhomogeneous. A special case is the unification of
inflation with dark energy in the presence of viscosity, a topic which is dealt with
most conveniently when  one introduces a scalar field. Additionally, we discuss
holographic dark energy in the presence of a viscous fluid.

In  Section \ref{Section4} we discuss various special topics, amongst them the
possibility for the viscous fluid to slide from the quintessence region into the phantom
region and then into a future singularity, if the magnitude of the present bulk viscosity
is large enough. Comparison with estimated values of the bulk viscosity derived from
observations, indicate that this may actually be a realistic scenario. In the same section
we also discuss the viscous Big Rip realization, and finally we see how the
Cardy-Verlinde formula becomes generalized when viscosity is accounted for, since the
thermodynamic (emergent) approach to gravity has become increasingly popular.

Finally, in Section \ref{Section5} we summarize the obtained results and we discuss on
the advantages of viscous cosmology.

\subsection{Basic formalism}
\label{subection1}

We begin by an outline of the general relativistic theory, setting, as usual, $k_B$ and
$c$ equal to one. The formalism below is taken from Ref.~\cite{Brevik:1994cd}. We adopt
the Minkowski metric in the form $(-+++)$, and we use Latin indices to denote the
spatial coordinates from 1 to 3, and Greek indices to denote spacetime ones, acquiring
values from 0 to 3.  $U^\mu=(U^0,U^i)$ denotes the four-velocity of the cosmic
fluid, and we have $U^0=1, U^i=0$ in a local comoving frame.

With $g_{\mu\nu}$ being a general metric tensor we introduce the projection tensor
\begin{equation}
h_{\mu\nu}=g_{\mu\nu}+U_\mu U_\nu, \label{1.1}
\end{equation}
and the rotation tensor
\begin{equation}
\omega_{\mu\nu}=h_\mu^\alpha h_\nu^\beta
U_{(\alpha;\beta)}=\frac{1}{2}(U_{\mu;\alpha}h_\nu^\alpha -
U_{\nu;\alpha}h_\mu^\alpha ). \label{1.2}
\end{equation}
The expansion tensor is
\begin{equation}
\theta_{\mu\nu}=h_\mu^\alpha h_\nu^\beta U_{(\alpha;\beta)} = \frac{1}{2}(U_{\mu;
\alpha}h_\nu^\alpha+U_{\nu; \alpha}h_\mu^\alpha ), \label{1.3}
\end{equation}
and has the trace $\theta \equiv \theta_\mu^\mu = {U^\mu}_{;\mu}$. The third tensor that
we shall
introduce is the shear tensor, namely
\begin{equation}
\sigma_{\mu\nu}=\theta_{\mu\nu}-\frac{1}{3}h_{\mu\nu}\theta, \label{1.4}
\end{equation}
which satisfies $\sigma_\mu^\mu=0$. Finally, it is often useful to make use of the three
tensors above in the following decomposition
of the covariant derivative of the fluid velocity:
\begin{equation}
U_{\mu;\nu}   =\omega_{\mu\nu}+\sigma_{\mu\nu}+\frac{1}{3}h_{\mu\nu}\theta - A_\mu U_\nu,
\label{1.5}
\end{equation}
where $A_\mu$ stands for the four-acceleration, namely $A_\mu={\dot{U}}_\mu=U^\nu
U_{\mu;\nu}$.

The above formalism is for a general geometry. In the following we will focus on
Friedmann-Robertson-Walker (FRW) geometry, which is of main interest in cosmology, whose
line element is
\begin{equation}
ds^2= -dt^2+a^2(t)\left[ \frac{dr^2}{1-kr^2}+r^2(d\theta^2+\sin^2 \theta
d\varphi^2)
\right], \label{1.6}
\end{equation}
where $a(t)$ is the scale factor and $k=1,0,-1$ the spatial curvature parameter. In this
case the coordinates $x^\mu$ are numerated as $(t,r,\theta,\varphi)$. In these coordinates
the covariant derivatives of the velocity acquire the simple form
\begin{equation}
 U_{\mu;\nu}    =Hh_{\mu\nu}, \label{1.7}
 \end{equation}
 with $H=\dot{a}/a$ the Hubble parameter. The rotation tensor, the shear tensor,
and the four-acceleration all vanish, i.e
 \begin{equation}
 \omega_{\mu\nu}=\sigma_{\mu\nu}=0,\quad A_\mu=0, \label{1.8}
 \end{equation}
and the relation between scalar expansion and Hubble parameter is simply
\begin{equation}
\theta=3H. \label{1.9}
\end{equation}

As a next step we consider the fluid's energy-momentum tensor $T_{\mu\nu}$ in the case
where viscosity as well as heat conduction are taken into account. If $K$ is
the thermal conductivity, considered in its nonrelativistic
framework, then for the spacelike heat flux density four-vector we have the expression
\begin{equation}
Q^\mu=-K h^{\mu\nu}(T_{,\nu}+TA_\nu), \label{1.10}
\end{equation}
with $T$ the temperature. The last term in this expression is of relativistic origin.
The coordinates used in (\ref{1.1}) are comoving, with freely moving reference particles
having vanishing four-acceleration. Thus, one obtains the usual expression
$Q_{\hat{i}}=-K T_{,\hat{i}}$ for
the heat flux density through a surface orthogonal to the unit vector
${\bf{e}}_{\hat{i}}$. Hence, assembling everything, in an FRW metric we can now introduce
the energy-momentum tensor as
\begin{equation}
T_{\mu\nu}=\rho U_\mu U_\nu+(p-3H\zeta)h_{\mu\nu}-2\eta \sigma_{\mu\nu}+Q_\mu U_\nu+Q_\nu
U_\mu, \label{1.11}
\end{equation}
with $\rho$ and $p$ the fluid's energy density and pressure respectively, and where
$\zeta$ is the bulk viscosity and $\eta$ the shear viscosity.

Taking all the above into consideration, we conclude that for a universe governed by
General Relativity  in the presence of a viscous fluid, in FRW geometry the two Friedmann
equations read as:
\begin{equation}
H^2+\frac{k}{a^2}=\frac{\kappa\rho}{3}
\label{fr1general}
\end{equation}
\begin{equation}
2 \dot{H}+3H^2=-\kappa p\,,
\label{fr2general}
\end{equation}
with $\kappa$ the gravitational constant. Note that these equations give
\begin{equation}
\dot{H}=-(\kappa/2)(\rho+p)
\end{equation}
for a flat universe. We mention that the energy density and pressure
can acquire a quite general form. For instance, a quite general parametrization of an
inhomogeneous viscous fluid in FRW geometry is
\cite{Capozziello:2005pa,Nojiri:2005sr,Nojiri:2006zh}
 \begin{equation}
p=w(\rho)\rho-B(a(t),H, \dot{H}...)\,,
\label{visceosinh00}
\end{equation}
where  $w(\rho)$  can depend on the energy density, and the bulk
viscosity  $B(a(t),H, \dot{H}...)$  can be a function of the scale
factor, and of the Hubble function and its derivatives. A usual subclass of the above
general equation of state is to assume that
\begin{equation}
B(a(t),H, \dot{H}...)=3 H\zeta(H)\,,
\label{eq.statebulk00}
\end{equation}
with $\zeta(H)>0$  the bulk viscosity,  which can be further  simplified to the subclass
with $\zeta(H)=\zeta=const.$.

Let us now focus on thermodynamics, and especially on the production of entropy. The
simplest way of extracting the relativistic formulae is to generalize the known formalism
from nonrelativistic  thermodynamics. We use $\sigma$ to denote the dimensionless
entropy per particle, where for definiteness as ``particle'' we mean a baryon. The
nonrelativistic entropy density thus becomes  $nk_B\sigma$,  where $n$ is the baryon
number density. Making use of the
relationship \cite{landau87}
 \begin{equation}
 \frac{dS}{dt}=\frac{2\eta}{T}(\theta_{ik}-\frac{1}{3}\delta_{ik}{\bf \nabla \cdot
u})^2+\frac{\zeta}{T}{\bf (\nabla \cdot u)}^2+\frac{K}{T^2}(\nabla T)^2, \label{1.12}
\end{equation}
where $\bf u$ denotes the nonrelativistic velocity and $\nabla$ the three-dimensional
Laplace operator, we can  generalize to a relativistic formalism
simply by imposing the effective substitutions
\begin{equation}
\theta_{ik} \rightarrow \theta_{\mu\nu}, \quad \delta_{ik} \rightarrow h_{\mu\nu}, \quad
{\bf \nabla \cdot u}\rightarrow 3H, \quad -K T_{,k} \rightarrow Q_\mu, \label{1.13}
\end{equation}
whereby we obtain the desired equation
\begin{equation}
{S^\mu}_{;\mu}=\frac{2\eta}{T}\sigma_{\mu\nu}\sigma^{\mu\nu}+\frac{9\zeta}{T}H^2+\frac{1}{
K
T^2}Q_\mu Q^\mu, \label{1.14}
\end{equation}
in which $S^\mu$ denotes the entropy current four-vector
\begin{equation}
S^\mu=nk_B\sigma U^\mu +\frac{1}{T}Q^\mu. \label{1.15}
\end{equation}
More detailed derivations of these results can be found, for instance, in
Refs. \cite{Weinberg:1971mx} and \cite{taub78}.

In summary, viscous cosmology is governed by the Friedmann equations (\ref{fr1general})
and (\ref{fr2general}), along with various considerations of the fluid's equation of
state. Hence, these relations will be the starting point of the discussion of this
manuscript. In the following sections we investigate viscous cosmology in detail.

\section{Inflation}
\label{Section2}

We start the investigation of viscous cosmology by focusing on early times, and in
particular on the inflationary realization. Inflation is considered to be a crucial
part of the universe cosmological history, since it can offer a solution to the flatness,
horizon and monopole problems \cite{Guthinflation,Linde:1981mu,Albrecht:1982wi}. In
order to obtain the inflationary phase one needs to consider a suitable mechanism, which
is either a scalar field in the framework of General Relativity
\cite{Linde:1983gd,Olive:1989nu,Bartolo:2004if}, or a degree of freedom arising from
gravitational modification \cite{Nojiri:2003ft,Capozziello:2011et}. In this section we
will see how inflation can be driven by a viscous fluid.

\subsection{Inflation: The basics}
\label{interinfa}

Before proceeding to the investigation of viscous inflation, let us briefly describe the
basic inflationary formulation and the relation to various observables. For convenience
we  review the scenarios of cold and warm inflation separately.

\begin{itemize}

\item{Cold Inflation}

We first start with the standard inflation realization, also called as ``cold'' inflation,
in which a scalar field $\phi$ plays the role of the inflaton field.  The
Friedmann equations are
\begin{equation}
  \label{2.1}
  H^2=\frac{\kappa}3\rho=\frac{\kappa}{3}\left( \frac {1}{2} \dot{\phi}^2+ V\right),
\end{equation}
\begin{equation}
\frac{\ddot a}{a}=- \frac{\kappa}{6} (\rho + 3p), \label{2.5}
\end{equation}
where $\rho$ and $p$ are respectively the energy density and pressure  of the inflaton
field, and
$V=V(\phi)$ is the corresponding potential. In (\ref{2.1}) we have used the fact that the
scalar field can be viewed as a perfect fluid with
  \begin{subequations}\label{eq:2.6}
    \begin{equation}
      \rho=\frac 12 \dot \phi^2+ V,
      \end{equation}
      \begin{equation}
        p=\frac 12 \dot \phi^2 - V,
      \end{equation}
\end{subequations}
and hence its equation-of-state (EoS) parameter reads
\begin{subequations}
\label{2.7}
  \begin{equation}
    p=w\rho,
  \end{equation}
  with
  \begin{equation}
    w= \frac{\tfrac 12\dot\phi^2- V}{\tfrac 12 \dot \phi^2+ V}. \label{2.8}
  \end{equation}
\end{subequations}
The fluid interpolates between an invariant vacuum energy  with $w=-1$ for a constant
inflaton field, and a stiff (Zel'dovich) fluid with $w=1$ and $V=0$.

The scalar-field equation of motion takes the
simple form
\begin{equation}
  \label{2.4}
  \ddot{\phi} + 3 H \dot \phi = - V',
\end{equation}
where $V'=dV/d \phi$, which can be re-written as
a  continuity equation
\begin{equation}\label{eq:2.2}
  \dot\rho + 3H(\rho+p)=0.
\end{equation}
Finally, we can define the quantity $N$,  i.e. the number of e-folds in the slow-roll
era,  as
the
logarithm of the ratio between the final value $a_f$ of the scale factor during inflation
and the initial value
$a(N)=a$, namely
\begin{equation}
  \label{2.26}
  N=\ln \left(\frac{a_f}{a}\right).
\end{equation}

In  inflationary theory it proves very convenient to define the so-called slow roll
parameters. One set of such parameters is defined via derivatives of the potential with
respect to the inflaton field. These ``potential'' slow roll parameters, conventionally
called $\varepsilon$, $\eta$, $\xi$, are defined as \cite{Martin:2013tda}
\begin{subequations}
\label{2.16}
\begin{align}
\label{2.16a}
\varepsilon&= \frac {1}{2\kappa} \left(\frac{V'}{V}
\right)^2,\\
\label{2.16b}
  \eta&= \frac 1\kappa \frac{V''}{V},\\
  \label{2.16c}
  \xi&=
   \frac{1}{\kappa^2}\frac{V' V'''}{V^2}.
\end{align}
\end{subequations}
Since these  should be small during the slow-roll period, the potential $V(\phi)$
must have a flat region.

One may also define the slow roll parameters in a different way, by taking the
derivatives of the Hubble parameter with respect to the  e-folding number (such an
approach has a more general application, since it can be also used in inflationary
realizations that are driven from modified gravity, where a  field and a potential
are absent) \cite{Martin:2013tda}. In particular,  these horizon-flow
\cite{liddle04,leach02,
schwarz04}
 parameters
$\epsilon_n$  (with $n$ a positive integer), are defined as
\begin{equation}
\epsilon_{n+1}\equiv \frac{d\ln |\epsilon_n|}{dN},
\end{equation}
with $\epsilon_0\equiv H_{ini}/H$ and $N$  the e-folding number,   and $H_{ini}$ the
Hubble parameter   at the beginning of inflation (inflation ends when $\epsilon_1=1$).
 Thus, the first three of them    are  calculated as
\begin{eqnarray}
\label{2.19a}
&&
\epsilon_1\equiv-\frac{\dot{H}}{H^2},
\\
&&
\epsilon_2 \equiv  \frac{\ddot{H}}{H\dot{H}}-\frac{2\dot{H}}{H^2},
\label{2.19b}\\
&&
\epsilon_3 \equiv
\left(\ddot{H}H-2\dot{H}^2\right)^{-1}\left[\dddot{H}- \frac{\ddot{H}^2}{\dot{H}} -
3\frac{\ddot{H}\dot{H}}{H} +
4\frac{\dot{H}^3}{H^2} \right].
\label{2.19c}
\end{eqnarray}

We now briefly review the formalism that is used to describe the temperature
fluctuations in the Cosmic Microwave Background (CMB) radiation. The power spectra of
scalar and tensor fluctuations are written as
\cite{Kinney:2003xf}
\begin{eqnarray}
\label{2.30}
  &&  P_s=A_s(k_*)\left( \frac{k}{k_*}\right)^{n_s- 1+ (1/2) \alpha_s\ln (k/k_*)}, \\
    &&P_T =A_T(k_*)\left(\frac{k}{k_*}\right)^{n_T+ (1/2) \alpha_T\ln(k/k_*)},
    \end{eqnarray}
    with
    \begin{eqnarray}
  &&  A_s=\frac{V}{24\pi^2\varepsilon M_p^4}=\left(\frac{H^2}{2\pi\dot\phi}\right)^2,
\\
  && A_T=\frac{2V}{3\pi^2 M_p^4}=\varepsilon\left(
\frac{2H^2}{\pi\dot\phi}\right)^2. \label{2.31}
   \end{eqnarray}
Here $k$ is the wave number of the perturbation, and $k_*$ is  a reference scale usually
chosen as the wave number at horizon crossing (the pivot scale). Often one  chooses
$k=\dot a= aH$, with $a$  the scale factor. The quantities $A_s$ and $A_T$ are amplitudes
at the
pivot scale, while $n_s$ and $n_T$ are called the  spectral indices of  scalar and
tensor fluctuations. Moreover, $-\delta_{ns}=n_s-1$ and
$n_T$ are called the tilts of the power spectrum, since they describe deviations from the
scale
invariant spectrum where
$\delta_{ns}=n_T=0$. The factors $\alpha_s$  and $\alpha_T$ are called  running
spectral
indices and are defined by
\begin{equation}
  \label{2.32}
    \alpha_s = \frac{dn_s}{d\ln k}, \quad
     \alpha_T= \frac{dn_T}{d\ln k}.
\end{equation}
Finally, the  tensor-to-scalar ratio $r$ is defined as
\begin{equation}
  \label{2.33}
  r =\frac{P_T(k_*)}{P_s(k_*)}=\frac{A_T}{A_s}.
\end{equation}

Analysis of the observations from the Planck satellite
give the result $n_s=0.
968(6) \pm 0.006$ \cite{Ade:2015tva,Ade:2015xua}. Furthermore,
the observations give
$\alpha_s= -0.
003 \pm 0.007$. The tilt of the curvature fluctuations is  $\delta_{ns}=0.032$.
  The combined  BICEP2/Planck and LIGO data
 give $n_T={-0.76}^{+1.37}_{-0.52}$  \cite{Huang:2015gka}, while  the BICEP/Planck data
alone constrain the tensor tilt
to be $n_T={0.66}^{+1.83}_{-1.44}$.

From the above equations we derive
\begin{equation}
  \label{2.35}
 \delta_{ns}=-\left[\frac{d\ln P_s (k)}{d\ln k}\right]_{k=aH}, \quad
    n_T=-\left[\frac{d\ln P_T (k)}{d\ln k}\right]_{k=aH},
\end{equation}
where the quantities  are evaluated at the horizon
crossing ($k=k_*$), and as we mentioned $k=aH$.
Hence, we can finally extract the expressions that relate the inflationary observables,
namely the tensor-to-scalar ratio, the scalar spectral index, the running of the
scalar spectral index, and the tensor spectral index, with the potential-related
slow-roll parameters    (\ref{2.19a})-(\ref{2.19c}), which read as
\cite{Martin:2013tda}:
\begin{eqnarray}
 r &\approx&16\epsilon ,
 \label{obsrv11}\\
 \delta_{ns}&\approx& 6\varepsilon -2\eta,
\label{obsrv22} \\
\alpha_s &\approx& 16\epsilon\eta-24\epsilon^2-2\xi^2 ,
\label{obsrv33}
\\
 n_\mathrm{T} &\approx& -2\epsilon .
\label{obsrv44}
\end{eqnarray}
Hence, a consistency relation between $r$ and $n_T$ follows from Eqs.~(\ref{2.30}),
(\ref{2.31}) and (\ref{2.35}), namely
$  n_T=-\frac{r}{8}$.
  The preferred BICEP2/Planck  value of $r=0.05$ then gives
$n_T = -0.006$.

Lastly, when the horizon flow  slow-roll parameters   are used,
the inflationary observables read as \cite{Martin:2013tda}
 \begin{eqnarray}
 r &\approx&16\epsilon_1 ,
 \label{rhubble11}\\
 \delta_{ns}&\approx&2(\varepsilon_1+\varepsilon_2),
\label{rhubble22} \\
\alpha_s &\approx& -2 \epsilon_1\epsilon_2-\epsilon_2\epsilon_3  ,
\label{rhubble33}\\
 n_\mathrm{T} &\approx& -2\epsilon_1  .
\label{rhubble44}
\end{eqnarray}
Definitely, in cases where both the potential slow-roll parameters and the horizon flow
slow-roll parameters can be used, the final expressions for the observables coincide.

\item{Warm Inflation}

Let us now briefly review the scenario of ``warm'' inflation. Usually, one is concerned
with cold inflationary models described above, for which dissipation arising
from the decay of inflaton energy to radiation is omitted. Nevertheless, this contrasts
the characteristic feature of the so-called warm inflation, where dissipation is
included as an important factor, and inflaton energy dissipates into heat
\cite{Berera:2008ar,campo11,Bartrum:2013oka,Panotopoulos:2015qwa}.
This implies in turn that the inflationary period lasts longer than it does in the cold
case. Additionally, no reheating at the end of the inflationary era is needed, and the
transition to   radiation era becomes a smooth one.

The main characteristic for the warm inflationary models is that the inflaton field
energy $\rho_\phi$ is considered to depend on the temperature $T$
\cite{Gron:2016zhz}, in a same way as the radiation
density $\rho_r$ depends on $T$. The first Friedmann equation writes as
\begin{equation}
  \label{2.49}
  H^2=\frac{\kappa}{3}(\rho_\phi+\rho_r),
\end{equation}
and the continuity equations for the two fluid components read
\begin{eqnarray}
  \label{2.50}
  &&  \dot \rho_\phi+3H(\rho_\phi+p_\phi)=-\Gamma\dot\phi^2,\\
   && \dot\rho_r+4H\rho_r=\Gamma\dot\phi^2,
    \label{2.50b}
\end{eqnarray}
where $\Gamma$ is a dissipation coefficient describing
the transfer of dark energy  into radiation and it is in general time dependent.
In warm inflation the inflaton energy is the dominating component,
$\rho_\phi\gg\rho_r$, and $H$, $\phi$ and $\Gamma$ vary slowly such that
$\ddot{\phi}\ll H\dot\phi$, $\dot\rho_r\ll 4H\rho_r$ and
$\dot\rho_r\ll\Gamma\dot\phi^2$. In the slow roll epoch, the radiation is produced by
dark energy dissipation. Thus
\begin{eqnarray}
  \label{2.51}
 &&   3H^2=\kappa\rho_\phi=\kappa V,\\
  &&  (3H+\Gamma)\dot\phi=-V'.
\end{eqnarray}
Defining the so-called dissipative ratio by
\begin{equation}
 \label{2.52}
  Q=\frac{\Gamma}{3H},
\end{equation}
we see that in the warm inflation era  Eq. (\ref{2.50b}) yields
\begin{equation}
  \label{2.53}
  \rho_r =\frac {3}{4} Q\dot\phi^2.
\end{equation}
During warm inflation $T>H$ (in geometric units), and it turns out that the
tensor-to-scalar ratio is modified in comparison to the cold inflation case, namely
\cite{BasteroGil:2009ec}
\begin{equation}
  \label{2.54}
  r=\frac{H/T}{(1+Q)^{5/2}}r,
\end{equation}
and thus this ratio is suppressed by a factor
$(T/H)(1+Q)^{5/2}$ compared to the  cold inflationary case.

The slow roll parameters in the present models are calculated at the beginning $t=t_i$
of the slow roll epoch. From the definition equation (\ref{2.16}) we acquire
\begin{equation}
  \label{2.55}
  \varepsilon=-(1+Q)\frac{\dot H}{H^2}.
\end{equation}
Comparing with (\ref{2.19a})   we see that the first  slow-roll parameter of the warm
inflation scenario is modified with the factor
$1+Q$ relative to the corresponding cold inflation parameter. Furthermore, manipulation
of the above equations then yields for the parameter $\eta$
\begin{equation}
  \label{2.55}
  \eta=\frac{Q}{1+Q}\frac 1\kappa\frac{\Gamma'V'}{\Gamma V}-
\frac{1+Q}{H}\frac{\ddot\phi}{\dot \phi}-\frac {\dot H}{H^2}.
\end{equation}
For convenience we introduce the quantity  $  \beta=
\Gamma'V'/(\kappa\Gamma V)$, and therefore this quantity appears in the expression
for the relative rate of change of the
radiation energy density, namely
\begin{equation}
  \label{2.58}
  \frac{\dot\rho_r}{H\rho_r}=-\frac{1}{1+Q}\left(
2\eta-\beta-\varepsilon+2\frac{\beta-\varepsilon}{
1+Q}\right).
\end{equation}
Introducing also
\begin{equation}
  \label{2.59}
  \omega=\frac{T}{H}\frac{2\sqrt 3 \pi Q}{\sqrt{3+4\pi Q}},
\end{equation}
one can find that \cite{Visinelli:2016rhn}
\begin{eqnarray}
  \label{2.60}
 && \delta_{ns}=\frac{1}{1+Q}
  \left\{4\varepsilon -2
\left(\eta-\beta+\frac{\beta-\varepsilon}{1+Q}\right)\right.\nonumber\\
&&
\ \ \ \ \ \ \ \,
\left.
+\frac{\omega}{1+\omega}\left[
\frac{2\eta+\beta-7\varepsilon}{4} + \frac{6+(3+4\pi)Q}{(1+Q)(
3+4\pi Q)}(\beta-\varepsilon)\right] \right\}.
\end{eqnarray}
When    warm inflation is strong, $Q\gg 1$ , $\omega\gg 1$, and thus
\begin{equation}
 \label{2.62}
  \delta_{ns}= \frac{2}{2Q}\left[\frac 32 (\varepsilon+\beta)-\eta\right],
\end{equation}
whereas when it is weak, $Q\ll 1$, and therefore
\begin{equation}
  \label{2.62}
  \delta_{ns}=2(3\varepsilon-\eta)-\frac{\omega/4}{1+\omega}(15\epsilon-2\eta-9 \beta).
\end{equation}
Finally, the  cold inflationary case  corresponds to the limit  $Q\to 0$ and $T\ll H$,
and then
$\omega\to 0$ and
\begin{equation}
  \label{2.61}
  \delta_{ns}\to2(3\varepsilon-\eta).
\end{equation}

Visinelli  found the following expression for  tensor-to-scalar ratio in warm inflation
\cite{Visinelli:2016rhn}:
\begin{equation}
  \label{2.63}
  r=\frac{16\varepsilon}{(1+Q)^2(1+\omega)}.
\end{equation}
Hence,   in the limit of  strong dissipative  warm inflation we have
\begin{equation}
  \label{2.65}
  r\to\frac{16}{Q^2\omega}\varepsilon\ll\varepsilon,
\end{equation}
while in  the limit of cold inflation we re-obtain the standard result  (\ref{obsrv11}),
namely $  r\to 16\varepsilon$. Thus, the warm inflation models with $Q \gg 1$ and
$\omega\gg 1$ yield a very small tensor-to-scalar ratio.

\item{Intermediate inflation}

Intermediate inflation scenario, introduced by  Barrow in 1990 \cite{Barrow:1990vx} (see
also \cite{Mohammadi:2015jka,Rezazadeh:2014fwa}), also uses a scalar field. We consider
the
scale factor to take the form
\begin{equation}
  \label{2.39}
  a(t)=\exp[ A(\hat{t}^\alpha -1)]',
\end{equation}
 with $ 0<\alpha<1$, and where $A$ is a positive dimensionless
constant, while $a_p$ refers to the
Planck time ($ \hat{t}=t/\sqrt{\kappa},\ t_p=\sqrt{\kappa}$). The reason that these
models are called intermediate, is that
the expansion is faster than the corresponding one in power-law inflation,  and
slower than an exponential inflation (the latter corresponding to
$\alpha=1$). It follows from  (\ref{2.39}) that
\begin{equation}
  \label{2.40}
  H=\frac{A\alpha}{t_s}\hat{t}^{\alpha-1}, \quad \dot{H}=
\frac{A\alpha}{t_p^2}(\alpha-1)\hat{t}^{\alpha-2},
 \end{equation}
and since $\dot H<0$ for $\alpha<1$, the Hubble parameter
decreases with  time. Inserting these equations into Eqs.~(\ref{2.1}) and (\ref{2.5}) we
obtain
\begin{equation}
  \label{2.41}
\rho=\frac{3A^2\alpha^2}{t_p^4}\hat{t}^{2\alpha-1}, \quad
    p= \frac{A \alpha}{t_p^4}\hat{t}^{\alpha-2}[2(1-\alpha)-3\alpha At_p^\alpha].
\end{equation}
Since $\rho+p=\dot \phi^2$ we obtain by integration, using the initial condition
$\phi(0)=0$, that
\begin{equation}
  \label{2.42}
  \phi(t)=\frac{2}{t_p}\sqrt{2A\frac{1-\alpha}{\alpha}}\hat{t}^{\frac{\alpha
}{2}},
\end{equation}
while since $V=\frac{1}{2}(\rho-p)$ we acquire
\begin{equation}
  \label{2.43}
  V(t) = \frac{A\alpha}{t_p^4}\hat{t}^{\alpha-2}[3A\alpha t_p^{-\alpha}-2(1-\alpha)].
\end{equation}
Hence, eliminating $t$ between (\ref{2.42}) and (\ref{2.43}) we can express the potential
as a function of the inflaton field:
\begin{equation}
  \label{2.44}
  V(\phi)
=\frac{A\alpha}{t_p^4}\left[\frac{\alpha}{2A(1-\alpha)}\right]^{\frac{\alpha-2}{\alpha}
} \left(\frac{t_p\phi}
{2}\right)^{\frac {2(\alpha-2)}\alpha}  \left[\frac {3\alpha^2}{2(1-\alpha)}\left(
\frac {t_p\phi}
  {2}\right)^2-2(1-\alpha) \right].
\end{equation}

For this class of models the spectral parameters are most easily calculated from the
Hubble slow
roll parameters
\begin{equation}
\varepsilon_H = -\frac{\dot{H}}{H^2}, \quad \eta_H=-\frac{1}{2}\frac{\ddot{H}}{\dot{H}H}.
\label{G2.
52}
\end{equation}
The optical parameters $\delta_{ns},\, n_r$ and $r$ can  be expressed in
terms of
the Hubble slow roll parameters  to lowest order as
\begin{equation}
\delta_{ns}=2(2\varepsilon_H-\eta_H), \quad n_r=-2\varepsilon_H, \quad r=16\varepsilon_H.
\label{G2.530000}
\end{equation}
This gives
\begin{equation}
\varepsilon_H =\frac{1-\alpha}{A\alpha}\hat{t}^{-\alpha}, \quad
\eta_H=\frac{2-\alpha}{2(1-\alpha)}\varepsilon_H. \label{G2.53}
\end{equation}
The slow roll parameter $\varepsilon_H$ can be expressed in terms of the inflaton field as
\begin{equation}
\varepsilon_H=8\left( \frac{1-\alpha}{\alpha}\right)^2\left( \frac{M_P}{\phi}\right)^2.
\label{G2.53e }
\end{equation}
In the intermediate inflation, the e-folding number becomes
\begin{equation}
N=A(\hat{t}_f^\alpha -\hat{t}_i^\alpha ), \label{G2.54}
\end{equation}
where $\hat{t}_i$ and $\hat{t}_f$ are the initial and final point of time of the
inflationary era,
respectively. In these models the beginning of the inflationary era is defined by the
condition $\varepsilon_H(\hat{t}_i)=1$, giving
\begin{equation}
t_i=\left( \frac{1-\alpha}{A\alpha}\right)^{1/\alpha}\, t_p. \label{G2.55}
\end{equation}
Hence, the inflationary era ends at a point of time
\begin{equation}
t_f=\left(\frac{N\alpha+1-\alpha}{A\alpha}\right)^{1/\alpha}\,t_p. \label{G2.56}
\end{equation}
The slow roll parameters are evaluated at this point of time, giving
\begin{equation}
\varepsilon_H=\frac{1-\alpha}{N\alpha+1-\alpha}, \quad
\eta_H=\frac{2-\alpha}{2(N\alpha+1-\alpha)}.
\label{G2.57}
\end{equation}
Inserting the above expressions into (\ref{G2.530000}) we can thus write
\begin{equation}
\delta_{ns}\equiv 1-n_s=\frac{2-3\alpha}{N\alpha+1-\alpha}, \quad
n_r=\frac{2(\alpha-1)}{N\alpha+1-\alpha}, \quad r=\frac{16(1-\alpha)}{N\alpha+1-\alpha}.
\label{G2.59}
\end{equation}
Note that the curvature spectrum is scale independent, corresponding to $n_s=1$, for
$\alpha=2/3$.
Furthermore, $n_s<1$ requires $\alpha<2/3$. Note that the expression for $n_s$ corrects
an error of Ref.~\cite{Mohammadi:2015jka}. For these models the $r,\, \delta_{ns}$
relation becomes
\begin{equation}
r=\frac{16(1-\alpha)}{2-3\alpha}\, \delta_{ns}. \label{G2.60}
\end{equation}
The constant $\alpha$ can be expressed in terms of $N$ and $\delta_{ns}$ as
\begin{equation}
\alpha=\frac{2-\delta_{ns}}{3+(N-1)\delta_{ns}} \approx \frac{2}{3+N\delta_{ns}}.
\label{G2.61}
\end{equation}
With the Planck values $\delta_{ns}=0.032$ and $N=60$ we get $\alpha=0.4$ giving $r=0.38$.
This value of $r$ is larger than permitted by   Planck observations. However, the more
general models with non-canonical inflaton fields studied in
Refs.~\cite{Mohammadi:2015jka} and \cite{Rezazadeh:2014fwa}, contain an adjustable
parameter in the expressions for the observables, leading to
agreement with observational data. Below we shall consider warm intermediate
inflation models, which lead naturally to   a suppression of the curvature perturbation,
resulting to  a small
value of $r$.

\end{itemize}

\subsection{Viscous Inflation}
\label{viscinfl}

Having described the basics of inflation, in this subsection we will see how
inflation can be realized in the framework of viscous cosmology, that is if instead of a
scalar field inflation is driven by a viscous fluid \cite{Bamba:2015sxa}. We start from
the two Friedmann equations  (\ref{fr1general})
and (\ref{fr2general}), namely
\begin{equation}
H^2+\frac{k}{a^2}=\frac{\kappa\rho}{3}
\label{fr1general22}
\end{equation}
\begin{equation}
2 \dot{H}+3H^2=-\kappa p\,.
\label{fr2general22}
\end{equation}
Concerning the viscosity of the fluid we consider a subclass of (\ref{visceosinh00})
 and   parametrize the equation of state
as
\begin{equation}
p = -\rho + A \rho^{\beta}+\zeta(H) \,,
\label{eq:2.8newsum}
\end{equation}
with $A$,$\beta$ constants, and $\zeta(H)$ the bulk viscosity considered with a dynamical
nature in general, i.e. being a function of the Hubble parameter. As a specific example
we consider
\begin{equation}
\zeta(H) = \bar{\zeta} H^{\gamma} \,,
\label{zetavisousinfl}
\end{equation}
with $\bar{\zeta}$, $\gamma$ parameters.

 From the Friedmann equation (\ref{fr1general22}) and for an expanding flat universe
($H>0, k=0$),
we
acquire
\begin{equation}
H=\sqrt{\frac{\kappa \rho}{3}} .
\label{Hubbleeq:2.5newsum}
\end{equation}
Therefore, $\zeta(H)$ can be expressed in terms of
$\rho$, i.e $\zeta(H) = \zeta(H(\rho))$. Thus,    comparing  the general expression
for the EoS of a fluid, namely
\begin{equation}
p = - \rho + f(\rho) \,,
\label{eq:2.5newsum}
\end{equation}
  with (\ref{eq:2.8newsum}) and  (\ref{zetavisousinfl}), we
deduce that
\begin{equation}
f(\rho) = A \rho^{\beta}+\zeta(H(\rho))
= A \rho^{\beta} + \bar{\zeta} \left(\sqrt{\frac{\kappa}{3}} \right)^{\gamma}
\rho^{\gamma/2} \,.
\label{eq:2.10newsum}
\end{equation}
We mention that  $f(\rho)$ is expressed as a series of  powers in $\rho$ due to the
imposed assumption that $\zeta(H)$ is a power of  $H$. Hence, this
allows us to find analytical solutions  and examine the behavior of various inflationary
observables.

Since in fluid inflation we do not have a potential, it proves convenient to use the
Hubble slow-roll parameters. Inserting the Hubble function
from (\ref{Hubbleeq:2.5newsum}) into
(\ref{2.19a})-(\ref{2.19c}) and then into the inflationary observables
(\ref{rhubble11})-(\ref{rhubble44}), after some algebra one can express the tilt,
 the tensor-to-scalar ratio
and the running spectral index as  \cite{Bamba:2015sxa}
 \begin{eqnarray}
(\delta_{ns}, r, \alpha_s) \approx
(6\frac{f(\rho)}{\rho(N)}\,, \,
24 \frac{f(\rho)}{\rho(N)}\,, \, -9 \left(\frac{f(\rho)}{\rho(N)}\right)^2)
\label{eq:2.6newsum} \\
= (6\left(w(N) +1\right)\,, \, 24\left(w(N) +1\right)\,, \, -9\left(w(N)
+1\right)^2)\,,
\label{eq:2.7newsum}
\end{eqnarray}
where  we have also used that $f(\rho)/\rho(N) = w(N) +1$. In these expressions all
quantities may be considered as functions of the e-folding number $N$.
Hence,  if we choose $f(\rho)/\rho(N) = 4.35 \times 10^{-3}$, we obtain $w = -0.996$, and
thus  $(n_{\mathrm{s}}, r, \alpha_s) =
(0.974, 0.104, -1.70 \times 10^{-4})$.
These results are consistent with the Planck data, namely $n_{\mathrm{s}} = 0.968 \pm
0.006\, (68\%\,\mathrm{CL})$,
$r< 0.11\, (95\%\,\mathrm{CL})$,
and $\alpha_s = -0.003 \pm 0.007\, (68\%\,\mathrm{CL})$,
  \cite{Ade:2015xua,Ade:2015lrj}.

Let us now use the required scalar spectral index in order to reconstruct the EoS of the
fluid through a corresponding effective potential, following
\cite{Chiba:2015zpa,Bamba:2015sxa}. In order to achieve this, we first express the
Friedmann equations using derivatives in terms of the
e-folding number $N$ as
\begin{eqnarray}
&&
\frac{3}{\kappa} \left[H (N)\right]^2 = \rho \,,
\label{eq:FR16-3-IIB1newsum} \\
&& - \frac{2}{\kappa } H(N) H'(N) = \rho + p \,,
\label{eq:FR16-3-IIB2newsum}
\end{eqnarray}
and similarly for the
slow-roll parameters (\ref{2.16a})-(\ref{2.16c}), namely
\begin{eqnarray}
&&\delta_{ns} = -\frac{d}{d N} \left[ \ln \left( \frac{1}{V^2(N)} \frac{dV(N)}{dN}
\right) \right]
\label{eq:3.1aaa}
\nonumber\\
&&r = \frac{8}{V(N)} \frac{dV(N)}{dN}
\label{eq:3.1bbb}
\nonumber\\
&&
\alpha_s = - \frac{d^2}{d N^2} \left[ \ln \left( \frac{1}{V^2(N)}
\frac{dV(N)}{dN} \right) \right] \,.
\label{eq:3.1ccc}
\end{eqnarray}
Hence, one can use these quantities in order to reconstruct the equation-of-state of the
corresponding fluid. In particular, having the  $\delta_{ns}(N)$ as a function of $N$,
using (\ref{eq:3.1aaa}) we can solve for $V(N)$, which will be the effective potential
in an equivalent scalar-field description. Then the Hubble function is related to
$V(N)$ through (\ref{eq:FR16-3-IIB1newsum}), and thus we obtain  $H=H(N)$. Finally, using
 (\ref{eq:FR16-3-IIB2newsum}) we can reconstruct $f(\rho)$ through (\ref{eq:2.5newsum}).

Let us give a specific example of the above method, in the case where
\cite{Bamba:2015sxa}
\begin{equation}
\delta_{ns} = \frac{2}{N} \,,
\label{eq:FR16-4-IIIB-1newsum}
\end{equation}
which is valid in Starobinsky inflation \cite{Starobinsky:1980te},
and it can be satisfied in chaotic inflation \cite{Linde:1983gd}, in new
Higgs inflation  \cite{Salopek:1988qh,Bezrukov:2007ep}, and in models of
$\alpha$-attractors \cite{Kallosh:2013hoa,Kallosh:2015lwa}, too.
Combining   (\ref{eq:FR16-4-IIIB-1newsum}) and (\ref{eq:3.1aaa})
 gives
\begin{equation}
V(N) = \frac{1}{\left( C_1 /N \right) + C_2} \,,
\label{eq:3.2aaa}
\end{equation}
where $C_1 (>0)$ and $C_2$ are constants. Hence, using
 (\ref{eq:3.2aaa}) and  (\ref{eq:3.1bbb}) we acquire
 \begin{equation}
r = \frac{8}{N \left[1+\left(C_2/C_1\right) N \right]} = \frac{4\delta_{ns}}{1+
\frac{C_2}{C_1}\frac{2}{\delta_{ns}}}=\frac{4\delta_{ns}^2}{\delta_{ns}+2C_2/C_1},
\label{newsumrelauz5}
\end{equation}
and thus
\begin{equation}
\frac{C_2}{C_1} = \left( 4\frac{\delta_{ns}}{r}-1\right) \frac{\delta_{ns}}{2}.
\end{equation}
If  $\delta_{ns}=0.032, \, r=0.05$ one gets $C_2/C_1 \approx 0.025$.
Finally, from (\ref{eq:3.1ccc}) we find that
\begin{equation}
\alpha_s = -\frac{2}{N^2} \,.
\label{alpharunnewsum}
\end{equation}
Thus, inserting a reasonable value  $N=60$ we obtain
$\alpha_s = -5.56 \times 10^{-4}$, in agreement with Planck analysis.

In the case of a fluid model one uses the equation-of-state parameter from
(\ref{eq:2.5newsum}) instead of the scalar potential. Hence, one can have
$\left(3/\kappa \right) \left(H (N)\right)^2 = \rho (N) \approx V (N)$, since
the last approximation arises from the slow-roll condition that the kinetic energy is
negligible comparing to the potential one. Therefore, using  (\ref{eq:3.2aaa})
we obtain
\begin{equation}
H(N) \approx
 \sqrt{\frac{\kappa}{3\left[ \left( C_1 /N \right) + C_2 \right]}} \,,
\label{eq:3.4newsum}
\end{equation}
with $\left( C_1 /N \right) + C_2 > 0$.
Additionally, inserting
$\rho \approx V$  into (\ref{eq:3.2aaa}) results to
\begin{equation}
N \approx \frac{C_1 \rho}{1-C_2 \rho}\,.
\label{eq:3.3newsum}
\end{equation}
 Thus, inserting (\ref{eq:3.4newsum}) into  (\ref{eq:FR16-3-IIB1newsum}) and
(\ref{eq:FR16-3-IIB2newsum}) gives
\begin{equation}
p = -\rho - \frac{2}{\kappa} H(N) H'(N)
\approx -\rho -\frac{3C_1}{N^2 \kappa^2} H^4 \,.
\label{eq:3.5newsum}
\end{equation}
Finally, comparing (\ref{eq:2.5newsum}) with (\ref{eq:3.5newsum}) leads to
\begin{equation}
f(\rho) \approx -\frac{3C_1}{N^2 \kappa^2} H^4
\approx -\frac{1}{3C_1} \left(1-2C_2\rho +C_2^2 \rho^2 \right) \,,
\label{eq:3.6newsum}
\end{equation}
where we have also used  (\ref{eq:FR16-3-IIB1newsum}) and
(\ref{eq:3.3newsum}).

We now focus on fluid inflationary models with $n_s$ and $r$ in agreement with
observations. From (\ref{eq:2.5newsum}) and (\ref{eq:2.10newsum}) we obtain
\begin{equation}
p = -\rho + f(\rho)
= -\rho + A \rho^{\beta}+ \bar{\zeta} \left(\sqrt{\frac{\kappa}{3}} \right)^{\gamma}
\rho^{\gamma/2}
\,.
\label{eq:3.7newsum}
\end{equation}
Therefore, we suitably choose the model parameters  $A$, $\bar{\zeta}$, $\beta$, and
$\gamma$,
in order for relation (\ref{eq:FR16-4-IIIB-1newsum}) to be satisfied. For convenience we
will focus in the regimes $\left| C_2\rho \right| \gg 1$ and $\left| C_2\rho \right| \ll
1$ separately following \cite{Bamba:2015sxa}.

\begin{itemize}

 \item {Case I: $\left| C_2\rho \right| \gg 1$}

In this case expression (\ref{eq:3.6newsum}) leads to
\begin{equation}
f(\rho) \approx \frac{2C_2}{3 C_1} \rho - \frac{C_2^2}{3 C_1} \rho^2 \,,
\label{eq:3.8newsum}
\end{equation}
with $C_2 <0$ in order to have a positive $N$ from  (\ref{eq:3.3newsum}).
From (\ref{eq:3.7newsum}) and (\ref{eq:3.8newsum})
we acquire
\begin{equation}
w = \frac{p}{\rho} \approx -1 - \frac{2}{3} \left( -\frac{C_2}{C_1} \right)
+ \frac{1}{3} \left( -\frac{C_2}{C_1} \right) \left(-C_2 \rho \right)
\approx -1 + \frac{1}{3N} \left(-2-C_2 \rho \right)\,,
\label{newauxrel1}
\end{equation}
where we have also used that $\left(-C_2 \right)/C_1 \approx 1/N$.
For instance, if $\left| C_2\rho \right| = \mathcal{O}(10)$,
$\left(-C_2 \right)/C_1 \approx 1/N$,  and  $N \gtrsim 60$, relation (\ref{newauxrel1})
leads to $w \approx -1$, and hence the de Sitter inflation can be realized, with a
scale-factor of the form
\begin{equation}
a(t) = a_\mathrm{i} \exp \left[ H_\mathrm{inf} (t-t_\mathrm{i}) \right] \,.
\label{newrel3bc}
\end{equation}
It should be noted that  for  $\left(-C_2 \right)/C_1 < 1/N$, relation
(\ref{newsumrelauz5})
for $N \gtrsim 73$ provides a tensor-to-scalar ratio $r>1$, in disagreement with
observations.

Comparing (\ref{eq:3.8newsum}) and (\ref{eq:2.10newsum}) we deduce that we obtain
equivalence for two combinations of parameters:
\begin{equation}
\mathrm{Model \ \   (A)}: \quad \quad
A = \frac{2C_2}{3 C_1}\,,
\quad
\bar{\zeta} = - \frac{3C_2^2}{C_1 \kappa^2}\,,
\quad
\beta =1\,,
\quad
\gamma=4 \,,
\label{eq:3.9newsum}
\end{equation}
and
\begin{equation}
\mathrm{Model \ \  (B)}: \quad \quad
A = -\frac{C_2^2}{3 C_1}\,,
\quad
\bar{\zeta} = \frac{2C_2}{C_1 \kappa}\,,
\quad
\beta =2\,,
\quad
\gamma=2 \,.
\label{eq:3.10newsum}
\end{equation}
Hence, the corresponding fluid equation of state can be  reconstructed.

\item {Case (II): $\left| C_2\rho \right| \ll 1$}

In this case expression (\ref{eq:3.6newsum}) leads to
\begin{equation}
f(\rho) \approx
-\frac{1}{3C_1} + \frac{2C_2}{3 C_1} \rho \,.
\label{auxrel543}
\end{equation}
Thus, (\ref{eq:3.3newsum}) with $\left| C_2\rho \right| \ll 1$
gives $C_1 \rho \approx N \gg 1$ and
thus $\left|C_2\right|/C_1 \ll 1$.
Hence, (\ref{eq:3.7newsum}) and (\ref{eq:3.8newsum}),
give
\begin{equation}
w = \frac{p}{\rho} \approx -1 - \frac{1}{3} \frac{1}{C_1 \rho}
+ \frac{2}{3} \left( \frac{C_2}{C_1} \right)
\approx -1+ \frac{1}{3} \left( -\frac{1}{N} + 2 \frac{C_2}{C_1} \right)\,,
\label{newsumaux787}
\end{equation}
where we have used that $C_1 \rho \approx N$. Similarly to the previous subcase,
 (\ref{newsumaux787}) with $1/N \ll 1$ and $\left|C_2\right|/C_1 \ll 1$,
leads to $w \approx -1$, i.e to the realization of the de Sitter inflation, with a scale
factor  given by (\ref{newrel3bc}). Moreover, for $C_2 >0$ and $C_2/C_1 \lesssim 1/N$,
and for $N \gtrsim 60$, relation (\ref{newsumrelauz5}) gives $r < 0.11$ in agreement with
Planck results. On the other hand, for $C_2 <0$ and $\left|C_2 \right|/C_1 < 1/N$,
we need to have  $N \gtrsim 73$ in order to get $r < 0.11$,
similarly to the previous Case (I).
Finally, comparing (\ref{auxrel543}) and (\ref{eq:2.10newsum}) we deduce that we
obtain equivalence for two combinations of parameters:
\begin{equation}
\mathrm{Model \ \ (C)}: \quad \quad
A = -\frac{1}{3 C_1}\,,
\quad
\bar{\zeta} = \frac{2C_2}{C_1 \kappa}\,,
\quad
\beta =0\,,
\quad
\gamma=2 \,,
\label{eq:3.12newsum}
\end{equation}
and
\begin{equation}
\!\!\!\!\!\!\!\!\!\!\!\!
\mathrm{Model \ \ (D)}: \quad \quad
A = \frac{2C_2}{3 C_1}\,,
\quad \ \ \,
\bar{\zeta} = -\frac{1}{3 C_1}\,,
\quad
\beta =1\,,
\quad
\gamma=0 \,.
\label{eq:3.13newsum}
\end{equation}

 \end{itemize}

Having analyzed the basic features of inflationary realization from a viscous fluid, let
us examine the crucial issue of obtaining a graceful exit and the
subsequent entrance to the reheating stage \cite{Bamba:2015sxa}. In particular, we will
investigate the instability of the de Sitter solution
 characterized by $H = H_\mathrm{inf}=const.$ under perturbations. One starts by
perturbing the Hubble function as \cite{Bamba:2015uxa}
\begin{equation}
H = H_\mathrm{inf} + H_\mathrm{inf} \delta(t) \,,
\label{eq:4.1newsum}
\end{equation}
where $\left| \delta(t) \right| \ll 1$.
Thus, the second Friedmann equation writes as a differential equation
in terms of the cosmic time $t$, namely
\begin{equation}
\ddot{H} -\frac{\kappa^4}{2}\left[\beta A^2 \left( \frac{3}{\kappa} \right)^{2\beta}
H^{4\beta-1}
+ \left(\beta + \frac{\gamma}{2} \right) A \bar{\zeta} \left( \frac{3}{\kappa}
\right)^{\beta} H^{
2\beta + \gamma - 1} +
\frac{\gamma}{2} \bar{\zeta}^2 H^{2\gamma - 1} \right] = 0 \,.
\label{eq:4.2newsum}
\end{equation}
Without loss of generality we choose
\begin{equation}
\delta(t) \equiv e^{\lambda t} \,,
\label{eq:4.3newsum}
\end{equation}
with $\lambda$ a constant, and therefore  a positive $\lambda$ would correspond to an
unstable de Sitter solution. This instability implies that the universe can exit from
inflation. On the other hand, a stable inflationary solution is just an eternal
inflation.

Inserting (\ref{eq:4.1newsum}) and (\ref{eq:4.3newsum})
into (\ref{eq:4.2newsum}), and keeping terms up to first order in $\delta(t)$, we obtain
\begin{equation}
\lambda^2 - \frac{1}{2} \frac{\kappa^2}{H_\mathrm{inf}^2} \mathcal{Q} = 0 \,,
\label{eq:4.4newsum}
\end{equation}
 with
\begin{eqnarray}
&&
\mathcal{Q} \equiv
\beta \left(4\beta - 1\right) A^2 \left( \frac{3}{\kappa} \right)^{2\beta}
H_\mathrm{inf}^{4\beta} + \left(\beta + \frac{\gamma}{2} \right)
\left(2\beta + \gamma - 1\right) A \bar{\zeta}
\left( \frac{3}{\kappa} \right)^{\beta} H_\mathrm{inf}^{2\beta + \gamma}\nonumber\\
&&
\ \ \ \ \ \
+
\frac{\gamma}{2} \left(2\gamma - 1\right)
\bar{\zeta}^2 H_\mathrm{inf}^{2\gamma} \,.
\label{eq:4.5newsum}
\end{eqnarray}
The solutions of (\ref{eq:4.4newsum}) read as
\begin{equation}
\lambda = \lambda_\pm \equiv \pm \frac{1}{\sqrt{2}} \frac{\kappa}{H_\mathrm{inf}}
\sqrt{\mathcal{Q}}\,,
\label{eq:4.6newsum}
\end{equation}
and therefore if $\mathcal{Q} > 0$  we may obtain
$\lambda = \lambda_+ > 0$, which implies the realization of a successful inflationary
exit.

Let us now check whether the four fluid models described in (\ref{eq:3.9newsum}),
(\ref{eq:3.10newsum}),
(\ref{eq:3.12newsum}),
and (\ref{eq:3.13newsum}) above,  can give rise to a graceful exit, i.e whether they can
give a positive
$\mathcal{Q}$ in (\ref{eq:4.5newsum}). Substituting the corresponding  values of $A$,
$\bar{\zeta}$, $\beta$, and $\gamma$ into  (\ref{eq:4.6newsum}), we obtain the
expressions of $\mathcal{Q}$ as \cite{Bamba:2015sxa}:
\begin{equation}
\mathrm{Model \ \  (A)}: \quad \quad
\mathcal{Q} = 2 \left(\frac{C_2}{C_1}\right)^2
\left(\frac{H_\mathrm{inf}}{\sqrt{\kappa}}\right)^4
\left[ 6 - 45 C_2 \left(\frac{H_\mathrm{inf}}{\sqrt{\kappa}}\right)^2
+ 63 C_2^2 \left(\frac{H_\mathrm{inf}}{\sqrt{\kappa}}\right)^4 \right] >0 \,,
\end{equation}
\begin{equation}
\mathrm{Model \ \  (B)}: \quad \quad
\mathcal{Q} = 6 \left(\frac{C_2}{C_1}\right)^2
\left(\frac{H_\mathrm{inf}}{\sqrt{\kappa}}\right)^4
\left[ 2 - 15 C_2 \left(\frac{H_\mathrm{inf}}{\sqrt{\kappa}}\right)^2
+ 21 C_2^2 \left(\frac{H_\mathrm{inf}}{\sqrt{\kappa}}\right)^4 \right] >0 \,.
\end{equation}
 \begin{equation}
 \mathrm{Model \ \ (C)}: \quad \quad
\mathcal{Q} = \left(\frac{C_2}{C_1}\right)^2
\left(\frac{H_\mathrm{inf}}{\sqrt{\kappa}}\right)^2
\left[ - \frac{1}{3C_2} + 12 \left(\frac{H_\mathrm{inf}}{\sqrt{\kappa}}\right)^2
\right] \,,
 \end{equation}
 \begin{equation}
 \mathrm{Model \ \ (D)}: \quad \quad
\mathcal{Q} = 2\left(\frac{C_2}{C_1}\right)^2
\left(\frac{H_\mathrm{inf}}{\sqrt{\kappa}}\right)^2
\left[ 6 \left(\frac{H_\mathrm{inf}}{\sqrt{\kappa}}\right)^2
- \frac{1}{3C_2} \right] \,.
 \end{equation}
Hence, Models (A) and (B) have always $\mathcal{Q} >0$. On the other hand, Models (C)
and (D) have $\mathcal{Q} >0$  for f $C_2 <0$, while for $C_2 >0$ they have $\mathcal{Q}
>0$ if
\begin{eqnarray}
C_2 > \frac{1}{36} \left(\frac{\sqrt{\kappa}}{H_\mathrm{inf}}\right)^2
\quad \quad
\mathrm{for} \,\,\,
\mathrm{Model  \ \ (C)}\,,
\label{eq:4.1newsum1} \\
C_2 > \frac{1}{18} \left(\frac{\sqrt{\kappa}}{H_\mathrm{inf}}\right)^2
\quad \quad
\mathrm{for}  \,\,\,
\mathrm{Model \ \ (D)}\,.
\label{eq:4.1newsum2}
\end{eqnarray}

In summary, we can see that the models of viscous fluid inflation can have a graceful
exit without any tuning. In Table \ref{table-2}  we  summarize   the obtained results.
 From the corresponding equation-of-state parameters, and comparing with
(\ref{eq:2.8newsum}), we can immediately see the term inspired by the bulk viscosity.
Finally, as we described in detail above, in these models the inflationary observables
are in agreement with observations. In particular,
the spectral index  from  (\ref{eq:FR16-4-IIIB-1newsum})
 is $n_s =0.967$ for $N=60$.
  The running of the spectral
index   is given by $\alpha_s = -2/N^2$  in (\ref{alpharunnewsum}), leading to
$\alpha_s = -5.56 \times 10^{-4}$.
\begin{table}[t]
\caption{The equation-of-state parameter of the reconstructed viscous inflationary models
of (\ref{eq:3.9newsum}), (\ref{eq:3.10newsum}), (\ref{eq:3.12newsum}),
and (\ref{eq:3.13newsum}), along with the conditions for a graceful exit.
The parameter  $C_1$ is always positive, while   $\left| C_2\rho \right| \gg 1$ and $C_2
<0$ for Models (A) and (B), and   $\left| C_2\rho \right| \ll 1$  for Models (C) and
(D). From \cite{Bamba:2015sxa}.}
{\small{\begin{center}
\begin{tabular}
{cccc}
\hline
\hline
Case
& Model
& EoS
& Conditions for   graceful exit
\\[0mm]
\hline
(i)
&(a)
& $p=-\rho+ \left[2C_2/\left(3 C_1\right) \right]\rho
- \left[3C_2^2/\left(C_1 \kappa^2 \right)\right] H^4$
& No condition
\rule{0mm}{3.5mm}
\\[0mm]
(i)
&(b)
& $p=-\rho -\left[ C_2^2/\left(3 C_1\right) \right]\rho^2
+ \left[2C_2/\left(C_1 \kappa \right) \right] H^2$
& No condition
\\[0mm]
(ii)
&(c)
& $p=-\rho - \left[1/\left(3 C_1\right) \right]
+ \left[2C_2/\left(C_1 \kappa \right) \right] H^2$
& $C_2 <0$ or $C_2 > \left(1/36\right)
\left(\sqrt{\kappa}/H_\mathrm{inf}\right)^2$
\\[0mm]
(ii)
&(d)
& $p=-\rho + \left[2C_2/\left(3 C_1\right) \right] \rho
- \left[1/\left(3 C_1\right) \right]$
& $C_2 <0$ or $C_2 > \left(1/18\right)
\left(\sqrt{\kappa}/H_\mathrm{inf}\right)^2$
\\[1mm]
\hline
\hline
\end{tabular}
\end{center}}}
\label{table-2}
\end{table}

We close this subsection by studying the  singular inflation in the above  viscous fluid
model. The finite-time  singularities are classified into four types
\cite{Nojiri:2005sx}, and hence one can see that Type IV singularity can be applied in
singular inflation since there are no divergences  in the scale factor and in the the
effective (i.e. total) energy density and pressure. In particular, in Type IV
singularity, as $t\to t_{\mathrm{s}}$, with $t_{\mathrm{s}}$ the singularity time, we
have   $a \to a_{\mathrm{s}}$,
$\rho \to 0$
and $\left| p \right| \to 0$.
Here, $a_{\mathrm{s}}$ is the value of $a$ at $t=t_{\mathrm{s}}$. Nevertheless,
the higher derivatives of the Hubble function diverge.

Let us consider the above viscous fluid inflationary realization, assuming that
\begin{eqnarray}
H = H_\mathrm{inf} + \bar{H}\left(t_{\mathrm{s}} -t \right)^{q} \,,
\quad q>1 \,,
\label{eq:5.1newsum}\\
a = \bar{a} \exp \left[ H_\mathrm{inf}t -\frac{\bar{H}}{q+1} \left(t_{\mathrm{s}} -t
\right)^{
q+1} \right] \,,
\label{eq:5.2newsum}
\end{eqnarray}
with $\bar{H}$, $q$, and $\bar{a}$ the model parameters. From the  two Friedmann
equations  (\ref{fr1general22}),(\ref{fr2general22}) we straightforwardly acquire
\begin{equation}
\rho = \frac{3H^2}{\kappa}\,,
\quad
p = -\frac{2\dot{H}+3H^2}{\kappa} \,.
\label{eq:5.3newsum}
\end{equation}
Therefore, a Type IV singularity appears at $t = t_{\mathrm{s}}$, since
 (\ref{eq:5.2newsum}) and
 (\ref{eq:5.3newsum}) imply that as $t\to t_{\mathrm{s}}$ the quantities $a$,
$\rho$, and $p$ asymptotically approach finite values, while from
 (\ref{eq:5.1newsum}) we deduce that higher derivatives of $H$ diverge.
From   (\ref{eq:5.1newsum}) and
 (\ref{eq:5.3newsum}) we find the following equation-of-state parameter of the cosmic
fluid:
\begin{equation}
p = -\rho + f(\rho),
\end{equation}
with
\begin{equation}
f(\rho) =
\frac{2q \bar{H}^{1/q}}{\kappa} \left( \sqrt{\frac{\kappa \rho}{3}} - H_\mathrm{inf}
\right)^{\left(q-1\right)/q} \,.
\end{equation}
In the case where $H_\mathrm{inf}/\sqrt{\kappa \rho/3} = H_\mathrm{inf}/H \ll 1$ we have
\begin{equation}
f(\rho) \approx \frac{2}{3^{\left(q-1\right)/\left(2q\right)}}
\frac{\bar{H}^{1/q}}{\kappa^{\left(q+1\right)/2q}}
\left[ \rho^{\left(q-1\right)/\left(2q\right)}
- \frac{\sqrt{3}\left(q-1\right)}{q}
\frac{H_\mathrm{inf}}{\sqrt{\kappa}} \rho^{-1/\left(2q\right)} \right] \,.
\label{eq:5.5newsum}
\end{equation}
 Thus, one can clearly see  from  (\ref{eq:5.5newsum}) that the function $f(\rho)$
includes a linear combination of two powers of $\rho$, as   in (\ref{eq:2.10newsum})
and  (\ref{eq:3.8newsum}). Hence, indeed  this scenario can be realized by the viscous
fluid models reconstructed above.

From (\ref{eq:5.5newsum}),  using   (\ref{eq:5.3newsum}),  we find
\begin{equation}
\frac{f(\rho)}{\rho} \approx
 = \frac{2q}{3} \left(\frac{\bar{H}}{H^{q+1}}\right)^{1/q}
\left[ 1 - \frac{\left(q-1\right)}{q} \frac{H_\mathrm{inf}}{H} \right]\,,
\label{eq:5.6newsum}
\end{equation}
and therefore  for $\bar{H}/H^{q+1} \ll 1$ we
get $f(\rho)/\rho \ll 1$. Thus,
$n_s$, $r$, and $\alpha_s$ can be approximately given
by (\ref{eq:2.6newsum}) and  be in agreement with observations, which act as an
additional advantage of singular inflation.

We now examine the limit $\bar{\zeta} = 0$ in  (\ref{eq:2.8newsum}),
in which  the fluid equation of state in (\ref{eq:2.8newsum}) becomes
$p=-\rho + A\rho^{\beta}$. In this limit from (\ref{eq:2.10newsum}) we deduce
that $f(\rho) = A\rho^{\beta}$,  i.e $f(\rho)$ has only one power of $\rho$.
However, from (\ref{eq:5.5newsum}) and (\ref{eq:5.6newsum}) we see that $f(\rho)$
consists of two $\rho$ powers. Thus, $f(\rho)$ can be given by (\ref{eq:5.5newsum}) and
(\ref{eq:5.6newsum}) only if the singular inflation is realized. Hence, for a non-viscous
fluid, i.e. for a fluid  without the $\zeta(H)$-term  in  (\ref{eq:2.8newsum}),
  singular inflation cannot be realized. From this feature we can see the importance of
the viscous term, and its significant effect on the dynamics of the early universe. This
important issue will be studied in more detail in the following subsection.

In summary, in the present subsection we studied the realization of inflation in a
fluid framework, whose equation-of-state parameter has an additional term corresponding
to bulk viscosity. Firstly, we saw that the obtained inflationary observables, namely
 $n_s$,  $r$ and $\alpha_s$, are in agreement with Planck data.
Secondly, we presented a reconstruction procedure of the fluid's equation of state, when
a specific  $n_s$ is given, while the tensor-to-scalar ratio is still
in agreement with observations. Thirdly, we analyzed the stability of the inflationary,
de Sitter phase, showing that a graceful exit and the pass to the subsequent thermal
history of the universe is obtained without fine tuning. Finally, we investigated the
realization of singular inflation, corresponding to Type IV singularity,
in the present viscous fluid model. Hence, viscous fluid inflation can be a candidate for
the description of early universe.

\subsection{Viscous warm and  intermediate inflation}

In this subsection we  show how the viscous cold inflationary models considered above can
be generalized to the warm case. These kind of models are most likely more physical
than the idealized cold ones, since they take into account the presence of massive
particles produced from the decaying inflaton field. Moreover, an important advantage of
warm scenarios is that they give rise to a much smaller tensor-to-scalar ratio than the
cold models, and hence are easier to be in agreement with the Planck data. The
presence of massive particles provides a natural way to explain why the cosmic fluid
can be associated with a bulk viscosity.

We abstain from using the simple equation of state $p=(1/3)\rho$ holding for
radiation, and we assume instead the more general form $p=w\rho$, where $w$ is  constant.
For convenience, one can  introduce the form $p=(\gamma -1)\rho$
with $\gamma=1+w$.  The effective pressure becomes $p_{\rm eff}=p+p_\zeta$, where
\begin{equation}
p_\zeta=-3H\zeta \label{2.67}
\end{equation}
is the viscous part of the pressure and $\zeta$ the bulk viscosity. In this case
 Eq. (\ref{2.50b})   generalizes to \cite{Setare:2014oka}
\begin{equation}
\dot{\rho}+3H(\rho+p-3\zeta H)=\Gamma {\dot{\phi}}^2.
\label{2.68}
\end{equation}
The usual condition about quasi-stationarity implies $\dot{\rho} \ll 3H(\gamma
\rho-3\zeta
H)$ and
$\dot{\rho} \ll \Gamma {\dot{\phi}}^2$.

We will henceforth follow the formalism  of \cite{Setare:2014oka}
for the strong dissipative case, namely for $Q\gg 1$ (see also \cite{Sharif:2014pba}). As
it
was mentioned in subsection \ref{interinfa} above, in intermediate inflation the scale
factor and the Hubble parameters are given by (\ref{2.39}) and (\ref{2.40}).
We will base the analysis on the basic assumptions
\begin{equation}
\Gamma(\phi)=\kappa^{3/2} V(\phi), \quad \zeta=\zeta_1 \rho, \label{2.71}
\end{equation}
where the proportionality of $\zeta$ to $\rho$ is a frequently used assumption (a similar
analysis can be performed for the case where $\Gamma$ and $\zeta$ are assumed
constants \cite{Setare:2014oka,Sharif:2014pba}, however we will not go into
further details and focus on the general case). From
Eqs.~(\ref{2.51}) and (\ref{2.52}) we then have $Q=\sqrt{\kappa}H$. Focusing on the
strong dissipative case $Q\gg 1$, manipulation of the equations gives the following
expression for the inflaton field as a
function of time:
\begin{equation}
\phi(t)=2\kappa^{-3/4}\sqrt{2(1-\alpha) t}. \label{2.72}
\end{equation}
This equation, predicting $\phi(t)$ to increase with time,  is seen to be different from
the corresponding Eq.~(\ref{2.42}) for cold intermediate inflation.

Taking into account the expression (\ref{2.40}) for $H$ we can express the potential as a
function of time:
\begin{equation}
V(t)=3A^2\alpha^2\kappa^{-2} (t/\sqrt{\kappa})^{2(\alpha-1)}, \label{2.73}
\end{equation}
which can alternatively be represented as a function of $\phi$ as
\begin{equation}
V(\phi)=3A^2\alpha^2 \kappa^{-2}
\left[\frac{\sqrt{\kappa}\phi}{2 \sqrt{2(1-\alpha)}}\right]^{ 4(\alpha-1)}.\label{2.74}
\end{equation}
Since
\begin{equation}
\rho=\frac{V\dot{\phi}^2}{3H(\gamma-3\zeta_1 H)}, \label{2.75}
\end{equation}
we see that it is necessary for the constant $\zeta_1$ in (\ref{2.71}) to satisfy the
condition $\zeta_1<\gamma/3H$ in order to make $\rho$ positive. The density varies with
time as
\begin{equation}
\rho(t)=\frac{2A\alpha (1-\alpha)\kappa^{-3/2}
(t/\sqrt{\kappa})^{\alpha-2}}{\gamma \sqrt{\kappa}-3\zeta_1A\alpha
(t/\sqrt{\kappa})^{2(\alpha-1)}}, \label{2.76}
\end{equation}
while when considered as a function of the inflaton field it reads
\begin{equation}
\rho(\phi)=
\frac{2A\alpha (1-\alpha)\kappa^{-3/2} \left[
\sqrt{2\kappa(1-\alpha)} \phi/2\right]^{2(\alpha-2) }}
{\gamma\sqrt{\kappa}-3\zeta_1A\alpha \left[ \sqrt{2\kappa(1-\alpha)}
\phi/2\right]^{2(\alpha-1)}}.
\label{2.77}
\end{equation}
Additionally, the number of e-folds becomes in this case
\begin{equation}
N=\frac{\sqrt{\kappa}}{\sqrt{3}}\int_{\phi_f}^\phi \frac{V^{3/2}}{V^\prime}d\phi
=A\frac{(1-\alpha)}{\alpha}-A\left[
\frac{\sqrt{\kappa}\phi}{2\sqrt{2(1-\alpha)}}\right]^{2\alpha},
\label{2.80}
\end{equation}
where  $\phi_f$ is the inflaton field at the end of the slow-roll epoch.
Finally, the slow-roll parameters   in the strong
dissipative epoch  ($Q\gg1$) become
\begin{equation}
\varepsilon=\frac{1}{2Q}\left( \frac{V^\prime}{V}\right)^2, \quad
\eta=\frac{1}{Q}\left[ \frac{
V''}{V}-\frac{1}{2}\left( \frac{{V^\prime}}{V}\right)^2\right], \label{2.78}
\end{equation}
giving in turn for the spectral parameter $\delta_{ns}$
\begin{equation}
\delta_{ns}=\frac{3\alpha-2}{1-\alpha}\varepsilon=\frac{3\alpha-2}{\alpha A}
\left[\frac{\sqrt{\kappa}\phi}{2 \sqrt{2(1-\alpha)}}\right]^{-2\alpha}. \label{2.79}
\end{equation}
Hence, the Harrison-Zel'dovich spectrum (independent of scale) corresponds to
$\alpha=2/3$.

\subsection{Singular inflation from fluids with generalized equation of state }

In the end of subsection \ref{viscinfl} we presented a brief discussion on the
possibility to realize singular inflation in the framework of viscous cosmology. Since
this is an important issue, in this subsection we investigate it in detail following
\cite{Nojiri:2015wsa}, considering more general viscous
equation of states. We consider an inhomogeneous viscous
equation-of-state parameter of the form
\begin{equation}\label{inhoeos}
p =-\rho -f(\rho )+G(H) \, ,
\end{equation}
which is a subclass of the general ansatz  (\ref{visceosinh00}). Thus, when the
function $G(H)$ becomes zero we re-obtain the homogeneous case.
An even more general equation of state would be to consider
\begin{equation}
\label{C5singevol}
p = f \left( \rho , H \right) \, .
\end{equation}
In the following we desire to investigate the realization of type IV singularity in
inflation driven by a fluid with the above EoS's.

As we mentioned earlier, a type IV singularity occurs at $t \to t_s$, if the
scale factor and the effective energy density and pressure remain finite, but the higher
derivatives of the Hubble function  diverge. A general form of the Hubble function which
can describe a Type IV singularity reads as
\begin{equation}
\label{IV1singevol}
H(t) = f_1(t) + f_2(t) \left( t_s - t \right)^\alpha\, ,
\end{equation}
with $f_1(t)$, $f_2(t)$ being arbitrary differentiable functions. Hence the type IV
singularity occurs when   $\alpha>1$, and without loss of generality we can consider it
to take the form
\begin{equation}
\label{IV2singevol}
\alpha= \frac{n}{2m + 1}\, ,
\end{equation}
with $n$, $m$ positive integers.

Let us start from a simple example of  type IV singularity realization, namely we
consider
 $f_1(t)=0$ and $f_2(t) = f_0$, with $f_0$ a positive parameter.
In this case  the two Friedmann equations, namely $\rho =\frac{3}{\kappa}H^2$ and
$ p = -\frac{1}{\kappa}\left(3H^2 +2\dot H\right)$,
become
\begin{eqnarray}
\label{C3singevol}
&&\rho =\frac{3f_0^2 }{\kappa}\left( t_s - t \right)^{2\alpha}\\
&&p
= -\frac{1}{\kappa}\left[ 3f_0^2 \left( t_s - t \right)^{2\alpha}
+ 2\alpha f_0 \left( t_s - t \right)^{\alpha -1}\right]\, ,
\end{eqnarray}
and hence eliminating $t_s-t$ we get the result 
\begin{equation}
\label{C4singevol}
p = - \rho - 2 \cdot 3^{- \frac{\alpha - 1}{2\alpha}}  \kappa^{-
\frac{\alpha + 1}{2\alpha}} f_0^{1/\alpha} \rho ^\frac{\alpha -
1}{2\alpha}\, .
\end{equation}
Hence, a viscous fluid with this equation of state can generate the Hubble function
(\ref{IV1singevol}) and hence the type IV singularity.
Defining $\tilde\alpha\equiv \frac{\alpha - 1}{2\alpha}$, a type IV singularity will
occur if   $0<\tilde\alpha<\frac{1}{2}$ (or equivalently,
$\alpha>1$).

Observing the equation-of-state parameter in (\ref{C4singevol}) we deduce that it can
 be seen either as a homogeneous one, of the form (\ref{inhoeos})
with   $G(H)=0$ and
\begin{equation}\label{view1singevol}
f(\rho )=- 2 \cdot 3^{- \frac{\alpha - 1}{2\alpha}}  \kappa^{-
\frac{\alpha + 1}{2\alpha}} f_0^{1/\alpha} \rho ^\frac{\alpha -
1}{2\alpha}\, ,
\end{equation}
or as an inhomogeneous one, of the form (\ref{inhoeos}) with $f(\rho )=0$ and
\begin{equation}
\label{ghsingevol}
G(H)=- \frac{2 \alpha}{\kappa} f_0^{1/\alpha} H^\frac{\alpha -
1}{\alpha} \,
\end{equation}
(since $\rho =\frac{3}{\kappa}H^2$).

Let us now consider a more general Hubble function inside the class (\ref{IV1singevol}),
namely
\begin{equation}
\label{hubnew}
H(t)=f_0 (t-t_1)^{\alpha } +c_0(t-t_2)^{\beta }\, ,
\end{equation}
where  $c_0$, $f_0$ are constants, and $\alpha$,$\beta
>1$. Thus, two type IV singularities appear at $t=t_1$ and $t=t_2$.
We choose $t_1$ to correspond to the inflation end and $t_2$ to lie at
late times. In order to simplify the expressions, we focus our analysis in the vicinity
of the type IV singularity. In this region, inserting (\ref{hubnew})   into the two
Friedmann equations leads to
 \begin{eqnarray}
 \label{energypress2sinevol}
& &\rho \approx \frac{3 c_0^2 (t-t_2)^{2 \beta }}{\kappa}\\
&&p \approx-\frac{3
c_0^2 (t-t_2)^{2 \beta }}{\kappa }-\frac{2 c_0 (t-t_2)^{-1+\beta } \beta
}{\kappa } \, ,
\end{eqnarray}
and therefore  the equation of state
reads 
\begin{equation}
\label{neweqna}
p =-\rho -\frac{2 c_0\beta }{\kappa } \left(\frac{\rho \kappa }{3
c_0^2}\right)^{\frac{\beta -1}{2 \beta }}\, .
\end{equation}
Interestingly enough, we observe that the late-time type IV singularity is related to the
early-time type IV singularity and the corresponding equation-of-state paramater.
In the same lines, the early-time singularity is related to the effective
equation of state that gives rise to the late time one.

One can proceed in similar lines, and study the scenario where
\begin{equation}
\label{hubnoj1}
H(t)=\frac{f_1}{\sqrt{t^2+t_0^2}}+\frac{f_2 t^2 (-t+t_1)^{\alpha
}}{t^4+t_0^4}+f_3 (-t+t_2)^{\beta } \, .
\end{equation}
In this case, in the vicinity of the early-time singularity at $t_1$ we obtain
\cite{Nojiri:2015wsa}
\begin{align}
\label{C8newsingevol}
\rho \simeq & \frac{3 f_1^2}{\left(t^2+t_0^2\right) \kappa }+\frac{6 f_1
f_3 (-t+t_2)^{\beta }}{\sqrt{t^2+t_0^2} \kappa }+\frac{3 f_3^2
(-t+t_2)^{2 \beta }}{\kappa } \, , \nonumber\\
p \simeq& \frac{2 f_1 t}{\left(t^2+t_0^2\right)^{3/2} \kappa }-\frac{3
f_1^2}{\left(t^2+t_0^2\right) \kappa }-\frac{6 f_1 f_3 (-t+t_2)^{\beta
}}{\sqrt{t^2+t_0^2} \kappa }-\frac{3 f_3^2 (-t+t_2)^{2 \beta }}{\kappa
}+\frac{2 f_3 (-t+t_2)^{-1+\beta } \beta }{\kappa } \, ,
\end{align}
where these relations are again determined by the late-time singularity.

Finally, one can study the scenario where
\begin{equation}\label{hnew1singevol}
H(t)=f_0+c \left(t-t_1\right)^{\alpha}\left(t-t_2\right)^{\beta}\, ,
\end{equation}
which is reproduced by
\begin{align}
\rho =& \frac{3 f_0^2}{\kappa }+\frac{6 c f_0 (-t+t_1)^{\alpha }
(-t+t_2)^{\beta }}{\kappa }+\frac{3 f_0^2 (-t+t_1)^{2 \alpha }
(-t+t_2)^{2 \beta }}{\kappa } \, , \nonumber\\
p =&-\frac{3 f_0^2}{\kappa }-\frac{6 c f_0 (-t+t_1)^{\alpha }
(-t+t_2)^{\beta }}{\kappa }-\frac{3 f_0^2 (-t+t_1)^{2 \alpha }
(-t+t_2)^{2 \beta }}{\kappa } \nonumber\\ &
+\frac{2 f_0 (-t+t_1)^{-1+\alpha } (-t+t_2)^{\beta } \alpha }{\kappa
}+\frac{2 f_0 (-t+t_1)^{\alpha } (-t+t_2)^{-1+\beta } \beta }{\kappa
}\, .
\end{align}
At both type IV singularities at $t_1$ and $t_2$, the  effective energy density and
pressure become
\begin{equation}
\rho =\frac{3 f_0^2}{\kappa } \, , \quad p =-\frac{3 f_0^2}{\kappa }\, ,
\end{equation}
and thus the corresponding equation-of-state parameter becomes $-1$.

Let us now proceed to the calculation of the slow-roll parameters, which as usual are
used for the calculation of the various inflationary observables, since the effect of the
type IV singularity can be significant. The starting point is that in flat FRW geometry
one can express the various quantities as a function of the number of
$e$-foldings  $N$, namely
\cite{Nojiri:2015wsa}
\begin{eqnarray}
\label{efold1sinfevol0}
&& \rho =\frac{3}{\kappa}\left(H(N)\right)^2 \\
&& p (N)+\rho (N)=
-\frac{2 H(N)H'(N)}{\kappa}\, ,
\label{efold1sinfevol}
\end{eqnarray}
where $H'(N)=\mathrm{d}H/\mathrm{d}N$. Assuming that the equation of state is given by
the general ansatz:
\begin{equation}
\label{SS2singevol}
p (N) = - \rho _\mathrm{mat}(N) + \tilde f (\rho (N))\, ,
\end{equation}
then   (\ref{efold1sinfevol}) gives
\begin{equation}\label{efold2}
\tilde f (\rho (N))= -\frac{2 H(N)H'(N)}{\kappa}\, .
\end{equation}
Since the usual  conservation equation  is valid, namely
\begin{equation}
\rho' (N)+3H(N)\left( \rho (N)+p
(N)\right)=0\, ,
\end{equation}
with $\rho' (N)=\mathrm{d}\tilde f (\rho (N))/\mathrm{d}N$, using
(\ref{efold2}) we find
\begin{equation}
\label{efold3singevol}
\rho'(N)+3\tilde f (\rho (N))=0\, .
\end{equation}
Finally, inserting (\ref{efold3singevol}) into (\ref{SS2singevol}) we acquire
\begin{equation}
\frac{2}{\kappa }\left[(H'(N))^2+H(N)+H''(N)\right]=3\tilde f'(\rho
)f(\rho ) \, ,
\end{equation}
with $\tilde f '(\rho
(N)\equiv\mathrm{d}\tilde f (\rho )/\mathrm{d}\rho $.

Now, for a given $H(t)$, the slow-roll parameters $\epsilon$, $\eta$ and
$\xi$ write as \cite{Nojiri:2015wsa}
\begin{eqnarray}
&&
\!\!\!\!\!
\epsilon =  - \frac{H^2}{4 \dot H} \left( \frac{6\dot H}{H^2} +
\frac{\ddot H}{H^3} \right)^2 \left( 3 + \frac{\dot H}{H^2} \right)^{-2}\,
, \nonumber\\
&&
\!\!\!\!\!
\eta =  - \frac{1}{2} \left( 3 + \frac{\dot H}{H^2} \right)^{-1} \left(
\frac{6\dot H}{H^2} + \frac{{\dot H}^2}{2 H^4} - \frac{\ddot H}{H^3}
- \frac{{\dot H}^4}{2 H^4} + \frac{{\dot H}^2 \ddot H}{H^5} - \frac{{\ddot
H}^2}{2H^2} + \frac{3 \ddot H}{H \dot H}
+ \frac{\dddot H}{H^2 \dot H} \right)\, ,
\nonumber\\
&&
\!\!\!\!\!
\xi^2 =  \frac{1}{4} \left( \frac{6\dot{H}}{H^2} + \frac{\ddot{H}}{H^3}
\right) \left( 3 + \frac{\dot{H}}{H^2} \right)^{-1} \left(
\frac{9\ddot{H}}{H {\dot H} }
+ \frac{3\dddot{H}}{{\dot{H}}^2} + \frac{2 \dddot{H}}{H^2 \dot{H}} +
\frac{4 {\ddot{H}}^2}{H^2 {\dot{H}}^2}\right.\nonumber\\
&&\left.
\ \ \ \ \ \ \ \ \ \ \ \ \ \ \ \ \ \ \ \ \ \ \ \ \ \ \ \ \ \ \ \ \ \ \ \ \ \ \ \ \ \ \ \ \
\ \,
- \frac{\ddot{H} \dddot{H}}{H {\dot{H}}^3} -
\frac{3{\ddot{H}}^2}{{\dot{H}}^3}
+ \frac{{\ddot{H}}^3}{H {\dot{H}}^4} + \frac{\ddddot{H}}{H {\dot{H}}^2}
\right)\, .
\end{eqnarray}
Hence, if $H(t)$ is given by  (\ref{IV1singevol}), and if $\alpha>1$, i.e when a type
IV singularity is obtained at $t\sim t_s$ , the slow-roll parameters at the vicinity of
the singularity become
\begin{align}
\label{S10}
\epsilon \sim & \left\{
\begin{array}
{ll}
- \frac{f_1(t_s)^2}{4 \dot f_1(t_s)} \left[ \frac{6\dot f_1(t_s) f_1(t_s)}{
f_1(t_s)^2} + \frac{\ddot f_1(t_s)}{ f_1(t_s)^3} \right]^2 \left[ 3 +
\frac{\dot f_1(t_s)}{ f_1(t_s)^2} \right]^{-2}\, ,
& \mbox{when $\alpha>2$} \\
- \frac{f_1(t_s)^2}{4 \dot f_1(t_s)} f_2(t_s) \alpha (\alpha - 1) \left(
t_s - t \right)^{\alpha-2} \left[ 3 + \frac{\dot f_1(t_s)}{ f_1(t_s)^2}
\right]^{-2}\, ,
& \mbox{when $2>\alpha>1$}
\end{array}
\right. \, , \nonumber\\
\eta \sim & \left\{
\begin{array}{ll}
- \frac{1}{2} \left[ 3 + \frac{\dot f_1(t_s)}{ f_1(t_s)^2} \right]^{-1}
 \left[ \frac{6\dot f_1(t_s)}{f_1(t_s)} + \frac{{\dot
f_1(t_s)}^2}{2 f_1(t_s)^4} - \frac{\ddot f_1(t_s)}{f_1(t_s)^3}
- \frac{{\dot f_1(t_s)}^4}{2 f_1(t_s)^4}
+ \frac{{\dot f_1(t_s)}^2 \ddot
f_1(t_s)}{f_1(t_s)^5}
\right.
&
\\
\left. \ \  \ \  \ \  \ \  \ \  \ \  \ \  \ \  \ \  \ \  \ \  \ \   \ \  \,
 - \frac{{\ddot f_1(t_s)}^2}{2f_1(t_s)^2} + \frac{3
\ddot f_1(t_s)}{f_1(t_s) \dot f_1(t_s)}
+ \frac{\dddot f_1(t_s)}{f_1(t_s)^2 \dot f_1(t_s)} \right]\, ,
& \mbox{when $\alpha>3$} \\
- \frac{1}{2} \left[ 3 + \frac{\dot f_1(t_s)}{ f_1(t_s)^2} \right]^{-1}
\frac{f_2 \alpha ( \alpha - 1 ) (\alpha -2 )}{f_1(t_s)^2 \dot f_1
(t_s)}\left(t_s - t \right)^{\alpha - 3}\, ,
& \mbox{when $3>\alpha>1$}
\end{array}
\right. \, , \nonumber\\
\xi^2 \sim & \left\{
\begin{array}
{ll}
\frac{1}{4} \left[\frac{6\dot f_1(t_s)}{f_1(t_s)^2} + \frac{\ddot
f_1(t_s)}{f_1(t_s)^3} \right] \left[ 3 + \frac{\dot f_1(t_s)}{f_1(t_s)^2}
\right]^{-1} \left[ \frac{9\ddot f_1(t_s)}{f_1(t_s) \dot f_1(t_s)}
+ \frac{3\dddot f_1(t_s)}{{\dot f_1(t_s)}^2} + \frac{2 \dddot
f_1(t_s)}{f_1(t_s)^2 \dot f_1(t_s)}\right.
&
\\
\left.\ \  \ \  \ \  \ \  \ \  \ \  \ \  \ \  \ \  \ \  \ \  \ \   \ \  \,\ \  \ \  \ \
\
\  \ \  \ \  \ \
   + \frac{4 {\ddot
f_1(t_s)}^2}{f_1(t_s)^2 {\dot f_1(t_s)}^2}
- \frac{\ddot f_1(t_s) \dddot f_1(t_s)}{f_1(t_s) {\dot f_1(t_s)}^3}
\right. & \\
 \left.
 \ \  \ \  \ \  \ \  \ \  \ \  \ \  \ \  \ \  \ \  \ \  \ \   \ \  \,\ \  \ \  \ \  \
\  \ \  \ \  \ \
- \frac{3{\ddot f_1(t_s)}^2}{{\dot f_1(t_s)}^3}
+ \frac{{\ddot f_1(t_s)}^3}{f_1(t_s) {\dot f_1(t_s)}^4} + \frac{\ddddot
f_1(t_s)}{f_1(t_s) {\dot f_1(t_s)}^2} \right]\, ,
& \mbox{when $\alpha>4$}
\\
\frac{1}{4} \left[ \frac{6\dot f_1(t_s)}{f_1(t_s)^2} + \frac{\ddot
f_1(t_s)}{f_1(t_s)^3} \right] \left[ 3 + \frac{\dot f_1(t_s)}{f_1(t_s)^2}
\right]^{-1}
\frac{f_2(t_s) \alpha ( \alpha - 1) ( \alpha - 2) (\alpha - 3)}{f_1(t_s)
{\dot f_1(t_s)}^2} \left( t_s - t \right)^{\alpha -4}\, ,
& \mbox{when $4>\alpha>2$} \\
\frac{1}{4} \left[ 3 + \frac{\dot f_1(t_s)}{f_1(t_s)^2} \right]^{-1}
\frac{f_2(t_s)^2 \alpha^2 ( \alpha - 1)^2 ( \alpha - 2) (\alpha -
3)}{f_1(t_s)^4 {\dot f_1(t_s)}^2} \left( t_s - t \right)^{2\alpha -6}\, ,
& \mbox{when $2>\alpha>1$} \\
\end{array}
\right. \, .
\end{align}
Therefore, we deduce that if $f_1(t)$ is a smooth function then the slow-roll parameter
$\epsilon$ diverges when $2>\alpha >1$, whereas it remains regular for
$\alpha>2$. Moreover,
$\eta$ diverges for $3>\alpha >1$. Finally,  $\xi^2$ diverges    when $2>\alpha >1$ and
when $4>\alpha >2$. In summary, when $\alpha>4$ all slow-roll
parameters are non-singular near the Type IV
singularity.

In order to provide a more concrete example, we consider the simplified case where
 $H(t)=f_0\left(t-t_s\right)^{\alpha}$. Thus, the slow-roll parameters become
\cite{Nojiri:2015wsa}
\begin{equation}\label{slwersingevol}
\epsilon=\frac{f_0 (t-t_s)^{-1+\alpha } \alpha (-1+6 t-6 t_s+\alpha )^2}{4
\left[3 f_0 (t-t_s)^{1+\alpha }+\alpha \right]^2}\, ,
\end{equation}
\begin{eqnarray}
\label{sletasingevol}
&&
\!\!\!\!\!\!\!\!\!\!\!\!\!\!\!\!\!\!\!\!
\eta = \left\{4 f_0 \left[3 f_0 (t-t_s)^{1+\alpha }+\alpha \right]\right\}^{-1}
\Big\{
 (t-t_s)^{-3-\alpha }\left[-t_s^2 \alpha ^2+2 \alpha ^3-2 \alpha ^4
 \right.\nonumber\\
&&\left.
\ \ \ \ \ \ \ \ \ \ \ \ \ \ \ \ \ \ \ \ \ \
-6
f_0 (t-t_s)^{3+\alpha } (-1+3 \alpha )
+f_0^2 (t-t_s)^{2 \alpha } \alpha ^2
(1-2 \alpha +2 \alpha^2 )\right]
\nonumber\\
&&
+
(t-t_s)^{-3-\alpha }
\left(-4 t^2+8 t t_s-4 t_s^2+4 t^2
\alpha -8 t t_s \alpha +4 t_s^2 \alpha -t^2 \alpha ^2+2 t t_s \alpha
^2\right)
\Big\} \, ,
\end{eqnarray}
\begin{eqnarray}
\label{slwxisingevol}
&&
\!\!\!\!\!\!\!\!\!\!\!\!\!\!\!\!\!\!\!\!\!\!\!\!\!
\xi^2=\frac{(t-t_s)^{-5-2 \alpha } (-1+\alpha )
(-1+6 t-6 t_s+\alpha ) }{4
f_0^2 \left[3 f_0 (t-t_s)^{1+\alpha }+\alpha \right]}
\nonumber\\
&&\!\!\!\!\!\!\!\!\!\!\!\!\!\!\!\!
\times \left[5 (t-t_s)^2 (\alpha-1 )^2+3 f_0^2 (t-t_s)^{2 \alpha }
(\alpha-2 )^2 (\alpha-1 ) \alpha +3 f_0 (t-t_s)^{3+\alpha } (1+2 \alpha
)\right].
\end{eqnarray}
Hence, we immediately observe that the slow-roll parameters exhibit singularities  at
$t=t_s$, as mentioned above. However, such singularities in the
slow-roll parameters can be viewed as rather unwanted features.

We close this subsection by making a comparison with observations. In order to achieve
this it proves convenient to express the various quantities in terms of the number of
$e$-foldings  $N$. In particular, for a given $H(N)$ the slow-roll parameters
read as \cite{Nojiri:2015wsa}
{\small{
\begin{equation}
\!\!\!\!\!\!\!\!\!\!\!\!\!\!\!\!\!\!\!\!\!\!\!\!\!\!\!\!\!\!\!\!\!\!
\!\!\!\!\!\!\!\!\!\!\!\!\!\!\!\!\!\!\!\!\!
\epsilon = - \frac{H(N)}{4 H'(N)} \left\{ \frac{\frac{6 H'(N)}{H(N)}
+ \frac{H''(N)}{H(\phi)} + \left[ \frac{H'(N)}{H(N)} \right]^2}
{3 + \frac{H'(N)}{H(N)}} \right\}^2
\end{equation}
\begin{eqnarray}
&&
\!\!\!\!\!\!\!\!\!\!\!\!\!\!\!\!\!\!\!\!\!
\eta =  -\frac{1}{2} \left[ 3 +
\frac{H'(N)}{H(N)} \right]^{-1} \left\{
\frac{9 H'(N)}{H(N)} + \frac{3 H''(N)}{H(N)}+ \frac{1}{2} \left[
\frac{H'(N)}{H(N)} \right]^2
\right.
\nonumber\\
&&\ \ \ \ \ \ \ \ \ \ \ \ \ \ \ \ \ \ \ \ \ \ \ \
\left.
-\frac{1}{2} \left[ \frac{H''(N)}{H'(N)}
\right]^2
+ \frac{3 H''(N)}{H'(N)} + \frac{H'''(N)}{H'(N)} \right\}
\end{eqnarray}
\begin{eqnarray}
&&\xi^2 =  \frac{ \frac{6 H'(N)}{H(N)} + \frac{H''(N)}{H(N)}
+ \left[ \frac{H'(N)}{H(N)} \right]^2 }{4 \left[ 3 + \frac{H'(N)}{H(N)}
\right]^2}
\left\{ \frac{3 H(N) H'''(N)}{H'(N)^2} + \frac{9 H'(N)}{H(N)}
- \frac{2 H(N) H''(N) H'''(N)}{H'(N)^3}
\right.
\nonumber\\
&&\ \ \ \ \ \ \ \ \ \ \ \ \ \ \ \ \ \ \ \ \ \ \ \ \ \ \ \ \ \ \ \ \ \ \ \ \ \
\left.
+ \frac{4 H''(N)}{H(N)}
+ \frac{H(N) H''(N)^3}{H'(N)^4} + \frac{5 H'''(N)}{H'(N)} - \frac{3 H(N)
H''(N)^2}{H'(N)^3}\right.
\nonumber\\
&&\ \ \ \ \ \ \ \ \ \ \ \ \ \ \ \ \ \ \ \ \ \ \ \ \ \ \ \ \ \ \ \ \ \ \ \ \ \
\left. - \left[ \frac{H''(N)}{H'(N)}
\right]^2
+ \frac{15 H''(N)}{H'(N)}
+ \frac{H(N) H''''(N)}{H'(N)^2} \right\}\, .
\end{eqnarray}}}
Therefore, we can use (\ref{efold1sinfevol0}) and (\ref{efold1sinfevol}) in order to
calculate the usual inflationary observables, namely the spectral index $n_s$, the
tensor-to-scalar ratio $r$ and the running spectral index $a_s$ as
 \cite{Nojiri:2015wsa}
\begin{align}
n_s - 1 =& - 9 \rho(N) \tilde f(\rho (N)) \left( \frac{\tilde
f'(\rho (N))-2}{2\rho (N)
- \tilde f(\rho (N))}\right)^2
+\frac{6\rho (N)}{2\rho (N) - \tilde f(\rho (N))}
\left\{ \frac{ \tilde f(\rho (N))}{\rho (N)} \right. \nonumber\\
& \left. + \frac{1}{2} \left(\tilde f'(\rho (N))\right)^2
+ \tilde f'(\rho (N)) -\frac{5}{2} \frac{\tilde f(\rho (N)) \tilde f'(\rho
(N))}{\rho (N)}
+ \left(\frac{f(\rho)}{\rho(N)}\right)^2+\frac{1}{3} \frac{\rho
'(N)}{\tilde f(\rho (N))}
\right. \nonumber \\
& \left.
\times \left[\left( \tilde f'(\rho (N)) \right)^2 + \tilde f(\rho (N))
\tilde f''(\rho (N))
-2 \frac{ \tilde f(\rho (N)) \tilde f'(\rho (N))}{\rho (N)}
+ \left( \frac{ \tilde f(\rho (N))}{\rho (N)} \right)^2 \right] \right\} \,,
\label{eq:2.32singevol} \\
r =& 24\rho (N) \tilde f(\rho (N))
\left( \frac{ \tilde f'(\rho (N))-2}
{2\rho (N) - \tilde f(\rho (N))}\right)^2 \,,
\label{eq:2.33singevol} \\
\alpha_s =& \rho (N) \tilde f(\rho (N)) \left( \frac{\tilde
f'(\rho (N))-2}
{2\rho (N) - \tilde f(\rho (N))}\right)^2 \left[ \frac{72\rho (N)}{2\rho (N)
- \tilde f(\rho (N))} J_1 \right. \nonumber\\
& \left. -54 \rho (N) \tilde f(\rho (N)) \left( \frac{\tilde f'(\rho (N))-2}
{2\rho (N) - \tilde f(\rho (N))}\right)^2 -\frac{1}{\tilde f'(\rho
(N))-2}J_2 \right] \,,
\label{eq:2.34singevol}
\end{align}
where
{\small{
\begin{align}
J_1 \equiv& \frac{\tilde f(\rho (N))}{\rho (N)} + \frac{1}{2}
\left(\tilde f'(\rho (N))\right)^2 + \tilde f'(\rho (N))
-\frac{5}{2} \frac{\tilde f(\rho (N)) \tilde f'(\rho (N))}{\rho (N)}
+ \left(\frac{\tilde f(\rho (N))}{\rho (N)}\right)^2 \nonumber+\frac{1}{3}
\frac{\rho '(N)}{\tilde f(\rho (N))} \nonumber\\
&
\times \left[\left(\tilde f'(\rho (N))\right)^2 + \tilde f(\rho (N)) \tilde
f''(\rho (N))
-2 \frac{\tilde f(\rho (N)) \tilde f'(\rho (N))}{\rho (N)} + \left(
\frac{\tilde f(\rho (N))}{\rho (N)}
\right)^2 \right]\, ,
\end{align}}}
and
{\small{
\begin{align}
J_2 \equiv &
\frac{45}{2} \frac{\tilde f(\rho (N))}{\rho (N)} \left(\tilde f'(\rho (N))
- \frac{1}{2} \frac{\tilde f(\rho (N))}{\rho (N)}\right)
+ 18\left(\frac{\tilde f(\rho (N))}{\rho (N)}\right)^{-1} \left\{
\left(\tilde f'(\rho (N))-\frac{1}{2} \frac{ \tilde f(\rho (N))}{\rho
(N)}\right)^2 \right. \nonumber\\
& \left. + \left( \tilde f'(\rho (N))-\frac{1}{2} \frac{\tilde f(\rho
(N))}{\rho (N)}\right)^3 \right\}
-9\left( \tilde f'(\rho (N)) - \frac{1}{2} \frac{\tilde f(\rho (N))}{\rho
(N)}\right)^2
-45 \tilde f'(\rho (N)) + 9\frac{\tilde f(\rho (N))}{\rho (N)}
\nonumber \\
&
+3 \left(4 \tilde f'(\rho (N)) -7\frac{ \tilde f(\rho (N))}{\rho (N)}
+2\right)
\left\{
-\frac{3}{2}\left(\tilde f'(\rho (N)) -\frac{1}{2}\frac{\tilde f(\rho
(N))}{\rho (N)}\right)
+ \left(\frac{\tilde f(\rho (N))}{\rho (N)}\right)^{-2}
\frac{\rho '(N)}{\rho (N)} \right. \nonumber\\
& \times \left[ \left( \tilde f'(\rho (N))\right)^2 + \tilde f(\rho (N))
\tilde f''(\rho (N))
\left.
- 2 \frac{ \tilde f(\rho (N)) \tilde f'(\rho (N))}{\rho (N)} + \left(
\frac{\tilde f(\rho (N))}{\rho (N)}
\right)^2 \right] \right\}
\nonumber \\
& +\left(\frac{ \tilde f(\rho (N))}{\rho (N)}\right)^{-2}
\left\{
-\frac{3}{2} \left(\frac{\tilde f(\rho (N))}{\rho (N)}\right)
\left(\frac{\rho '(N)}{\rho (N)}\right)
\left[
3\left( \tilde f'(\rho (N)) \right)^2
+2 \tilde f(\rho (N)) \tilde f''(\rho (N)) \right. \right. \nonumber\\
& \left. -\frac{11}{2}\frac{ \tilde f(\rho (N)) \tilde f'(\rho (N))}{\rho
(N)}
+\frac{5}{2} \left( \frac{\tilde f(\rho (N))}{\rho (N)} \right)^2
\right]
\nonumber \\
& \left.
+ \left(\frac{\rho ''(N)}{\rho (N)}\right)
\left[
\left( \tilde f'(\rho (N)) \right)^2 + \tilde f(\rho (N)) \tilde f''(\rho
(N))
-2\frac{ \tilde f(\rho ) \tilde f'(\rho )}{\rho (N)}
+ \left( \frac{ \tilde f(\rho )}{\rho } \right)^2
\right]
\right.
\nonumber \\
& \left.
+ \left(\frac{\rho '(N)}{\rho (N)}\right)^2
\left[ \left(3\tilde f'(\rho (N)) \tilde f''(\rho (N)) + \tilde f(\rho (N))
\tilde f'''(\rho (N)) \right) \rho (N)
-3\left( \tilde f'(\rho (N)) \right)^2 \right. \right. \nonumber\\
& \left. \left. - 3 \tilde f(\rho (N)) \tilde f''(\rho (N))
+6\frac{\tilde f(\rho (N)) \tilde f'(\rho (N))}{\rho (N)}
-3\left( \frac{\tilde f(\rho (N))}{\rho (N)} \right)^2
\right]
\right.
\,.
\end{align}}}

In order to obtain a qualitative picture of the above observables we consider
an equation of state of the form $p =-\rho +f(\rho )$, with
$f(\rho )=A \rho ^{\alpha}$. Since according to  (\ref{efold1sinfevol0}) and
(\ref{efold1sinfevol})  the scale factor  reads as
\begin{equation}
a(t)=a_0 e^{\frac{\rho ^{1-\alpha}}{3(1-\alpha)A}}\, ,
\end{equation}
we can express $\rho$ as a function of $N$ as
\begin{equation}
\label{reffsingevol}
\rho(N)
=\left[3(1-\alpha)A\right]^{\frac{1}{1-\alpha}}N^{\frac{1}{1-\alpha}}\, .
\end{equation}
Additionally, since a Type IV singularity is obtained when $0<\alpha<\frac{1}{2}$, we can
choose
 $\frac{f(\rho )}{\rho }\ll 1$. Inserting these into
(\ref{eq:2.32singevol})-(\ref{eq:2.34singevol})
we finally acquire
 \begin{equation}
n_s\simeq 1-\frac{2}{N (1-\alpha )}\, , \quad r\simeq \frac{8}{N (1-\alpha
)},{\,}{\,}\alpha_s\simeq -\frac{1}{N^2 (1-\alpha )^2}\, .
\end{equation}
Now, the 2015 Planck results \cite{Ade:2015xua} provide the following values:
\begin{equation}
n_s=0.9644\pm 0.0049\, , \quad r<0.10\, , \quad a_s=-0.0057\pm 0.0071\, .
\end{equation}
Hence, if in the scenario at hand we choose $(N,\alpha)=(60,1/20)$, we obtain
\begin{equation}
n_s\simeq 0.96491\, , \quad r\simeq 0.1403\, , \quad a_s=-0.000307\, .
\end{equation}
Therefore, concerning  the spectral index we acquire a good agreement, however the values
of the tensor-to-scalar ration and of the running spectral index  are not inside the
observational bounds. Nevertheless, we can obtain satisfactory agreement in more
sophisticated models instead of the simple example $f(\rho )=A \rho ^{\alpha}$.

\section{Late-time acceleration}
\label{Section3}

According to the concordance  model of cosmology the universe is currently accelerating,
while it entered this era after  being in a long matter-dominated epoch. This behavior,
similarly to the early accelerated era of inflation, cannot be reproduced within the
standard framework of general relativity and flat $\Lambda$CDM-model with dust and vacuum
energy,
and therefore
extra degrees of freedom should be introduced. One can attribute these extra degrees of
freedom to new, exotic forms of matter, such as the inflaton field at early times
and/or the dark energy concept at late times (for reviews see
\cite{bamba12e,Cai:2009zp}). Alternatively, one can consider the extra
degrees of freedom to have a gravitational origin, i.e. to arise from a gravitational
modification that possesses general relativity as a particular limit (see
\cite{Nojiri:2006ri,Capozziello:2011et,Nojiri:2010wj,Nojiri:2017ncd,Cai:2015emx} and 
references therein).
In this
section we will show how the late-time acceleration can be driven by the fluid viscosity
\cite{Weinberg:1971mx,Brevik:1994cd,Zimdahl:1996fj,Caldwell:2003vq,
Nojiri:2003vn,Nojiri:2005sr,
Nojiri:2005pu, Cardone:2005ut,Lepe:2008eu,
Mostafapoor:2013kha,Brevik:2011mm,Brevik:2012nt,Brevik:2015xsa,Brevik:2014eya,
Sasidharan:2015ihq,Bamba:2015sxa,Normann:2016jns,Laciana:2016txp}.

\subsection{Late-time viscous cosmology}

We start our investigation by studying the basic scenario of late-time viscous cosmology,
presenting the main properties of the viscous cosmic
fluid, following \cite{Brevik:2014cxa}. As usual we assume a homogeneous and
isotropic FRW universe with geodesic fluid flow, and thus the two Friedmann equations are
given by (\ref{fr1general}),(\ref{fr2general}).

Let us very briefly discuss the non-viscous case. According to the standard model the
total
energy density
and pressure are
\begin{equation}
\rho_{\rm tot}=\rho+\rho_\Lambda, \quad p_{\rm tot}=p+p_\Lambda=-\rho_\Lambda,
\label{1.16}
\end{equation}
where   $\rho_\Lambda=\Lambda/8\pi G$ is
the Lorentz invariant vacuum energy density   and  $p_\Lambda=-\Lambda/8\pi G$ is
the vacuum pressure corresponding to a positive tensile stress. With the critical energy
density $\rho_c$, the matter density parameter $\Omega_M$, and the Einstein gravitational
constant $\kappa$ defined as
\begin{equation}
\kappa \rho_c=3H^2,  \quad \Omega_M=\frac{\rho}{\rho_c}, \quad \kappa=8\pi G, \label{1.17}
\end{equation}
we obtain for the scale factor  \cite{Gron:2008ah}
\begin{equation}
a(t)=K_s^{1/3}\sinh^{2/3}\left( \frac{t}{t_\Lambda}\right),
\end{equation}
with $t_\Lambda
=\frac{2}{3H_0 \sqrt{\Omega_{\Lambda 0}}}, \quad K_s=\frac{1-\Omega_{\Lambda
0}}{\Omega_{\Lambda 0}}$, where the subscript zero refers to the present time $t=t_0$ (as
usual  we impose $a(t_0)=1$). The
present age
of the universe is
\begin{equation}
t_0=t_\Lambda \rm{arctanh} \sqrt{\Omega_{\Lambda 0}}, \label{G3.4}
\end{equation}
which leads to $t_\Lambda =11.4\times 10^9$ years if we insert that $t_0=13.7\times 10^9$
years and $\Omega_{\Lambda 0}=0.7$. In terms of these quantities  for the Hubble
parameter we obtain
\begin{equation}
H=\frac{2}{3t_\Lambda}\coth \left( \frac{t}{t_\Lambda}\right), \label{1.20}
\end{equation}
whereas the deceleration parameter
becomes
\begin{equation}
q\equiv  -1-\frac{\dot{H}}{H^2} =\frac{1}{2}\left[ 1-3\tanh^2
\left(\frac{t}{t_\Lambda}\right)\right]. \label{1.22}
\end{equation}
Inserting Eq.~(\ref{G3.4}) the present value for for the deceleration parameter of the
$\Lambda$CDM-
universe is
\begin{equation}
\hat{q}_0=\frac{1}{2}(1-3\Omega_{\Lambda 0}). \label{G3.7}
\end{equation}
With $\Omega_{\Lambda 0}=0.7$ we obtain $\hat{q}_0=-0.55$.

It is of interest to determine the time $t=t_1$ when deceleration turns into
acceleration.
The condition for this is $q(t_1)=0$, and leads to
$t_1=t_\Lambda \rm{arctanh}\frac{1}{\sqrt{3}}$,
with corresponding redshift
\begin{equation}
z_1=\frac{1}{a(t_1)}-1=\left( \frac{2\Omega_{\Lambda 0}}{1-\Omega_{\Lambda
0}}\right)^{1/3}-1, \label{1.24}
\end{equation}
that is $t_1=7.4\times 10^9~$years and $z_1=0.67$.
Finally, let $t_e$ be the time of emission of a signal that arrives at the time $t_0$.
Considering time in units of Gyr and inserting $t_0=13.7$ and
$\Omega_{\Lambda 0}=0.7$, we acquire the useful expression
\begin{equation}
t_e=11.3 ~\rm{arctanh}[ 1.53(1+z)^{-1.5}]. \label{1.25}
\end{equation}

After this brief introduction we now proceed to the investigation of the viscous case,
that is we switch on the viscosity in the equation-of-state parameter of the  cosmic
fluid. For convenience we assume a flat geometry. Without loss of generality we consider
the simplest ansatz (\ref{eq.statebulk00}), and thus the two Friedmann
equations (\ref{fr1general}),(\ref{fr2general}) become
\begin{equation}
3H^2=\kappa \rho +\Lambda, \label{1.29}
\end{equation}
\begin{equation}
\dot{H} +H^2= \frac{\kappa}{6}\left(9\zeta H-\rho-3p\right)+\frac{1}{3}\Lambda,
\label{1.27}
\end{equation}
while the conservation equation  reads
\begin{equation}
\dot{\rho}+3H(\rho+p)=9\zeta H^2, \label{1.26}
\end{equation}
with
\begin{equation}
p=w\rho, \label{1.28}
\end{equation}
where in its simplest version $w$ is a constant.
Finally, similarly to $\Omega_M=\rho/\rho_c$, it proves convenient to
introduce the  density parameters
\begin{equation}
\Omega_\zeta=\frac{\kappa \zeta}{H}, \quad
\Omega_\Lambda=\frac{
\Lambda}{\kappa \rho_c}, \label{1.32}
\end{equation}
where the critical   density follows  from $3H^2=\kappa \rho_c$. Thus,   we can express
the current deceleration
parameter as
\begin{equation}
q_0=\frac{1}{2}(1+3w)-\frac{3}{2}[\Omega_{\zeta 0}+(1+w)\Omega_{\Lambda 0}]. \label{1.33}
\end{equation}
If the cosmic fluid is cold, i.e. with $w=0$, as is often assumed, we obtain
\begin{equation}
\Omega_{\zeta 0}=\frac{1}{3}(1-2q_0)-\Omega_{\Lambda 0}. \label{1.34}
\end{equation}
In principle, this equation enables one to estimate the viscosity parameter
$\Omega_{\zeta 0}$ if one has at hand accurate measured values of $q_0$ and
$\Omega_{\Lambda 0}$. It follows from Eqs.~(\ref{G3.7}) and (\ref{1.34}) that
\begin{equation}
\Omega_{\zeta 0}=\frac{2}{3}(\hat{q}_0-q_0). \label{G3.16}
\end{equation}
Hence $\Omega_{\zeta 0}$ is proportional to the deviation of the measured deceleration
parameter from the standard $\Lambda$CDM-value as given in Eq.~(\ref{G3.7}). This means
that one needs to measure the deceleration parameter very accurately in order to obtain
information about the viscosity coefficient from its relation to the deceleration
parameter. One has so far not been able
to determine $\Omega_{\zeta 0}$ in this way.

However, we can indicate its present status. Ten years ago D. Rapetti et al.
\cite{Rapetti:2007} gave kinematical constraints on the deceleration parameter using type
Ia supernovae- and X-ray cluster gas mass fraction measurements, obtaining $q_0=-0.81 \pm
0.14$ at the 1$\sigma$ confidence
level. Inserting $q_0>-0.95$ in Eq.~(\ref{G3.16}) we obtain $\Omega_{\zeta 0}<0.27$.
However, some years later Giostri et al. \cite{Giostri:2012} used SN Ia and BAO/CMB
measurements and found $-0.42 < q_0 < -0.20$ with one light curve fitted, and $-0.66 <
q_0<-0.36$ with another. Note that if
measurements give $q_0<-0.55$ then $\Omega_{\zeta 0}<0$, which is unphysical.

 We mention though an interesting study of Mathews et al.
\cite{Mathews:2008hk}, in which the production of viscosity was associated with the decay
of dark matter particles into relativistic particles in a recent epoch with redshift
$z<1$.

Let us review the simplest viscous model in some detail. It was proposed by Padmanabhan
and Chitre already in 1987 \cite{Padmanabhan:1987dg}, and is based upon a dust model for
matter, vanishing cosmological constant, and constant viscosity
coefficient $\zeta=\zeta_0$.   Equation (\ref{1.27}) gives
\begin{equation}
\dot{H}=-\frac{3}{2}H^2+\frac{3}{2}\Omega_{\zeta 0}H_0 H, \label{1.35}
\end{equation}
which upon integration with $H( t_0)=H_0$ leads to
\begin{equation}
H=\frac{\Omega_{\zeta 0}H_0}{1-(1-\Omega_{\zeta 0})e^{\frac{3}{2}\Omega_{\zeta
0}H_0(t_0-t)}}. \label{1.36}
\end{equation}
Another integration with $a(t_0)=1$ gives
\begin{equation}
a=\left[ \frac{e^{\frac{3}{2}\Omega_{\zeta 0}H_0(t-t_0)}-(1-\Omega_{\zeta
0})}{\Omega_{\zeta 0}}\right]^{\frac{2}{3}}. \label{1.36a}
\end{equation}
This implies that the age of the universe when expressed in terms of the present
Hubble parameter $H_0$ becomes
\begin{equation}
t_0=\frac{4}{3\Omega_{\zeta 0}H_0}\, \rm{ arctanh}\left( \frac{\Omega_{\zeta
0}}{2-\Omega_{\zeta 0}}\right). \label{1.37}
\end{equation}
Hence, it is seen that for early times, in which $\Omega_{\zeta 0}H_0t \ll 1$, the
viscosity can
be neglected, and we obtain
\begin{equation}
a \approx \left[ 1+\frac{3}{2}H_0(t-t_0)\right]^{\frac{2}{3}}, \label{1.37a}
\end{equation}
corresponding to  the evolution of a dust universe. At late times $\Omega_{\zeta 0}H_0t
\gg 1$, the expansion becomes exponential with $H=\kappa \zeta_0, \, a \propto \exp
(\kappa
\zeta_0), \, \rho=3\kappa \zeta_0^2$, and thus the universe enters into a late
inflationary era with accelerated expansion. A drawback of this model is however that the
time when the bulk viscosity becomes dominant is predicted to be unrealistically large.

Let us now consider briefly the following model, which has attracted attention, namely the
one where viscosity is considered to be \cite{Ren:2005nw,Hu:2005fu,Mostafapoor:2013kha}
\begin{equation}
\zeta = \zeta_0+\zeta_1\frac{\dot{a}}{a}+\zeta_2\frac{\ddot{a}}{a}. \label{1.38}
\end{equation}
It is based on the physical idea that the dynamic state of the fluid influences its
viscosity. We
then obtain
\begin{equation}
a\dot{H}=-bH^2+cH+d, \label{1.39}
\end{equation}
where
\begin{equation}
a=1-\frac{3\kappa \zeta_2}{2}, \quad b=\frac{3}{2}[1+w-\kappa (\zeta_1+\zeta_2)], \quad
c=\frac{3\kappa \zeta_0}{2}, \quad d=\frac{1}{2}(1+w)\Lambda. \label{1.40}
\end{equation}
Integrating this equation  with  $a(0)=0, \, a(t_0)=1$  and assuming
$\kappa(\zeta_1+\zeta_2)<1$
and $w \geq 0$, which lead to $b>0$ and $4bd+c^2>0$, we obtain
\begin{equation}
   H(t)=\frac{c}{2b}+\frac{a}{b}\hat{H}\coth (\hat{H}t), \label{1.41}
\end{equation}
with $ \quad
\hat{H}^2=\frac{bd}{a^2}+\frac{c^2}{
4a^2}$.
  The age of the universe  in this model becomes
\begin{equation}
t_0=\frac{1}{\hat{H}}\, {\rm arctanh}\left( \frac{2a\hat{H}}{2bH_0-c}\right), \label{1.42}
\end{equation}
and thus   viscosity increases the age of the universe.  Hence, assuming that  $\kappa
\zeta_0 \ll H_0$  the increase of the age due to viscosity is roughly
$\Omega_{\zeta 0}^2 \,t_0$.

Let us return to the solution (\ref{1.41}) and apply it to the case where the universe
does not contain any matter  but only dark energy with $w=-1$. Moreover, we assume
a linear viscosity $(\zeta_1=\zeta_2=0)$ and therefore $ b=0$. Cataldo et al.
\cite{Cataldo:2005qh} found that in this case
\begin{equation}
\dot{H}=\frac{3\kappa \zeta_0}{2}H, \label{1.44}
\end{equation}
and thus integration with $a(t_0)=1$ gives
\begin{equation}
H(t)=H_0 \exp\left[\frac{3\Omega_{\zeta 0}H_0}{2}(t-t_0)\right], \label{1.45}
\end{equation}
\begin{equation}
 a(t)=\exp\left\{ \frac{2}{3\Omega_{\zeta 0}}\left[ e^{\frac{3\Omega_{\zeta
0}H_0}{2}(t-t_0)}-1\right]\right\}.  \label{1.46}
\end{equation}
Hence, a universe dominated by viscous dark energy with constant viscosity coefficient
expands exponentially faster comparing to the corresponding universe without
viscosity.

One may now ask the question how does the introduction of a bulk viscosity confront with
the observed acceleration of the universe. There have been several works dealing with
this issue, for  instance see Refs. \cite{Kremer:2002hz,Fabris:2005ts,Avelino:2008ph}. In
the model of Avelino and Nucamendi \cite{Avelino:2008ph} it was considered that
$\zeta_1=\zeta_2=0$, $w=0$, $\Omega_M=1$, $\Omega_\Lambda=0$, and therefore the
scale factor can be written as
\begin{equation}
a(t)=\left(\frac{1-\Omega_{\zeta 0}}{\Omega_{\zeta
0}}\right)^{2/3}\left(e^{\frac{3}{2}\Omega_{\zeta 0}H_0t}-1\right)^{2/3}, \label{1.47}
\end{equation}
which satisfies the boundary conditions  $a(0)=0, a(t_0)=1$.
The age of the universe in this model becomes
\begin{equation}
t_0=\frac{4}{3\Omega_{\zeta 0}H_0}\,{\rm arctanh} \left(\frac{\Omega_{\zeta
0}}{2-\Omega_{\zeta0}}\right)=-\frac{
2}{3\Omega_{\zeta 0}H_0}\ln (1-\Omega_{\zeta 0}). \label{1.48}
\end{equation}
Such a universe model was actually  considered earlier, by Brevik and Gorbunova
\cite{Brevik:2004pm,Brevik:2005bj} and by Gr{\o}n  \cite{gron10}, and is also similar to
the model of Padmanabhan and Chitre considered
above \cite{Padmanabhan:1987dg}. The Hubble parameter reads as
\begin{equation}
H(t)=\frac{\Omega_{\zeta 0}H_0}{1-e^{-(3/2)\Omega_{\zeta 0}H_0t}}, \label{1.49}
\end{equation}
and it approaches a de Sitter phase for $t \gg 1/\Omega_{\zeta 0}H_0$, with a
constant Hubble parameter equal to $\Omega_{\zeta 0}H_0$. The deceleration parameter  is
\begin{equation}
q=\frac{3}{2\exp[(3/2)\Omega_{\zeta 0} H_0 t]}-1, \label{1.50}
\end{equation}
and its value at present is
\begin{equation}
q(t_0)=(1-3\Omega_{\zeta 0})/2. \label{1.51}
\end{equation}
Hence
\begin{equation}
\Omega_{\zeta 0}=\frac{1}{3}(1-2q_0). \label{G3.34}
\end{equation}
Assuming that accurate measurements will verify the $\Lambda$CDM model, so that
$q_0=-0.55$, this
equation implies that $\Omega_{\zeta 0}=0.7$. This means that for the universe model to
be
realistic, there must exist a physical mechanism able to produce a viscosity of this
magnitude.

The expansion thus starts from a Big Bang with an infinitely large velocity, but
decelerates to a finite value. As usual, when $t=t_1$ determined by $q(t_1)=0$ there is a
transition to an accelerated eternal expansion, namely at
\begin{equation}
t_1=\frac{2\ln (3/2)}{3\Omega_{\zeta 0}H_0}, \label{1.52}
\end{equation}
at which time the scale factor is
\begin{equation}
a(t_1)=\left(\frac{1-\Omega_{\zeta 0}}{2\Omega_{\zeta 0}}\right)^{2/3}, \label{1.53}
\end{equation}
and the corresponding  redshift is
\begin{equation}
z_1=\left( \frac{2\Omega_{\zeta 0}}{1-\Omega_{\zeta 0}}\right)^{2/3}-1. \label{1.54}
\end{equation}
Under the assumption that this model contains a mechanism producing viscosity so that
$\Omega_{\zeta 0}=0.7$, this equation gives $z_1=0.8$. This is larger than the
corresponding value in the $\Lambda$CDM model. Hence the transition to accelerated
expansion happens earlier if the acceleration
of the expansion is driven by viscosity than by dark energy.

We deduce that the bulk viscosity must have been  sufficiently large, namely
$\Omega_{\zeta 0} > 1/3$, in order for this transition to have been realized in the past,
i.e. at  $a(t_1)<1$. Finally, note that for this universe model, with spatial curvature
$k=0$, the matter density is equal to the
critical density, namely
\begin{equation}
\rho=\frac{3H^2}{\kappa}=\frac{3\Omega_{\zeta 0}^2H_0^2}{\kappa \left[
1-e^{-(3/2)\Omega_{\zeta 0}H_
0 t}\right]^2}, \label{1.55}
\end{equation}
and thus the matter density approaches a constant value, $\rho \rightarrow
(3/\kappa)\Omega_{\zeta 0}^2 H_0^2$.

In the aforementioned study of Avelino and Nucamendi \cite{Avelino:2008ph} supernova data
were used in order to estimate the value of $\Omega_{\zeta 0}$,  giving the best fit for a
universe containing dust with constant viscosity coefficient. The result was that
$\Omega_{\zeta 0}=0.64$ had to be several orders of magnitude greater than estimates
based upon kinetic gas theory \cite{Brevik:1994cd}. However, as an unorthodox idea we may
mention here the probability for producing larger viscosity via dark
matter particles decaying into relativistic products \cite{Singh:2008zzj}.
Additionally, the comparison between the magnitude of bulk viscosity and astronomical
observations were also performed in a recent paper by Normann and Brevik
\cite{Normann:2016jns}, using the analyses of various experimentally-based sources
\cite{Wang:2013uka,Chen:2013vea}. Various ansatzes for the bulk viscosity were analyzed:
(i) $\zeta=$constant,  (ii) $\zeta \propto \sqrt{\rho}$, and (iii) $\zeta \propto \rho$.
The differences between the predictions of the options were found to be small.  As a
simple estimate based upon this analysis, we suggest that \begin{equation}
\zeta_0 \sim 10^6~\rm{Pa~s} \label{1.56}
\end{equation}
can serve as a reasonable mean estimate for the present viscosity. With
$H_0=67.7\,
\rm{km}\,\rm{s}^{-1}
\, \rm{Mpc}^{-1}$
this corresponds to
$\Omega_{\zeta 0}=0.01$.

The behavior of a viscous universe in its final stages has been discussed in
\cite{Brevik:2004pm,Brevik:2005bj} and in
\cite{Cataldo:2005qh,Lepe:2017yvs}. Consider
first a universe
without viscosity and dark energy, containing only a non-viscous fluid with $p=w\rho$. In
this case Eq. (\ref{1.39})    reduces to
\begin{equation}
\dot{H}=-bH^2,  \label{1.57}
\end{equation}
where $b=\frac{3}{2}(1+w)$.
For such a universe there is a Big Rip at
\begin{equation}
t_{R0}=t_0+\frac{2}{3(1+w)H_0}. \label{1.58}
\end{equation}

On the other hand, in Ref.~\cite{Cataldo:2005qh}  a fluid was considered to have $w<-1$
and constant viscosity coefficient $\xi_0$, implying   $b<0$ and $d=0$. In this case
(\ref{1.39})  reduces to
\begin{equation}
\dot{H}=-\frac{3}{2}(1+w)H^2+\frac{3}{2} \Omega_{\zeta 0} H_0 H. \label{1.59}
\end{equation}
The Hubble parameter, scale factor and density for this universe   are respectively
extracted to be
\begin{equation}
H=\frac{H_0}{\frac{1+w}{\Omega_{\zeta 0}}+\left( 1-\frac{1+w}{\Omega_{\zeta
0}}\right)e^{-\frac{3}{
2}\Omega_{\zeta 0}(t-t_0)}}, \label{1.60}
\end{equation}
\begin{equation}
a=\left[ 1-\frac{1+w}{\Omega_{\zeta
0}}+\frac{1+w}{\Omega_{\zeta_0}}e^{\frac{3}{2}\Omega_{\zeta 0}H_
0(t-t_0)}\right]^{\frac{2}{3(1+w)}}, \label{1.61}
\end{equation}
and
\begin{equation}
\rho=\frac{\rho_0}{\left[ \frac{1+w}{\Omega_{\zeta 0}}+\left( 1-\frac{1+w}{\Omega_{\zeta
0}}\right)
e^{-\frac{3}{2}\Omega_{\zeta 0}(t-t_0)}\right]^2}. \label{1.62}
\end{equation}
Thus, in this case there is a Big Rip singularity at
\begin{equation}
t_R=t_0+\frac{2}{3\Omega_{\zeta 0}H_0}\ln \left( 1-\frac{\Omega_{\zeta 0}}{1+w}\right).
\label{1.63}
\end{equation}
Similar models, with variable gravitational and cosmological ``constants'' have
been investigated by Singh et al. \cite{Singh:1998bz,Singh:2007ht}. Furthermore, one can
go beyond  isotropic geometry and study viscous fluids in spatially anisotropic spaces,
belonging to  the Bianchi type-I class. The interested reader might consult, for instance,
the discussion in Ref.~\cite{Brevik:2014cxa}.

\subsection{Inhomogeneous equation of state of the universe: phantom era and
singularities}
\label{subsectsingularities22}

In this subsection we examine the appearance of singularities in viscous cosmology.
It is well-known that in FRW geometry, when the equation of state modeling the
matter content is a linear equation with an equation of state parameter greater than
$-1$, the  Big Bang singularity appears at early times, where the energy density of
the universe diverges. Moreover, dealing with nonlinear equations of state one can see
that other kind of singularities such as Sudden singularity
\cite{Barrow:2004xh,Nojiri:2004ip,Barrow:2004hk} or Big Freeze
\cite{Nojiri:2005sx,Nojiri:2009pf, Astashenok:2012kb,BouhmadiLopez:2007qb} appear.

In fact, the future
singularities are classified as follows \cite{Nojiri:2005sx} (see
also \cite{Fernandez-Jambrina:2014sga} for a more detailed classification):
\begin{itemize}
\item Type I (Big Rip): $t\to t_s$, $a\to\infty$, $\rho\to\infty$ and $|p|\to\infty$.
\item Type II (Sudden): $t\to t_s$, $a\to a_s$, $\rho\to\rho_s$ and $|p|\to\infty$.
\item Type III (Big Freeze): $t\to t_s$, $a\to a_s$, $\rho\to\infty$ and $|p|\to\infty$.
\item Type IV (Generalized Sudden): $t\to t_s$, $a\to a_s$, $\rho\to 0$, $|p|\to 0$ and
derivatives of $H$ diverge.
\end{itemize}
Similarly to the future ones, one can define the past singularities:
\begin{itemize}
\item Type I (Big Bang): $t\to t_s$, $a\to0$, $\rho\to\infty$ and $|p|\to\infty$.
\item Type II (Past Sudden): $t\to t_s$, $a\to a_s$, $\rho\to\rho_s$ and $|p|\to\infty$.
\item Type III (Big Hottest): $t\to t_s$, $a\to a_s$, $\rho\to\infty$ and $|p|\to\infty$.
\item Type IV (Generalized past Sudden): $t\to t_s$, $a\to a_s$, $\rho\to 0$, $|p|\to 0$
and derivatives of $H$ diverge.
\end{itemize}

For the simple case of a linear equation of state $p=w\rho$ it is
well-known that for a non-phantom fluid ($w>-1)$ one obtains a Big Bang singularity,
while for a phantom fluid ($w<-1)$
\cite{Caldwell:1999ew,Chen:2008ft,Saridakis:2008fy,Elizalde:2009gx,elizalde:2004mq} the
singularity
is a
future Type I (Big Rip). Hence, in order to obtain the other type of singularities one has
to consider phantom fluids modeled by
non-linear equations of state of the form
\begin{equation}
p=-\rho-f(\rho),
\end{equation}
 where $f$ is a positive
function. The simplest model is obtained taking $f(\rho)=A\rho^{\alpha}$ with $A>0$. In
this case from the conservation equation $\dot{\rho}=-3H(\rho+p)$ and the Friedmann
equation $H^2=\frac{\kappa \rho}{3}$ one
obtains the dynamical equation
\begin{equation}\label{4.59}
\dot{\rho}=\sqrt{3\kappa}A\rho^{\alpha+\frac{1}{2}},
\end{equation}
whose solution is
\begin{eqnarray}\label{4.60}
\rho=\left\{\begin{array}{ccc}
\left[ \frac{\sqrt{3\kappa}A}{2}(t-t_0)(1-2\alpha)+
\rho_0^{\frac{1}{2}-\alpha}\right]^{\frac{2}
{1-2\alpha}}&\mbox{when}& \alpha\not=\frac{1}{2}\\
\rho_0e^{\sqrt{3\kappa}A(t-t_0)}&\mbox{when}& \alpha=\frac{1}{2}.
\end{array}\right.
\end{eqnarray}
Furthermore, in order to obtain the evolution of the scale factor we will integrate the
conservation equation,
resulting in
\begin{equation}\label{4.61}
a=a_0\mbox{exp}\left( \frac{1}{3}\int_{\rho_0}^{\rho}\frac{\bar\rho
d\bar\rho}{f(\bar\rho)}
\right),
\end{equation}
which using (\ref{4.60}) leads to
\begin{eqnarray}\label{4.62}
a=\left\{\begin{array}{ccc}
a_0\mbox{exp}\left[ \frac{1}{3A(1-\alpha)} (\rho^{1-\alpha}-\rho_0^{1-\alpha})\right]
&\mbox{when}&
\alpha\not=1\\
a_0\left( \frac{\rho}{\rho_0} \right)^{\frac{1}{3A}}&\mbox{when}& \alpha=1.
\end{array}
\right.\end{eqnarray}

Once we have calculated these quantities, we have the following different situations (see
also \cite{Nojiri:2005sx}):
\begin{enumerate}
\item When $\alpha<0$ we have a past singularity of Type II, since the energy density
vanishes for
$t_s=t_0-\frac{2}{\sqrt{3\kappa}A}\frac{\rho_0^{\frac{1}{2}-\alpha}}{1-2\alpha}<t_0$,
implying that
the pressure diverges at $t=t_s$.
\item When $\alpha=0$  there are no singularities. The dynamics is defined from
$t_s=t_0-\frac{2}{\sqrt{3\kappa}A}\sqrt{\rho_0}$ (where the energy density is zero) up to
 $t\rightarrow\infty$.
\item When $0<\alpha<\frac{1}{2}$  there are two different cases:
\begin{enumerate}
\item  $\frac{1}{1-2\alpha}$  is not a natural number. One has a past Type IV singularity
at $t_s=t_
0-\frac{2}{\sqrt{3\kappa}A}\frac{\rho_0^{\frac{1}{2}-\alpha}}{1-2\alpha}$, since higher
derivatives
 of $H$ diverge at $t=t_s$.
\item $\frac{1}{1-2\alpha}$  is  a natural number. In that case there are not any
singularites and the
dynamics is defined from
$t_s=t_0-\frac{2}{\sqrt{3\kappa}A}\frac{\rho_0^{\frac{1}{2}-\alpha}}{1-2\alpha}$  to
 $t\rightarrow\infty$.
\end{enumerate}
\item When $\alpha=\frac{1}{2}$ there are no singularities in cosmic time.
\item When $\frac{1}{2}<\alpha<1$, one has  future  Type I singularities, since in this
case $\rho$, $p$ and $a$ diverge at
 $t_s=t_0-\frac{2}{\sqrt{3\kappa}A}\frac{\rho_0^{\frac{1}{2}-\alpha}}{1-2\alpha}>t_0$.
 \item When $\alpha=1$ the equation of state is linear, and thus we obtain a Big Rip
singularity.
 \item When $\alpha>1$, the energy density and the pressure diverge but the scale factor
remains
finite at $t=t_s$, implying  that we have a  future Type III singularity.
\end{enumerate}

The remarkable case appears when $0<\alpha<\frac{1}{2}$ and with $\frac{1}{1-2\alpha}$
being a natural number. In this case, from the Friedmann equation
$H^2=\frac{\kappa\rho}{3}$ and the solution (\ref{4.60}) one obtains
\begin{equation}\label{4.63}
H=\sqrt{\frac{\kappa}{3}}\left[ \frac{\sqrt{3\kappa}A}{2}(t-t_0)(1-2\alpha)+
\rho_0^{\frac{1}{2}
-\alpha}\right]^n,
\end{equation}
with $n=\frac{1}{1-2\alpha}$. As we have already seen, this solution describes a universe
in the expanding phase driven by a phantom fluid, which is defined from
$t_s=t_0-\frac{2}{\sqrt{3\kappa}A}\frac{\rho_0^{\frac{1}{2}-\alpha}}{1-2\alpha}$  (where
$H=0$) up to $t\rightarrow\infty$. However,   solution (\ref{4.63}) could be extended
analytically back in time.
There are two different cases: When $n$ is odd, this extended solution describes a
universe driven by a phantom field that goes from the contracting to expanding phase,
bouncing at time $t_s$. On the contrary, when $n$ is even the universe moves always in
the expanding phase, and before $t_s$ it is driven by a non-phantom field, while after
$t_s$ the universe enters in a phantom era. We will explain this phenomenon in more
detail in the next subsection.

Motivated by the introduction of  bulk viscous terms in an ideal fluid
one can consider a subclass of the general equation of state of (\ref{visceosinh00}) of
the form
\begin{equation}\label{4.64}
p=-\rho-f(\rho)+G(H).
\end{equation}
Then, the conservation equation becomes $\dot{\rho}=3H[f(\rho)-G(H)]$, and using the
Friedmann equation (\ref{fr1general}) in the expanding phase
leads to
\begin{equation}\label{4.65}
\dot{\rho}=3H\left[f(\rho)-G\left(\sqrt{\frac{\kappa\rho}{3}}\right)\right]\equiv
3HF(\rho),
\end{equation}
which reveals  that this formalism is equivalent with considering a fluid with an
effective equation of state given by
\begin{eqnarray}\label{4.66}
p=-\rho-F(\rho)=-\rho-f(\rho)+G\left(\sqrt{\frac{\kappa\rho}{3}}\right).
\end{eqnarray}

It is clear that, in general, the equation of state (\ref{4.64}) does not lead to a
universe crossing the phantom barrier. A simple way to obtain transitions from the
non-phantom to the phantom regime is  to explicitly consider   an
inhomogeneous equation of state of the form
$F(\rho,p, H)=0$, for example \cite{Nojiri:2005sr,Nojiri:2005pu}
\begin{eqnarray}\label{4.67}
(\rho+p)^2-C_0\rho^2\left(1-\frac{H_0}{H}\right)=0,
\end{eqnarray}
with $C_0$ and $H_0$ some positive constants. Inserting this into the square
of the equation $\dot{H}=-\frac{\kappa}{2}(\rho+p)$, one obtains
 the bi-valued dynamical equation
\begin{eqnarray}\label{4.68}
\dot{H}^2=\frac{9}{4}C_0 H^4\left(1-\frac{H_0}{H}\right).
\end{eqnarray}
From this equation,  since  there are two  square roots and  the effective equation of
state parameter is given by $w_{eff}\equiv -1-\frac{2\dot{H}}{3H^2}$,
one can see that there are  two different dynamics: one which corresponds to the
branch with $\dot{H}<0$ describing a universe in a non-phantom regime, and  one
corresponding
to the branch $\dot{H}>0$ describing a universe in the phantom era.
In fact,     (\ref{4.68}) can be integrated as
\begin{eqnarray}\label{4.69}
H(t)=\frac{16}{9C_0^2H_0(t-t_-)(t_+-t)},
\end{eqnarray}
where we have introduced the notation $t_{\pm}=\pm \frac{4}{3C_0H_0}$.
It is easy to check that $H(t)$ is only defined for times between $t_-$ and $t_+$,
since at $t_{\pm}$ the Hubble function $H$ diverges (we obtain a Big Bang at $t_-$ and a
Big Rip at $t_+$). Moreover, it is a decreasing function for $t\in (t_-,0)$ and
an increasing one for $t\in (0, t_+)$, implying that at $t=0$ the universe
crosses the phantom divide (it passes from the non-phantom to the phantom era).

Another interesting example arises from the equation of state
\begin{eqnarray}\label{4.70}
(\rho+p)^2 +\frac{16H_1}{\kappa^2 t_0^2}(H_0-H)\ln\left(\frac{H_0-H}{H_1}  \right)=0,
\end{eqnarray}
where $t_0,H_0, H_1$ are   parameters satisfying $H_0>H_1>0$. The corresponding
bi-valued dynamical equation is
\begin{eqnarray}\label{4.71}
\dot{H}^2=-\frac{4H_1}{t_0^2}(H_0-H)\ln\left(\frac{H_0-H}{H_1}  \right),
\end{eqnarray}
which has two fixed points, namely $H_0$ and $H_0-H_1$.
As we have already explained, when $\dot{H}<0$  (resp. $\dot{H}>0$ ) the universe is in a
non-phantom (resp. phantom) era. When the universe is in the branch with $\dot{H}<0$
 it moves from $H_0$ to $H_0-H_1$, it reaches $H=H_0$ and then it enters in the other
branch ($\dot{H}>0$) going from $H_0-H_1$ to $H_0$. In fact, in
\cite{Nojiri:2005sr,Nojiri:2005pu} the authors found the following
solution:
\begin{eqnarray}\label{4.72}
H(t)=H_0-H_1 \mbox{exp}\left(-\frac{t^2}{t_0^2}\right),
\end{eqnarray}
which satisfy all the properties described above.

 A final remark is in order: One can indeed consider the  more general equation of
state given in  (\ref{visceosinh00}), namely of the form
$F(\rho,p,H, \dot{H},\ddot{H},\cdots)=0 $, containing higher order derivatives of the
Hubble parameter. In this case, using the Friedmann equations the equation of state
becomes the dynamical equation
\begin{eqnarray}\label{4.74}
F\left(\frac{3H^2}{\kappa},
-\frac{2\dot{H}}{\kappa}-\frac{3H^2}{\kappa},\dot{H},\ddot{H},
\cdots\right)=0  .
\end{eqnarray}
A non-trivial example is the following equation of state
\cite{Nojiri:2005sr,Nojiri:2005pu}:
\begin{equation}\label{4.75}
p=w\rho-G_0-\frac{2}{\kappa}\dot{H}+G_1\dot{H}^2,
\end{equation}
where $G_0$ and $G_1$ are constant.  Then, the dynamical equation becomes
\begin{eqnarray}\label{4.76}
-\frac{3H^2(1+w)}{\kappa}=-G_0+G_1\dot{H}^2.
\end{eqnarray}
We look for periodic solutions of the form $H(t)=H_0\cos(\Omega t)$ depicting an
oscillatory universe. Inserting this expression into (\ref{4.76}) we obtain the
algebraic system:
\begin{eqnarray}\label{4.77}
G_0=G_1\Omega^2H_0^2,\qquad G_0=\frac{3H_0^2(1+w)}{\kappa},
\end{eqnarray}
whose solution is given by
\begin{eqnarray}\label{4.78}
H_0=\sqrt{\frac{\kappa G_0}{3(1+w)}},\qquad \Omega=\sqrt{\frac{3(1+w)}{\kappa
G_1}},\end{eqnarray}
provided that $G_0(1+w)>0$ and $G_1(1+w)>0$.
On the other hand, when $G_1(1+w)<0$, one can look for solutions of the form
$H(t)=H_0\cosh(\Omega
t)$, obtaining
\begin{eqnarray}\label{4.79}
H_0=\sqrt{\frac{\kappa G_0}{3(1+w)}},\qquad \Omega=\sqrt{-\frac{3(1+w)}{\kappa
G_1}}.\end{eqnarray}

\subsection{Unification of inflation with dark energy  in viscous cosmology}

In this subsection we analyze how one can describe in a unified way the early
(inflationary) and late time acceleration, in the framework of viscous cosmology.
The simplest way to unify early inflationary epoch with the current cosmic acceleration
is by using scalar fields \cite{Hossain:2014zma}. Starting with the action
\begin{eqnarray}\label{h1}
S=\int d^4x\sqrt{-g}\left\{
\frac{1}{2\kappa}R-\frac{1}{2}\omega(\phi)\partial_{\mu}\phi\partial^{\mu}\phi-V(\phi)
\right\},
\end{eqnarray}
where $\omega$ and $V$ are   functions of the scalar field $\phi$, and focusing on
flat FRW geometry, one obtains the following dynamical equation
\begin{eqnarray}\label{h2}
\omega(\phi)\ddot{\phi}+\frac{1}{2}\omega'(\phi)\dot{\phi}^2+3H\omega(\phi)\dot{\phi}
+V'(\phi)=0.
\end{eqnarray}
The relevant fact, is that given a function $f(\phi)$ the equation (\ref{h2}) has always
the solution $\phi=t$ and $H=f(t)$, provided that  (for details see
\cite{Nojiri:2005sr,Nojiri:2005pu})
\begin{eqnarray}\label{h3}
&&\omega(\phi)=-\frac{2}{\kappa}f'(\phi), \\
&&V(\phi)=
\frac{1}{\kappa}\left[3f^2(\phi)+f'(\phi)  \right].
\end{eqnarray}

An interesting example is obtained when one considers the function
\begin{eqnarray}\label{h4}
f(\phi)=H_0 \left(\frac{\phi_s}{\phi}+\frac{\phi_s}{\phi_s-\phi}\right),
\end{eqnarray}
where $H_0$ and $\phi_s$ are the two positive parameters of the model. In this case one
has
\begin{eqnarray}\label{h5}
&&\omega(\phi)= \frac{2H_0\phi_s^2(\phi_s-2\phi)}{\kappa\phi^2(\phi_s-\phi)^2},
\\
&&V(\phi)=\frac{H_0\phi_s^2}{\kappa \phi^2(\phi_s-\phi)^2}(3H_0\phi_s^2-\phi_s+2\phi),
\end{eqnarray}
whose dynamics is given by
\begin{eqnarray}\label{h6}
H=\frac{H_0t_s^2}{t(t_s-t)}, \quad a=a_0\left(\frac{t}{t_s-t}\right)^{H_0t_s},
\end{eqnarray}
where we have introduced the notation $t_s=\phi_s$.
Since $H$ diverges at $t=0$ and $t=t_s$, the dynamics is defined in $(0,t_s)$. In fact at
$t=0$ one has $a=0$, which means that we obtain a Big Bang singularity, while at
$t=t_s$ the scale factor diverges, implying that we have a Big Rip singularity.
On the other hand, the derivative of the Hubble parameter reads as
 \begin{eqnarray}\label{h7}
 \dot{H}=\frac{H_0t_s^2}{t^2(t_s-t)^2}(2t-t_s),
 \end{eqnarray}
 that is the universe lies in the non-phantom regime when $0<t<t_s/2$, while it lies
in the phantom phase for $t_s/2<t<t_s$.  Hence, we conclude that this model could
describe the current cosmic acceleration.

In order to examine the behavior at early times, we note that near $t=0$ one can make
the approximation $a=a_0\left(\frac{t}{t_s}\right)^{H_0t_s}$, and thus its second
derivative at early times is approximately
\begin{eqnarray}\label{h8}
\ddot{a}=a\frac{H_0t_s(H_0t_s-1)}{t^2}.
\end{eqnarray}
From this we deduce that if one chooses $H_0t_s>1$  the universe will have an early
period of acceleration.

Another  example is to consider
\begin{eqnarray}\label{h9}
f(\phi)=H_0\sin(\nu \phi),
\end{eqnarray}
with $H_0$ and $\nu$  positive parameters. A straightforward calculation leads to
\begin{eqnarray}
&&\omega(\phi)=-\frac{2H_0\nu}{\kappa}\cos(\nu \phi)\\
&&
V(\phi)= \frac{2}{\kappa}\left[H_0\nu\cos(\nu\phi)+H_0^2\sin^2(\nu\phi)\right].
\end{eqnarray}
In this case one obtains a non-singular oscillating universe, whose dynamics is given by
\begin{eqnarray}\label{h10}
&&H=H_0\sin(\nu t),\\
&&a=a_0\mbox{exp}\left[-\frac{H_0}{\nu}\cos(\nu t)  \right].
\end{eqnarray}
This solution depicts a universe that bounces at time $t=\frac{n\pi}{\nu}$ where $n$ is
an integer,
and since $\dot{H}=H_0\nu\cos(\nu t)$ one can easily check that
the universe lies in the phantom regime when $\frac{\pi}{\nu}\left(-\frac{1}{2}+2n
\right)<t<
\frac{\pi}
{\nu}\left(\frac{1}{2}+2n  \right)$, while it is in the non-phantom phase when
$\frac{\pi}{\nu}\left(\frac{1}{2}+2n  \right)<t< \frac{\pi}{\nu}\left(\frac{3}{2}+2n
\right)$.

We proceed by considering viscosity, taking $\Lambda=0$ and $w=1$ in (\ref{1.27}). Based
on the equivalence between bulk viscous and open cosmology, where isentropic particle
production is allowed  \cite{prigogine}, we choose the following viscosity coefficient
\cite{Haro:2015ljc}:
\begin{eqnarray}
\label{h11}
 \zeta(H)=\frac{1}{\kappa}
    \left(-\xi_0+2H+\frac{\xi_0^2}{8H}\right),
\end{eqnarray}
where $\xi_0>0$ is a constant. Hence, the second Friedmann equation (\ref{1.27}) becomes
\begin{eqnarray}\label{h12}
\dot{H}=-\frac{3}{2}H\xi_0+\frac{3}{16}\xi_0^2,
\end{eqnarray}
which only has $H=\frac{\xi_0}{8}$ as a fixed point. If one considers the dynamics in the
domain $\frac{\xi_0}{8}\leq H\leq  \infty$, it is easy to check that the effective
equation of state parameter is greater than $-1$, which implies that the Hubble parameter
varies from infinity to $\frac{\xi_0}{8}$.
Moreover, since
\begin{eqnarray}\label{h13}
w_{eff}=-1+\frac{\xi_0}{H}- \frac{\xi_0^2}{8H^2},\end{eqnarray}
$w_{eff}\cong -1$ at early ($H\gg \xi_0$) and late ($H\cong \frac{\xi_0}{8}$) times,
from which we deduce that this viscous fluid model unifies inflation with the current
cosmic acceleration. Additionally, $w_{eff}$ is positive when
$\xi_0\frac{\sqrt{2}-1}{2\sqrt{2}}<H<\frac{\sqrt{2}+1}
{2\sqrt{2}}$, having the maximum value $w_{eff}=1$ at $H=\frac{\xi_0}{4}$. Thus, in
summary, in the scenario at hand the universe starts from an inflationary epoch,  it
evolves through a Zel'dovich fluid ($w_{eff}=1$), radiation- ($w_{eff}=1/3$) and
matter-  ($w_{eff}=0$) dominated epochs, and finally it enters into late-time acceleration
tending towards a de Sitter phase.

The solution of   equation (\ref{h12}) is
\begin{eqnarray}\label{h14}
H=\frac{\xi_0}{8}\left( e^{-\frac{3}{2}\xi_0 t}+1\right),
\end{eqnarray}
and the scalar field  that induces this dynamics, if one chooses  $\omega(\phi)\equiv
1$, has the
following Higgs-style potential (for details see \cite{Haro:2015ljc}):
\begin{eqnarray}\label{h15}
 V(\phi)=
  \frac{27\xi_0^2\kappa}{256}\left(\phi^2-\frac{2}{3\kappa}\right)^2.
\end{eqnarray}
 We stress here that this unified model for inflation and late-time acceleration
leads to inflationary observables, namely    the spectral index,
its running and the  ratio of tensor to scalar perturbations, that match at $2\sigma$
Confidence Level with the observational data provided by Planck 2015 announcements
\cite{Ade:2015xua} (for a detailed discussion see \cite{Haro:2015ljc,deHaro:2016hpl}).

We close this section by considering a very simple quintessential-inflation potential
which unifies
inflation with late time acceleration, namely \cite{deHaro:2016cdm}
\begin{eqnarray}\label{h16}
V(\phi)=\left\{\begin{array}{ccc}
\frac{9}{2}\left(H_E^2-\frac{\Lambda}{3}\right)\left( \phi^2-\frac{2}{3\kappa} \right)&
\mbox{for} &
 \phi\leq \phi_E\\
\frac{\Lambda}{\kappa}& \mbox{for} & \phi\geq \phi_E,
\end{array}\right.
\end{eqnarray}
where $\phi_E\equiv -\sqrt{\frac{2}{3\kappa}}\frac{H_E}{\sqrt{H_E^2-\frac{\Lambda}{3}}}$,
and with $H_E>0$ the parameter of the model.
This model leads to the following dynamics
\begin{eqnarray}\label{h17}
\dot{H}=\left\{\begin{array}{ccc}
-3H_E^2+\Lambda& \mbox{for}& H\geq H_E\\
-3H^2+\Lambda& \mbox{for}& H\leq H_E,
\end{array}\right.
\end{eqnarray}
whose solution has the following expression
\begin{eqnarray}\label{h18}
 H(t)=\left\{\begin{array}{cc}
      \left(-3H_E^2+{\Lambda}  \right)t +1 & t\leq 0\\
      \sqrt{\frac{\Lambda}{3}}\frac{3H_E+\sqrt{3\Lambda}\tanh(\sqrt{3\Lambda}
t)}{3H_E\tanh(\sqrt{3\Lambda} t)+\sqrt{3\Lambda}} & t\geq 0,
             \end{array}\right.
\end{eqnarray}
with the corresponding scale factor
\begin{eqnarray}\label{h19}
 a(t)=\left\{\begin{array}{cc}
      a_E e^{\left[\left(-3H_E^2+{\Lambda}  \right)\frac{t^2}{2}+t\right]}
       & t\leq 0\\
    a_E \left[\frac{3H_E}{\sqrt{3\Lambda}}\sinh(\sqrt{3\Lambda} t)+\cosh(\sqrt{3\Lambda}
t)\right]^{
\frac{1}{3}}  & t\geq 0.
             \end{array}\right.
\end{eqnarray}
We mention that this dynamics arises also from a universe filled with a fluid with
the simple linear equation of state of the form
\begin{eqnarray}\label{h20}
 p=\left\{\begin{array}{cc}
      -\rho+2{\rho_E} -\frac{2\Lambda}{\kappa} & \rho\geq \rho_E\\
      \rho-\frac{2\Lambda}{\kappa} &  \rho\leq \rho_E ,
             \end{array}\right.
\end{eqnarray}
where $\rho_E=\frac{3H_E^2}{\kappa}$. Equivalently it can arise from a viscous fluid,
since effectively, choosing $w=1$ and  the  following viscosity coefficient
\begin{eqnarray}\zeta=\left\{\begin{array}{cc}
\frac{2}{\kappa}\left(H-\frac{H_E^2}{H}\right)& H\geq H_E\\
0& H\leq H_E,
\end{array}\right.
\end{eqnarray}
and inserting it into (\ref{1.27}) one obtains the dynamics (\ref{h17}).
Finally, for this model the effective equation-of-state parameter is
given by
\begin{eqnarray}\label{h21}
 w_{eff}=\left\{\begin{array}{cc}
      -1+\frac{2}{3H^2}\left(3H_E^2-\Lambda\right)   & H\geq H_E\\
      1- \frac{2\Lambda}{3H^2}&  H\leq H_E,
             \end{array}\right.
\end{eqnarray}
which shows that for $H\gg H_E$ one has
$w_{eff}(H)\cong -1$ (early quasi - de Sitter period). When $H\cong H_E$, the
equation-of-state parameter satisfies
$w_{eff}(H)\cong 1$ (deflationary period dominated by a Zel'dovich fluid), and lastly for
$H\cong
\sqrt{\frac{\Lambda}{3}}$ one also acquires  $w_{eff}(H)\cong -1$ (late quasi -
de Sitter period).

\subsection{Generalized holographic dark energy with a viscous fluid}

In this subsection we investigate the cosmological scenario of holographic dark energy
with the presence of a viscous fluid. Holographic dark energy
\cite{Li:2004rb,Hsu:2004ri} is a scenario in the direction of
incorporating the nature of dark energy using some basic quantum gravitational principles.
It is based on black hole thermodynamics \cite{Myers:1986un} and the connection of the
ultraviolet cut-of of a quantum field theory, which induces the vacuum energy, with the
largest distance of this theory \cite{Cohen:1998zx}.
Determining suitably an IR cut-off  $L$, and imposing
that the total vacuum energy in the maximum volume
cannot be greater than the mass of a black hole of the same
size,   one obtains the holographic dark energy, namely
\begin{equation}
\rho_{DE}=\frac{3c^2}{\kappa L^2},
\label{HFEKurd}
\end{equation}
with $\kappa$ the gravitational constant, set to $ \kappa=1$ in the following for
simplicity, and $c$ a parameter. The holographic dark energy scenario has interesting
cosmological applications
\cite{Nojiri:2005pu,Elizalde:2005ju,Saridakis:2007cy,Saridakis:2007ns,Saridakis:2007wx}.
Concerning the ultraviolet cut-off one uses the future event horizon $L_{f}$
\begin{equation}
L_{f} = a\int_{t}^{\infty}{\frac{dt}{a}}.
\label{HFEKurdLfdef}
\end{equation}
However, one can generalize the model using the quadratic  Nojiri-Odintsov cut-off $L$
defined
as
\citep{Nojiri:2005pu,Khurshudyan:2016uql,Khurshudyan:2016zse}
\begin{equation}
\label{NojOdcuof}
\frac{c}{L} = \frac{1}{L_{f}} \left ( \alpha_{0} + \alpha_{1} L_{f} + \alpha_{2}
L_{f}^{2}\right)
\end{equation}
with $c$, $\alpha_{0}$, $\alpha_{1}$ and $\alpha_{2}$ constants, or the generalized
Nojiri-Odintsov cut-off defined in Refs.~\cite{Nojiri:2005pu} and \cite{Nojiri:2017opc}.

In this subsection we will consider the scenario in which generalized holographic dark
energy interacts with a viscous fluid, following \cite{Khurshudyan:2016gmb}. In
particular, we consider a dark matter sector with a viscous equation of state of the form
\begin{equation}
\label{viscousPDM22}
p_{DM} = - \rho_{DM} + \rho_{DM}^{\alpha} + \chi H ^{\beta},
\end{equation}
where $\rho_{DM}$ and $p_{DM}$ are respectively the dark matter energy density and
pressure, and with $\alpha$, $\chi$ and $\beta$ the model parameters. Furthermore, we
allow for an interaction between viscous dark matter and holographic dark energy:
\begin{equation}
\label{rhoDEevolKurd}
\dot{\rho}_{DE} + 3 H \rho_{DE}(1+w_{DE}) = -Q,
\end{equation}
\begin{equation}
\label{rhoMevolKurd}
\dot{\rho}_{DM} + 3 H \rho_{DM} (1+ w_{DM}) = Q,
\end{equation}
with $w_{DE}$ and $w_{DM}$  respectively the equation-of-state parameters of the dark
energy and dark matter sectors, and where $Q$ is a function that determines the
interaction. One can impose the following form for $Q$ \cite{Khurshudyan:2016gmb}
\begin{equation}
\label{Qformmod1}
Q =  3 H b (\rho_{DE} + \rho_{DM}),
\end{equation}
where $b$ is a constant,
although more complicated forms could also be used \cite{Chen:2011cy}. Finally,  the
first Friedmann equation reads as
\begin{equation}\label{H2visHDE1}
H^{2} = \frac{1}{3} \rho_{eff},
\end{equation}
where the effective (total) energy density is given by $\rho_{eff} = \rho_{DE} +
\rho_{DM}$.

We start our analysis by investigating the non-interacting scenario, that is setting $Q$
(i.e. $b$) to zero. In this case, using
(\ref{HFEKurd}),(\ref{HFEKurdLfdef}),(\ref{viscousPDM22}) and (\ref{H2visHDE1}),  the
deceleration parameter $q\equiv  -1-\dot{H}/H^2$ is found to be
\begin{equation}
\label{qkurdmod1}
q = \frac{-2 \sqrt{\Omega _{de}} \dot{L}_{f} (\alpha_{1}+2 \alpha_{2}
L_{f})-\hat{q}_{0}}{2 H^2 L_{
f}},
\end{equation}
where $\hat{q}_{0} = H \Omega _{de} (\dot{L}_{f}+1)+L_{f} \left(H^2+p_{DM}\right)$.
Moreover, the evolution of the dark matter density parameter $\Omega_{DM}$ is determined
by the differential equation
\begin{equation}
\Omega^{\prime}_{DM} = \frac{2 \Omega _{DE}^{3/2} \hat{L}_{f} - 2 \sqrt{\Omega _{DE}}
\hat{L}_{f} -
\hat{A}_{0}}{H^2 L_{f}},
\end{equation}
with $\hat{L}_{f} = \dot{L}_{f} (\alpha_{1} + 2 \alpha_{2} L_{f})$ and $\hat{A}_{0} =
\Omega _{DE} [H ( \dot{L}_{f}+1)+L_{f} p_{DM}] + H \Omega _{DE}^2 ( \dot{L}_{f}+1)$, and
where primes denote differentiation with respect to $N = \ln a$.
Finally, for the non-interacting case (\ref{HFEKurd}) and  (\ref{rhoDEevolKurd}) lead to
\begin{equation}
\label{wDEKurfmod1}
 w_{DE}=-1+\frac{2\dot{L}}{3HL}=-1+\frac{2\dot{L}_f}{3HL_f  }\left[
1
 -\frac{ L_f   \left ( \alpha_1 + 2\alpha_2 L_f    \right) }{\left (
\alpha_{0} + \alpha_{1} L_{f} + \alpha_{2}
L_{f}^{2}\right) }
  \right] .
\end{equation}
As one can see, the deceleration parameter $q$ starts from positive values, it decreases,
and it becomes negative marking the passage to late-time accelerated phase
\cite{Khurshudyan:2016gmb}. The role of the viscosity parameter $\chi$ is significant,
since larger positive $\chi$ leads the transition redshift $z_{tr}$ (from deceleration to
acceleration) to smaller values and the present deceleration parameter to negative values
closer to zero. Hence, the larger the fluid viscosity is the more difficult  it is for the
universe to  exhibit accelerated expansion. Lastly, an important feature is that the dark
energy equation of state parameter can exhibit the phantom divide-crossing, as can be seen
from (\ref{wDEKurfmod1}) \cite{Nojiri:2005pu,Khurshudyan:2016gmb,Setare:2008pc}.
In Tables \ref{tTable1Kurd1} and \ref{tTable1Kurd2} we present the various
calculated values for different choices of the model parameters, where the features
described above are obvious.
\begin{table}[ht!]
  \centering
  \caption{The present-day values of the deceleration parameter $q$, of the dark energy
equation-of-state parameter $w_{DE}$ and its derivative $w^{\prime}_{DE}$, the
statefinder parameters $(r,s)$ and the value of the transition redshift $z_{tr}$,
for the non-interacting model, for several values of the viscosity parameter $\chi$ in
(\ref{viscousPDM22}), and with $\alpha=1.15$, $\alpha_0 = 0.15$, $\alpha_1 = 0.2$,
$\alpha_2= 0.25$. We have set   $H_{0}=0.7$ and $\Omega_{DM} = 0.27$. From
\cite{Khurshudyan:2016gmb}.}
    \begin{tabular}{ | l | l | l | l | p{1cm} |}
    \hline
 $\chi$ & $q$ & $(w^{\prime}_{DE},w_{DE})$ & $(r,s)$ & $z_{tr}$ \\
      \hline
 $-0.25$ & $-0.766$ & $(0.433, -0.952)$ & $(1.356,  -0.094)$ & $1.21$ \\
          \hline
 $-0.1$ & $-0.666$ & $(0.487, -0.954)$ & $(1.797,  -0.228)$ & $0.82$ \\
    \hline
 $0.0$ & $-0.599$ & $(0.529, -0.953)$ & $(2.111  -0.337)$ & $0.67$ \\
     \hline
 $0.1$ & $-0.533$ & $(0.572, -0.952)$ & $(2.441,  -0.464)$ & $0.53$ \\
     \hline
 $0.25$ & $-0.433$ & $(0.635, -0.952)$ & $(2.965,  -0.701)$ & $0.4$ \\
     \hline
    \end{tabular}
  \label{tTable1Kurd1}
\end{table}
\begin{table}[ht!]
  \centering
    \caption{The present-day values of the deceleration parameter $q$, of the dark energy
equation-of-state parameter $w_{DE}$ and its derivative $w^{\prime}_{DE}$, the
statefinder
parameters $(r,s)$ and the value of the transition redshift $z_{tr}$,
for the non-interacting model, for several values of the parameter $\alpha$ in
(\ref{viscousPDM22}), and with
$\chi=-0.1$, $\alpha_0 = 0.15$, $\alpha_1 = 0.2$, $\alpha_2= 0.25$. We have set
$H_{0}=0.7$ and $\Omega_{DM} = 0.27$. From  \cite{Khurshudyan:2016gmb}.}
    \begin{tabular}{ | l | l | l | l | p{1cm} |}
    \hline
 $\alpha$ & $q$ & $(w^{\prime}_{DE},w_{DE})$ & $(r,s)$ & $z_{tr}$ \\
      \hline
 $0.75$ & $-0.508$ & $(0.587, -0.952)$ & $(1.894,  -0.296)$ & $-$ \\
          \hline
 $0.85$ & $-0.554$ & $(0.559, -0.952)$ & $(1.977,  -0.309)$ & $-$ \\
    \hline
 $0.95$ & $-0.595$ & $(0.534, -0.952)$ & $(1.967  -0.295)$ & $1.2$ \\
     \hline
 $1.15$ & $-0.666$ & $(0.487, -0.952)$ & $(1.797,  -0.227)$ & $0.82$ \\
     \hline
 $1.2$ & $-0.682$ & $(0.477, -0.952)$ & $(1.736,  -0.208)$ & $0.82$ \\
     \hline
    \end{tabular}
  \label{tTable1Kurd2}
\end{table}

Let us now study the interacting scenario, i.e. considering a non-zero $Q$ in
(\ref{rhoDEevolKurd}),(\ref{rhoMevolKurd}). In this case, using
(\ref{HFEKurd}),(\ref{HFEKurdLfdef}),(\ref{viscousPDM22}) and (\ref{H2visHDE1}), for the
deceleration and matter density parameters we find
\begin{equation}
q = \frac{L \left[(1-3 b) H^2+ p_{DM}\right]-2 \sqrt{\Omega _{DE}} \hat{L}_{f}-H \Omega
_{DE} (\dot{
L}_{f}+1)}{2 H^2 L},
\end{equation}
and
\begin{equation}
\Omega_{DM}^{\prime} = \frac{ A_{1} + 2 \Omega _{DE}^{3/2} \hat{L}_{f} - 2 \sqrt{\Omega
_{DE}} \hat{
L}_{f} + H \Omega _{DE}^2 (\dot{L}_{f} + 1)}{H^2 L_{f}},
\end{equation}
with $A_{1} = \Omega _{DE} \left\{H [(3 b-1) H L_f-2]-L_f p_{DM}\right\}$, while for the
dark-energy equation-of-state parameter we obtain
\begin{equation}
w_{DE} = -\frac{3 b H^2 L+2 \sqrt{\Omega _{DE}} \hat{L}_{f} + H \Omega _{DE} (
\dot{L}_{f} + 1)
}{3 H^2 L_{f} \Omega _{DE}}.
\end{equation}
 As we observe, the deceleration parameter $q$ exhibits the transition from deceleration
to acceleration, and the role of the positive interaction parameter $b$ in
(\ref{Qformmod1}) is to make $z_{tr}$ larger and the present value of $q$ more
negative \cite{Khurshudyan:2016gmb}. This is expected since larger positive $b$ implies
larger positive $Q$ in (\ref{rhoDEevolKurd}),(\ref{rhoMevolKurd}) and thus larger energy
transfer to the dark energy sector. Moreover, the role of the viscosity parameter $\chi$
is as in the non-interacting case, i.e  the larger the $\chi$  is the more
difficult it is for the universe to  exhibit accelerated expansion. Lastly, the dark
energy
equation-of-state parameter $w_{DE}$ can exhibit the phantom divide-crossing, too.
In Table \ref{tTable1Kurdinter}  we present the various
calculated values for different choices of the model parameters, where the features
described above are obvious.
\begin{table}[ht!]
  \centering
   \caption{The present-day values of the deceleration parameter $q$, of the dark energy
equation-of-state parameter $w_{DE}$ and its derivative $w^{\prime}_{DE}$, the
statefinder parameters $(r,s)$ and the value of the transition redshift $z_{tr}$,
for the interacting model, for several values of the interaction parameter $\beta$ of
(\ref{Qformmod1}), and with
$\chi=0.1$, $\alpha=1.15$, $\alpha_0 = 0.15$, $\alpha_1 = 0.2$, $\alpha_2= 0.25$. We have
set
$H_{0}=0.7$ and $\Omega_{DM} = 0.27$. From  \cite{Khurshudyan:2016gmb}.}
    \begin{tabular}{ | l | l | l | l | p{1cm} |}
    \hline
 $b$ & $q$ & $(w^{\prime}_{DE},w_{DE})$ & $(r,s)$ & $z_{tr}$ \\
      \hline
 $0.0 $ & $-0.533$ & $(0.572, -0.952)$ & $(2.441,  -0.465)$ & $0.54$ \\
          \hline
 $0.01$ & $-0.548$ & $(0.572, -0.966)$ & $(2.342,  -0.427)$ & $0.58$ \\
    \hline
 $0.03$ & $-0.578$ & $(0.572, -0.993)$ & $(2.152,  -0.356)$ & $0.65$ \\
     \hline
 $0.05$ & $-0.608$ & $(0.568, -1.021)$ & $(1.972,  -0.292)$ & $0.72$ \\
     \hline
 $0.07$ & $-0.634$ & $(0.561, -1.048)$ & $(1.800,  -0.234)$ & $0.82$ \\
     \hline
    \end{tabular}
  \label{tTable1Kurdinter}
\end{table}

In summary, as we saw, one can study the scenario of generalized holographic dark energy
in the framework of viscous cosmology, allowing additionally for an interaction term
between viscous dark matter and holographic dark energy. As one can show, the role of
viscosity is to make the the transition to late-time acceleration more difficult, while
the role of interaction has the opposite effect. Lastly, the scenario at hand allows for
the phantom-divide crossing, which can be an additional advantage revealing its
capabilities.

\section{Special topics}
\label{Section4}

In this section we discuss various topics of viscous cosmological theory, focusing on
investigations in which the present authors have taken part. As a brief  remark to the
material covered below we think it is appropriate to underscore the great power of the
hydrodynamical formalism when applied to quite different problems in cosmology.
The formalism robustness is in general striking. Definitely, in view of the
considerably large activity in the field of viscous cosmology, there are many  aspects
that
cannot be discussed here. For instance, instead of assuming a one-component fluid model,
one might consider an extension of the model in order to encompass two different fluid
components. We may here mention the recent study of Ref.~\cite{Normann:2016zby}, where the
cosmic fluid was considered to be constituted of a dark matter component endowed with a
constant bulk viscosity, and a non-viscous dark energy component. In other related
works \cite{Brevik:2014eya,Brevik:2015cya}, viscous coupled-fluid models were
investigated when the equation of state was assumed to be inhomogeneous. Furthermore,
in \cite{Brevik:2016kwq} the authors studied the important self-reproduction problem of
the universe, namely the graceful exit from inflation, where it was shown how inflation
without self-reproduction can actually be obtained by imposing restrictions on the value
of the thermodynamic parameter in the equation of state. Finally, we mention the very
different approach which consists in applying particle physics theory and the
relativistic
Boltzmann equation in order to derive expressions for the bulk and the shear viscosities,
and the corresponding entropy production, in the specific lepton-photon era, where the
temperature dropped from $10^{12}~$K to $10^{10}~$K. Calculations of this kind were
recently given in Ref.~\cite{Husdal:2016pfd}, \cite{Husdal:2016ofa}
and \cite{hoogeveen86}.

\subsection{Estimate for the present bulk viscosity and remarks on the future
universe}
\label{estimatess}

A significant amount of research has been spent in order to study the behavior of the
cosmic fluid in the far future. In such an examination, and as we discussed in detail in
subsection \ref{subsectsingularities22} above, there may appear various kinds
of singularities: the Big Rip \cite{Caldwell:2003vq,Nojiri:2003vn}, the Little Rip
\cite{Frampton:2011sp,Brevik:2011mm,Frampton:2011rh}, the Pseudo-Rip
\cite{Frampton:2011aa}, the Quasi-Rip \cite{Wei:2012ct}, as well as other
kinds of soft singularities (for instance the so-called type IV finite time singularities
\cite{Brevik:2016kuy}).

In the framework of viscous cosmology, the value of the (effective) bulk viscosity at
present time is naturally an important ingredient of such investigation. Recent
observations from the Planck satellite have given us a better ground for estimating the
 bulk viscosity value $\zeta=\zeta_0$ at   present time  $t=t_0$. As discussed
already in the previous Sections, referring to \cite{Normann:2016jns}, as well as to
several other theoretical and experimental manuscripts, the
estimate
\begin{equation}
\zeta_0 \sim 10^6~\rm{Pa~s}
\end{equation}
was suggested as a reasonable (logarithmic) mean value. However, the corresponding
uncertainty is quite large; there have appeared  proposals ranging from about
$10^4~\rm{Pa~s}$ to about $10^7~\rm{Pa~s}$, depending on analyses of different sources.

We will follow the discussion of \cite{Normann:2016zby}, in which two different
cosmological models were analyzed: (1) a one-component dark energy
model where the bulk viscosity $\zeta$ was associated with the cosmic fluid as a whole,
and (2) a two-component model where $\zeta$ was associated with a dark matter component
$\rho_m$ only, the latter component assumed to be non-viscous. For convenience, we focus
on the one-component scenario.

We assume the simple equation of state $p=w\rho$, with $  w=const.$, and hence the two
viscous Friedmann equations acquire the usual form, namely
\begin{equation}
3H^2=\kappa \rho, \quad 2\dot{H}+3H^2=-\kappa [p-3H\zeta(\rho)], \label{3.3}
\end{equation}
 and the energy conservation equation reads as
\begin{equation}
\dot{\rho}+3H(\rho+p)=9H^2\zeta(\rho).
\end{equation}
Solving this equation in the regime around $w=-1$, i.e  expanding as $w=-1+\alpha$ and
assuming that $\alpha$ is small, we obtain
 \begin{equation}
 t=\frac{1}{\sqrt{3\kappa}}\int_{\rho_0}^\rho
\frac{d\rho}{\rho^{3/2}[-\alpha+\sqrt{3\kappa}\,\zeta(
\rho)/\sqrt{\rho}]}, \label{3.5}
 \end{equation}
where $t_0=0$, and the integration extends into the future. For the bulk viscosity we will
consider the form adopted in the literature, namely
\begin{equation}
\zeta=\zeta_0\left(\frac{H}{H_0}\right)^{2\lambda}=\zeta_0\left(\frac{\rho}{\rho_0}
\right)^\lambda,
\label{3.6}
\end{equation}
with $\lambda$ a constant. In the following we examine two options for the value of
$\lambda$, which are both physically reasonable.

\begin{itemize}

\item {\it Case (i): $\lambda=1/2 ~(\zeta \propto \sqrt{\rho})$.}

In this case, from Eq.~(\ref{3.5}) we obtain
\begin{equation}
t=\frac{2}{3H_0 X_0}\left( 1-\frac{1}{\sqrt{\Omega}}\right), \label{3.7}
\end{equation}
where for convenience we have introduced the dimensionless quantities
\begin{equation}
X_0= \Omega_{\zeta 0}-\alpha,  \quad
\Omega=\frac{\rho}{\rho_0}. \label{3.8spectop1}
\end{equation}

The point that worths attention here is that even if the fluid is initially in the
quintessence
region $\alpha>0$ at $t=0$ it will, if $X_0>0$, inevitably be driven into a Big Rip
singularity $(\rho \rightarrow\infty)$ after a finite time
\cite{Brevik:2004pm,Brevik:2005bj,Brevik:2015xsa,Normann:2016jns}
\begin{equation}
t_s=\frac{2}{3H_0X_0}, \quad (\zeta \propto \sqrt{\rho}). \label{3.9}
\end{equation}

If on the other hand the combination of equation-of-state parameter $\alpha$ and
viscosity $\zeta_0$ is such that $X_0<0$, then the cosmic fluid becomes gradually diluted
as $\rho \propto 1/t^2$ in
the far future.


\item  {\it Case (ii): $\lambda=0 ~(\zeta= \zeta_0=const.)$.}

In this case we obtain the solution
\begin{equation}
t=\frac{2}{3\Omega_{\zeta 0}H_0}\ln \left[ \frac{X_0}{-\alpha +
\Omega_{\zeta_0}/\sqrt{\Omega}}\right], \quad
(\zeta=\zeta_0),
\end{equation}
which implies an energy density of the form
\begin{equation}
\Omega=\frac{\rho}{\rho_0}=\left\{ \frac{\Omega_{\zeta 0}}{\alpha+ (\Omega_{\zeta_0
}-\alpha)\exp\left[ -(
3/2)\Omega_{\zeta 0}H_0t\right]}\right\}^2.
\end{equation}
Hence in the far future $\rho \rightarrow const.$, which implies $H \rightarrow
const.$, which is just the de Sitter solution.  Let us denote the limiting value of the
density by $\rho_{\rm dS}$. Then
\begin{equation}
\rho_{\rm dS}=\rho_0\left( \frac{ \Omega_{\zeta 0}}{\alpha}\right)^2
=\frac{3\kappa
\zeta_0^2}{\alpha^2}. \label{3.12}
\end{equation}
From this expression we deduce that both $\alpha$ and $X_0$ are   important for the
future fate of the cosmic fluid.

Thus, this case may be defined as a pseudo-Rip in accordance with the definition given
by Frampton et al. \cite{Frampton:2011aa}, since the limiting value of the density
reached after an infinite span of time is finite.

\end{itemize}

We close this subsection by providing some values for the inflationary observables, in
order to compare with the 2015 Planck observations. In particular, from Table 5 of
 \cite{Ade:2015xua} we have $w=-1.019^{+0.075}_{-0.080}$. Thus, $\alpha
=1+w$ will be lying within two limits, i.e. between
\begin{equation}
\alpha_{\rm min}=-0.099, \quad \alpha_{\rm max}=+0.056. \label{3.13}
\end{equation}
As mentioned above, we took $\zeta_0=10^6~$Pa s, i.e. $\Omega_{\zeta 0}=0.01$,  to be a
reasonable mean value of the present viscosity.
 Then, according to (\ref{3.8spectop1}) we have
 \begin{equation}
 X_0(\alpha_{\rm max})=-0.046, \quad  X_0(\alpha_{\rm min})=+0.109.
\label{3.14}
 \end{equation}
Hence, we recover the cases 2 and 3 above: the future de Sitter energy density will
become lower than $\rho_0$.

\subsection{Is the bulk viscosity large enough to permit the phantom divide crossing?}

This subsection is a continuation of the previous one, and is motivated by the following
question: is the value of $\zeta_0$, as inferred from the analysis of recent observations,
actually large enough to permit the crossing of the phantom divide, i.e. the transition
from the quintessence region to the phantom region?  To analyze this question we have to
consider more carefully the uncertainties in the data found from  different sources. We
will present some material discussing this point, following the recent work
\cite{Brevik:2015xsa}.

Assume that the bulk viscosity varies with energy density as $\zeta \propto \sqrt{\rho}$.
The condition for phantom divide crossing, as noted above, is that the quantity $X_0$
defined in Eq.~(\ref{3.8spectop1}) has to be positive. In the analysis of Wang and Meng
\cite{Wang:2013uka} various assumptions for the bulk viscosity in the early universe were
considered, and the corresponding theoretical curves for $H=H(z)$ were compared with a
number of observations.
The detailed comparison is quite complicated, but for our purpose it is sufficient
to note that the preferred value of the magnitude  $  \Omega_{\zeta 0}$  is (compare also
with the
discussion in
 \cite{Normann:2016jns}):
\begin{equation}
\Omega_{\zeta 0}=0.5,
\end{equation}
 corresponding to
\begin{equation}
\zeta_0 \sim 5\times 10^7~ {\rm Pa~ s},
\end{equation}
which is a rather high value. In this context, we may compare
with the formula for the bulk viscosity in a photon fluid \cite{Weinberg:1971mx}, namely
\begin{equation}
\zeta = 4a_{\rm rad}T^4\tau_f\left[ \frac{1}{3}-\left( \frac{\partial p}{\partial
\rho}\right)_n\right]^2,
\end{equation}
where $a_{\rm rad}=\pi^2 k_B^4/15\hbar^3c^3$ is the radiation constant and $\tau_f$ the
mean free time. If we estimate $\tau_f = 1/H_0$ (the inverse Hubble radius), we
obtain $\zeta \sim 10^4~$Pa s, which is considerably lower. In summary, it seems that one
has to allow for a quite wide span in the value of the present bulk viscosity. All
suggestions in the literature can be encompassed if we write
\begin{equation}
10^4~{\rm Pa~s} <\zeta_0 < 10^7~{\rm Pa~s}, ~i.e. ~ 10^{-4}<\Omega_{\zeta 0}<0.1.
\label{3.18spetop2}
\end{equation}

We can now rewrite the condition for phantom divide crossing as
\begin{equation}
\zeta_0> \frac{H_0}{\kappa}\, \alpha= (1.18\times 10^8) \,\alpha,
\end{equation}
where we have inserted  $H_0=67.80$ km s$^{-1}$ Mpc$^{-1}=2.20\times 10^{-18}~{\rm
s}^{-1}$. As noted above, from the observed data we derive the maximum value of $\alpha$
to be $\alpha_{\rm max}=0.056$. This yields
\begin{equation}
\zeta_0 > \frac{H_0}{\kappa}\,\alpha_{\rm max}=  6.6\times 10^6~\rm{ Pa\,
s}, ~\rm{or}~ \Omega_{\zeta 0}>0.066. \label{3.20spetop2}
\end{equation}
Thus, comparison between (\ref{3.18spetop2}) and (\ref{3.20spetop2})
implies that, on the basis of available
data, a phantom divide crossing is actually possible even if $\alpha
=\alpha_{\rm max}$.

\subsection{Bounce universe with a
viscous fluid}

In this subsection we investigate the realization of bouncing solutions in the framework
of viscous cosmology following \cite{Myrzakulov:2014hva} (see also \cite{Singh:2016fnh}).
Bouncing cosmological
evolutions offer a solution to the initial singularity problem  \cite{Mukhanov:1991zn}.
Such models have been constructed in modified gravity constructions,  such as in the
Pre-Big-Bang \cite{Veneziano:1991ek}
and in the Ekpyrotic \cite{Khoury:2001wf} scenarios, in $f(R)$ gravity
\cite{Bamba:2013fha,Nojiri:2014zqa,Odintsov:2014gea}, in $f(T)$
gravity \cite{Cai:2011tc}, in braneworld scenarios
\cite{Shtanov:2002mb,Saridakis:2007cf},  in  loop quantum cosmology
\cite{Bojowald:2001xe,Odintsov:2015uca}  etc. Additionally,
non-singular bounces can be obtained using  matter
forms that  violate the null energy condition
\cite{Cai:2007qw,Nojiri:2015fia}.

In order to be more general, in the following we will allow also for a spatial curvature,
and hence the two Friedmann equations write as
\begin{eqnarray}
&&H^2+\frac{k}{a^2}=\frac{\kappa\rho}{3}
\label{frsbounc0}\\
&&-\frac{(2\dot
H+3H^2)}{\kappa}=p\,,
\label{frsbounc}
\end{eqnarray}
with  $k=-1,0,1$  corresponing to  open, flat or closed  geometry.
Additionally, concerning the fluid's equation of state we  will consider a general
inhomogeneous viscous one of the form (\ref{visceosinh00}), namely
\begin{equation}
p=w(\rho)\rho-B(a(t),H, \dot{H}...)\,,
\label{visceosinh}
\end{equation}
where  $w(\rho)$  can depend on the energy density, but the bulk
viscosity  $B(a(t),H, \dot{H}...)$  is allowed to be a function of the scale
factor, and of the Hubble function and its
derivatives.
Thus, the fluid stress-energy tensor writes as
\begin{equation}
T_{\mu\nu}=\rho u_{\mu}u_{\nu}+\left[w(\rho)\rho+B(\rho,a(t),H,
\dot{H}...)\right](g_{\mu\nu}
+u_{\mu}u_{\nu})\,,
\end{equation}
with  $u_{\mu}=(1,0,0,0)$ the four velocity.
Hence, the standard conservation law $\dot\rho+3H(\rho+p)=0$ leads to
\begin{equation}
\dot{\rho}+3H\rho(1+w(\rho))=
3 H B(\rho,a(t),H, \dot{H}...)
\end{equation}

We now proceed to the investigation of simple bounce solutions in the above framework,
and we discuss the properties of the viscosity of the fluids that drive such solutions.
A first example is the bounce with an exponential scale factor of the form
\begin{eqnarray}
&&a(t)=a_0\text{e}^{\alpha (t-t_0)^{2n}}\,\\
&&
H(t)=2n\alpha\,(t-t_0)^{2n-1}\,,
\label{expbouncscale}
\end{eqnarray}
with $n$ a positive integer and  $a_0\,,\alpha$   positive parameters. We consider
$t_0>0$  to be the bounce point, i.e. for $t<t_0$ we have a contracting universe
  and when $t>t_0$ expansion takes place. We mention that if $n$ is non-integer then
 singularities may arise, while the simplest case $n=1/2$ corresponds to just the
Sitter solution  $H(t)=const.$ (in general for $n=m/2$, with $m$ an odd integer, the
bounce is absent). Finally,  note that for the ansatz  (\ref{expbouncscale}) we have
\begin{equation}
\frac{\ddot a}{a}=H^2+\dot H=
2n\alpha(t-t_0)^{2(n-1)}\left[2n\alpha(t-t_0)^{2n}+(2n-1)
\right]\,,
\end{equation}
and hence we obtain (early-time) acceleration after the bounce, which is a significant
phenomenological advantage.

Inserting the bouncing scale factor (\ref{expbouncscale}) into (\ref{frsbounc0})
we acquire
\begin{equation}
\rho=\frac{3}{\kappa}\left[4n^2\alpha^2(t-t_0)^{2(2n-1)}+\frac{k}{a_0^2\text{e}^{
2\alpha(t-t_0)^{
2n}}}\right]\,.
\label{rhobouncrel1}
\end{equation}
Since for $k=-1$ the above quantity might become negative, we focus on the $k=0$ and
$k=+1$ cases where it is always positive definite. As we observe, in the flat case $\rho$
 decreases in the contracting phase, it becomes zero at $t=t_0$, and it increases in the
expanding regime. On the other hand, for $k=+1$, and when   $n>1$, there is a region
around the bouncing point where $\rho$ increases in the contracting phase, it reaches the
value  $\rho=3/(a_0\kappa)$ at $t=t_0$, and then it decreases (these can be seen by
examining the derivatives of (\ref{rhobouncrel1})). However, for $t\gg t_0$, the
energy density starts to increase. This behavior may have an important effect on the
cosmological parameter $
\Omega=1+\frac{k}{a^2 H^2}$, which for the bouncing scale factor  (\ref{expbouncscale})
becomes
\begin{equation}
\Omega=1+\frac{k}{a_0^2\alpha^2(t-t_0)^{2(2n-1)} \, e^{2\alpha(t-t_0)^{2n}}}\,,
\end{equation}
and thus it exhibits  a decreasing behavior. Such a post-bounce acceleration, with the
simultaneous decrease of $\rho$ and of $\Omega$ may be compatible with the inflationary
phenomenology, in which  at the end of inflation  $\Omega$ is very close to $1$.
Definitely, in order to stop the aforementioned early-time acceleration we need to add
additional fluids that could become dominant and trigger the transition to the matter era.

Let us now analyze what kind of fluids with equation of state given by (\ref{visceosinh})
can produce the bouncing solution  (\ref{expbouncscale}). We first consider an
inhomogeneous but non-viscous fluid, namely  we assume $B(a(t),H,
\dot{H}...)=0$. In this case, for the flat geometry, equations (\ref{frsbounc0}) and
(\ref{visceosinh}) lead to
\begin{equation}
p=-\rho-\rho^{\frac{(n-1)}{(2n-1)}}\left[
\frac{3}{\kappa}(2n\alpha)^2
\right]^{\frac{2n}{4n-2}}\left(\frac{2n-1}{3n\alpha}\right)\,,
\end{equation}
and thus to
\begin{equation}
w(\rho)=-1-\rho^{\frac{-n}{(2n-1)}}\left[
\frac{3}{\kappa}(2n\alpha)^2
\right]^{\frac{2n}{4n-2}}\left(\frac{2n-1}{3n\alpha}\right)\,.
\label{rhobouncsol1}
\end{equation}
As we mentioned earlier, if the exponent of $\rho$  in (\ref{rhobouncsol1}) is negative
then we obtain the bounce realization, however if it is positive then we have the
appearance of a singularity.

As a second example we switch on viscosity, considering
\begin{equation}
B(a(t),H, \dot{H}...)=3 H\zeta(H)\,,
\label{eq.statebulkvoscoex2}
\end{equation}
with $\zeta(H)>0$  the bulk viscosity, and for simplicity and without
loss of generality we consider   $w=-1$.  In this case, for the flat geometry, equations
(\ref{frsbounc0}) and
(\ref{visceosinh}) lead to
\begin{eqnarray}
&&p=-\rho-3H\zeta(H)\,,\nonumber\\
&&
\zeta(H)=\left(\frac{3}{\kappa}\right)^{\frac{2n-1}{2n-1}}
\left(2n\alpha
\right)^{\frac{1}{2n-1}}\left(\frac{2n-1}{3}\right)H^{-\frac{1}{2n-1}}\,.
\label{bounczetaex2}
\end{eqnarray}
Nevertheless, for $k=+1$ the equation of state for the fluid becomes complicated and
therefore it is necessary to  go beyond (\ref{eq.statebulkvoscoex2}) and consider a
viscosity that depends on the scale factor too. Such a case could be
\begin{equation}
p=-\rho-3H\zeta(H, a(t))
\,,
\end{equation}
which then leads to
\begin{equation}
\zeta(H,a(t))=\left(\frac{3}{\kappa}\right)^{\frac{2n-1}{2n-1}}
\left(2n\alpha
\right)^{\frac{1}{2n-1}}\left(\frac{2n-1}{3}\right)H^{-\frac{1}{2n-1}}
-\frac{2k}{3\kappa H a(t)^2}\,.
\end{equation}
As we can see, and as expected, for large scale factors the above relation  coincides
with (\ref{bounczetaex2}), and therefore we can treat the closed geometry as the flat one.

Since we have analyzed the exponential bounce, we now proceed to the
investigation of other bouncing solutions. In particular, we will focus on the  power-law
bouncing scale factor of the form
\begin{eqnarray}
&&
a(t)=a_0+\alpha(t-t_0)^{2n}\,,\\
&&
H(t)=\frac{2n\alpha(t-t_0)^{2n-1}}{a_0+\alpha(t-t_0)^{2n}}\,,
\label{powbounce}
\end{eqnarray}
with $n$ a positive integer,  $a_0\,,\alpha$   positive parameters, and
$t_0>0$  the bounce point.  This relation  leads to
\begin{equation}
\frac{\ddot a}{a}=\frac{2n(2n-1)\alpha(t-t_0)^{2(n-1)}}{a_0+\alpha(t-t_0)^{2n}}\,,
\end{equation}
which implies that the post-bounce expansion is accelerated.
Inserting (\ref{powbounce}) into the first Friedmann equation   (\ref{frsbounc0})
we acquire
\begin{equation}
\rho=\frac{3}{\kappa\left[a_0+\alpha(t-t_0)^{2n}\right]}\left[\frac{4n^2\alpha^2(t-t_0)^
{4n-2}+k}{
a_0+\alpha(t-t_0)^{2n}}\right]\,.
\label{rhobounceexmp2}
\end{equation}
Since for the open universe case $\rho$ can become negative (in particular,
$\rho=-3/(a_0\kappa)^2$ at $t=t_0$), we focus on the $k=0$  and $k=+1$ cases, where $\rho$
 is positive definite. Taking the derivative of (\ref{rhobounceexmp2}) we find
\begin{equation}
\dot{\rho}=
-\frac{4n(t-t_0)^{2n-3}\alpha[2n(t-t_0)^{2n}\alpha(a_0(1-2n)+(t-t_0)^{2n}
\alpha)+k(t-t_0)^2]}
{3(a_0+\alpha(t-t_0)^{2n})^3}\kappa\,,
\end{equation}
thus near the bounce point we have
\begin{equation}
\dot{\rho}(t\rightarrow t_0)\simeq
\frac{8n^2(t-t_0)^{4n-3}\alpha^2(2n-1)}
{3a_0^2}\kappa\,,
\end{equation}
from which we deduce that the energy density decreases in the contracting phase before
the bounce and increases immediately after it. Nevertheless, for $|t|\gg t_0$ we have
\begin{equation}
\dot{\rho}(|t|\gg t_0)=
-\frac{4n(t-t_0)^{-4n-3}[2n(t-t_0)^{4n}\alpha^2+k(t-t_0)^2]}
{3\alpha^2}\kappa\,,
\end{equation}
which implies that after a suitable amount of time in the expanding phase $\rho$ starts
decreasing again.  Finally, the
cosmological parameter $\Omega=1+\frac{k}{a^2 H^2}$
behaves as
\begin{equation}
\Omega=1+\frac{k}{4n^2\alpha^2(t-t_0)^{4n-2}}\,,
\end{equation}
and therefore it exhibits  a decreasing behavior.
Hence, similarly to the case of the exponential bounce analyzed
earlier, such behaviors could be interesting for the description of the post-bouncing
universe and the correct subsequent  thermal history, since it will leave a universe with
$\Omega$ very close to $1$ and a decreasing $\rho$.

Lastly, let us investigate  what kind of fluids with equation of state given by
(\ref{visceosinh})  can produce the power-law bouncing solution (\ref{powbounce}).
Considering  an inhomogeneous viscous fluid with equation of state
\begin{equation}
p=-\frac{\rho}{3}-3H\zeta(a(t), H)\,,
\end{equation}
and inserting (\ref{powbounce}), we find the bulk viscosity as
\begin{equation}
\zeta(a(t), H)=\frac{(2n-1)a(t)}{3n(a(t)-a_0)\kappa}\,.
\end{equation}
Note that away from the bouncing point, namely
 when $a(t)\gg a_0$, the bulk viscosity becomes
\begin{equation}
\zeta(H,a(t)\gg a_0)\simeq\frac{(2n-1)}{3n\kappa}=const.\,.
\end{equation}
Hence, if $0<\zeta<2/3$, which corresponds to $n>1/2$, then the bounce can be realized. On
the other hand if  $2/3<\zeta$ then, as we mentioned earlier, singularities might appear.

In summary, in this subsection we saw that viscous fluids can offer the mechanism to
violate the null energy condition, which is the necessary requirement for the bounce
realization. Hence, various bouncing solutions can be realized, driven by fluids with
suitably reconstructed viscosity. As specific examples we studied the exponential and the
power-law bounces, which are also capable of describing the accelerated post-bouncing
phase, with the additional establishment of the spatial flatness. These features reveal
the capabilities of viscosity.

\subsection{Inclusion of isotropic turbulence}

In this subsection we discuss turbulence issues in the framework of viscous cosmology.
From hydrodynamics point of view the inclusion of turbulence in the theory of the
cosmic fluid seems most natural, at least in the final stage of the universe's evolution
when the fluid motion may well turn out to be quite vigorous.
The local Reynolds number must then be expected to be very high. On a local scale this
brings the {\it shear} viscosity concept into consideration, as it has to furnish the
transport of eddies over the wave number spectrum until the local Reynolds number becomes
of order unity, marking the transfer of kinetic energy into heat. Due to the assumed
isotropy in the fluid, we must expect that the type of turbulence is isotropic when
looked upon on a large scale. According to standard theory of isotropic turbulence in
hydrodynamics we then expect to find a Loitziankii distribution for low wave numbers
(energy density varying as $k^4$), whereas for higher $k$ we expect an inertial subrange
in which the energy distribution is
 \begin{equation}
E(k)=\alpha \epsilon^{2/3}k^{-5/3}, \label{3.21spectop}
\end{equation}
where $\alpha$ denotes the Kolmogorov constant and $\epsilon$ is the mean energy
dissipation per unit mass and unit time. When $k$ reaches the inverse Kolmogorov length
$\eta_K$, i.e.
\begin{equation}
k  \rightarrow  k_L=\frac{1}{\eta_L}=\left(\frac{\epsilon}{\nu^3}\right)^{1/4},
\label{3.22}
\end{equation}
with $\nu$ the kinematic viscosity, then the dissipative region is reached.

In the following we will consider a dark fluid developing into the future from the
present time $t=0$, when turbulence is accounted for. We will perform the analysis in
two different ways: either assuming a two-fluid model with  one turbulent
constituent, or assuming simply a one-component fluid, following
\cite{Brevik:2014cxa,Brevik:2012nt,Brevik:2013qka,Brevik:2010db}.

We start by considering a two-component model, where the effective energy is written as a
sum of two parts, namely
\begin{equation}
\rho_{\rm eff}=\rho +\rho_{\rm turb},
\end{equation}
with $\rho$ denoting the conventional energy density. Taking $\rho_{\rm turb}$ to be
proportional to the scalar expansion $\theta=3H$, and calling the proportionality factor
$\tau$, we acquire
\begin{equation}
\rho_{\rm eff}=\rho(1+3\tau H).
\end{equation}
Additionally, the effective pressure $p_{\rm eff}$ is split in a similar way as
\begin{equation}
p_{\rm eff}=p+p_{\rm turb}.
\end{equation}
For both components we assume homogeneous equations of state, namely
\begin{equation}
p=w\rho, \quad p_{\rm turb}=w_{\rm turb}\,\rho_{\rm turb}, \label{3.26}
\end{equation}
The Friedmann equations can thus be written (recall that $\kappa=8\pi G$) as
\begin{equation}
H^2=\frac{1}{3}\kappa \rho(1+3\tau H),
\end{equation}
\begin{equation}
\frac{2\ddot{a}}{a}+H^2=-\kappa  \rho(w+3\tau H w_{\rm turb}),
\end{equation}
 leading to the following governing equation for $H$:
\begin{equation}
(1+3\tau H)\dot{H}+\frac{3}{2}\gamma H^2+\frac{9}{2}\tau \gamma_{\rm turb}H^3=0,
\label{3.29spectop}
\end{equation}
where we used the standard notation
\begin{equation}
\gamma=1+w, \quad \gamma_{\rm turb}=1+w_{\rm turb}. \label{3.30}
\end{equation}
Finally, when the energy dissipation is
assumed to be
 \begin{equation}
\epsilon=\epsilon_0(1+3\tau H),
\end{equation}
the energy balance may be written as
\begin{equation}
\dot{\rho}+3H(\rho+p)= -\rho \epsilon_0(1+3\tau H). \label{3.32spetop3}
\end{equation}

In summary, the input parameters in this model are $\{ w, w_{\rm turb}, \tau \}$, all of
them assumed to be constants. In the following we analyze the cases of two specific
choices for $w$ and $w_{\rm turb}$.

\begin{itemize}
\item {The case $w_{\rm turb}=w <-1$}

This assumption implies that we equalize the ordinary and turbulent components as far as
the EoS is concerned. From Eq.~(\ref{3.29spectop}) we acquire
\begin{equation}
H=\frac{H_0}{Z}, \quad Z=1+\frac{3}{2}\gamma H_0t.
\end{equation}
Hence, we have a Big Rip singularity after a finite time
\begin{equation}
t_s=\frac{2}{3|\gamma|H_0},
\end{equation}
and we obtain correspondingly
\begin{equation}
a=a_0Z^{2/3\gamma}, \quad \rho=\frac{3H_0^2}{\kappa}\,\frac{1}{Z}\,\frac{1}{Z+3\tau H_0}.
\end{equation}
In the vicinity of  $t_s$, using that $Z=1-t/t_s$, we find
\begin{equation}
H \sim \frac{1}{t_s-t}, \quad a \sim \frac{1}{(t_s-t)^{2/3|\gamma|}}, \label{3.36}
\end{equation}
\begin{equation}
\rho \sim \frac{1}{t_s-t}, \quad \frac{\rho_{\rm turb}}{\rho} \sim \frac{1}{t_s-t},
\label{3.37}
\end{equation}
which reveal the same kind of behavior for $H$ and $a$ as in conventional cosmology,
nevertheless the singularity in $\rho$ has become more weak. The physical
reason for this is obviously the presence of the factor $\tau$.

It is interesting to see how these solutions compare with our assumed form
(\ref{3.32spetop3}) for the energy equation. The left hand side of
Eq.~(\ref{3.32spetop3})
can be calculated, and we obtain in the limit $t\rightarrow t_s$ (details omitted here)
the following expression for the present energy dissipation:
\begin{equation}
\epsilon_0=\frac{1}{2}\frac{|\gamma|}{\tau}. \label{3.38}
\end{equation}
This result could hardly have been seen without calculation; it implies that the specific
dissipation $\epsilon_0$ is closely related to the EoS parameter $\gamma$ and the
parameter $\tau$.

\item{The  case $w<-1, ~ w_{\rm turb}>-1 $}

In general, the turbulent component is accordingly  not only a passive component in the
fluid. The  assumption of the present case, namely $w<-1, ~ w_{\rm turb}>-1 $,
encompasses the  region $-1<w_{\rm turb}<0$, in which the
turbulent pressure will be negative as before. However, it also covers the region
$w_{\rm turb}>0$, where the turbulent
pressure becomes positive as in ordinary hydrodynamics.

The governing equation (\ref{3.29spectop}) can be solved with respect to $t$ as
\begin{equation}
t=\frac{2}{3|\gamma|}\left(\frac{1}{H_0}-\frac{1}{H}\right)-\frac{2\tau}{|\gamma|}\left(
1+\frac{\gamma_{\rm turb}}{|\gamma|}\right)
 \ln \left[ \frac{|\gamma|-3\tau \gamma_{\rm turb}H}{|\gamma|-3\tau \gamma_{\rm
turb}H_0}\frac{H_0}{
H}\right], \label{3.39}
\end{equation}
showing that the kind of singularity encountered in this case is of the  Little Rip
type. As $t\rightarrow \infty$, the Hubble function $H$ approaches the finite value
\begin{equation}
H_{\rm crit}=\frac{1}{3\tau}\frac{|\gamma|}{\gamma_{\rm turb}}. \label{3.40}
\end{equation}
Physically, $\gamma_{\rm turb}$ plays the role of softening the evolution
towards the future singularity.

\end{itemize}

We close this subsection by investigating the case of a one-component scenario. In
particular, instead of assuming the fluid to consist of two components as above, we can
introduce a one-component model in which the fluid starts from $t=0$ as an ordinary
viscous non-turbulent fluid, and then after some time, marked as $t=t_*$, it enters a
turbulent state of motion. This picture is definitely closer to ordinary
hydrodynamics.

Let us follow the development of such a fluid, assuming as previously that  $w<-1$, in
order for the fluid to develop towards a future singularity. After the
sudden transition to turbulent
motion at $t_*$, we have that $w\rightarrow w_{\rm turb}$ and correspondingly  $p_{\rm
turb}=w_{\rm
turb} \,\rho_{\rm turb}$. Similarly to the two-component scenario, we assume $w_{\rm
turb} >-1$, and for simplicity we
assume   that $\zeta$ is a constant.

We can now easily solve the Friedmann equations, requiring the density of the fluid to be
continuous
 at $t=t_*$. It is convenient to introduce the ``viscosity time'', namely
\begin{equation}
t_c=\left(\frac{3}{2}\kappa\zeta \right)^{-1}.
\end{equation}
Hence,  for $0<t<t_*$ we obtain \cite{Brevik:2004pm,Brevik:2005bj}:
\begin{equation}
H=\frac{H_0\,e^{t/t_c}}{1-\frac{3}{2}|\gamma|H_0t_c(e^{t/t_c}-1)}, \label{3.42}
\end{equation}
\begin{equation}
a=\frac{a_0}{\left[1-\frac{3}{2}|\gamma|H_0t_c(e^{t/t_c}-1) \right]^{2/3|\gamma|}},
\label{3.43}
\end{equation}
\begin{equation}
\rho=\frac{\rho_0\, e^{2t/t_c}}{\left[1-\frac{3}{2}|\gamma|H_0t_c(e^{t/t_c}-1)
\right]^2}, \label{3.
44}
\end{equation}
whereas for $t>t_*$ we acquire:
\begin{equation}
H=\frac{H_*}{1+\frac{3}{2}\gamma_{\rm turb}H_*(t-t_*)}, \label{3.45}
\end{equation}
\begin{equation}
a=\frac{a_*}{\left[ 1+\frac{3}{2}\gamma_{\rm turb}H_*(t-t_*) \right]^{2/3\gamma_{\rm
turb}}}, \label{3.46}
\end{equation}
\begin{equation}
\rho=\frac{\rho_*}{\left[ 1+\frac{3}{2}\gamma_{\rm turb}H_*(t-t_*) \right]^{2}}.
\label{3.47}
\end{equation}
Thus, the density $\rho$ at first increases with time, and then decreases again until it
goes to zero as  $t^{-2}$ when $t\rightarrow \infty$. Note that in the turbulent region,
 $p_* =w_{\rm turb}\,\rho_* $ will even be greater than zero in the case where $w_{\rm
turb}>0$.

As a final remark of this subsection, we mention that the presence of turbulence may
alternatively be dealt with in terms of a more general equation
of state of the form (\ref{visceosinh00}), admitting inhomogeneity terms too.

\subsection{Viscous Little Rip cosmology}

As discussed in detail in subsection \ref{subsectsingularities22} above, it is well known
that there exist several theories for singularities in the future
universe  \cite{Nojiri:2005sx,Fernandez-Jambrina:2014sga}. Amongst them, the Little Rip
scenario proposed by Frampton et al. \cite{Frampton:2011sp,Frampton:2011rh} (for
nonviscous fluids) is an elegant solution, which we will consider in more detail
in this subsection, generalized to the case of viscous fluids. The essence of the
original model, as well as of its viscous counterpart, is that the dark energy is
predicted to increase with time in an asymptotic way, and therefore an infinite
span of time is required to reach the singularity. This implies that the
equation-of-state parameter is always $w<1$, but $w\rightarrow -1$ asymptotically.  In
the following we will survey the essentials of this  theory, as were developed by Brevik
et al. \cite{Brevik:2011mm}. In most cases the appearance of a bulk viscosity turns out
to promote the future singularity.

For concreteness we assume an equation of state of the form
\begin{equation}
p=-\rho-A\sqrt{\rho}-\xi(H), \label{3.47a}
\end{equation}
where $A$ is a constant and $\xi(H)$ a viscosity   function  (not the viscosity
itself). This is
an inhomogeneous equation of state. Assuming a spatially flat FRW universe the first and
second
Friedmann equations write as
\begin{equation}
H^2=\frac{\kappa}{3}\rho, \quad
\frac{\ddot{a}}{a}+\frac{1}{2}H^2=\frac{\kappa}{2}[\rho+A\sqrt{\rho}
 +\xi(H)], \label{3.47b}
\end{equation}
while the conservation equation for energy, namely ${T^{0\nu}}_{;\nu}=0$,  becomes
\begin{equation}
\dot{\rho}-3A\sqrt{\rho}H=3\xi(H)H. \label{3.47c}
\end{equation}
Let us study separately the non-viscous and viscous cases.

\begin{itemize}

\item   (I) Non-viscous case.

For comparison, we start from the non-viscous case
$\xi(H)=0$ \cite{Frampton:2011sp}. Setting the present scale factor $a_0$ equal to one,
we obtain
\begin{equation}
t=\frac{1}{\sqrt{3\kappa}}\frac{1}{A}\ln \frac{\rho}{\rho_0}. \label{3.47d}
\end{equation}
This relation reveals the Little Rip property: the singularity $\rho \rightarrow
\infty$ is not reached in a finite time. Additionally, the density $\rho$ can be
expressed as a function of the scale factor as
\begin{equation}
\rho(a)=\rho_0\left( 1+\frac{3A}{2\sqrt{\rho_0}}\ln a \right)^2. \label{3.47e}
\end{equation}
Using the first Friedmann equation we can also express $a$ as a function of $t$, namely
\begin{equation}
a(t)=\exp\left\{ \frac{2\sqrt{\rho_0}}{3A}\left[ \exp \left(
\frac{\sqrt{3\kappa}}{2}At\right) -1\right]\right\}. \label{3.47f}
\end{equation}

\item   (II) Viscous case.

Let us now switch on the viscous term in (\ref{3.47a}). In this case the second Friedmann
equation, as well as the energy conservation equation, will change. We shall consider
here only the simplifying ansatz where the viscosity function is constant, namely
\begin{equation}
\xi(H) \equiv \xi_0=\rm{const.} \label{3.47g}
\end{equation}
This choice is motivated mainly from mathematical reasons. Then, from the
governing equations above, it follows that
\begin{equation}
t=\frac{2}{\sqrt{3\kappa}}\frac{1}{A}\ln \frac{\xi_0+A\sqrt{\rho}}{\xi_0+A\sqrt{\rho_0}}.
\label{3.
47h}
\end{equation}
Inverting this equation we acquire
\begin{equation}
\rho(t)=\left[ \left( \frac{\xi_0}{A}+\sqrt{\rho_0}\right) \exp\left(
\frac{\sqrt{3\kappa}}{2}At\right) -\frac{\xi_0}{A}\right]^2. \label{3.47i}
\end{equation}
Hence, the state $\rho \rightarrow \infty$ can indeed be reached, however it requires an
infinite time interval. This is precisely the Little Rip characteristic, now met under
viscous conditions. The term $\xi_0/A$ multiplying the exponential tends to promote the
singularity, as mentioned. The influence from the last term $\xi_0/A$ becomes
negligible at large times.

\end{itemize}

\subsection{Viscous cosmology and the Cardy-Verlinde formula}

In this subsection we will discuss the connection of viscous cosmology with
thermodynamics. The apparent deep connection between general relativity, conformal field
theory (CFT), and thermodynamics, has aroused considerable interest for several  years. In
the following we will consider one specific aspect of this subject, namely to what
extent the Cardy-Verlinde entropy formula remains valid if we allow for bulk viscosity in
the cosmic fluid. For simplicity we will assume a one-component fluid
model,
and we
assume the bulk viscosity $\zeta$ to be constant. For more details, the reader may
consult
Refs.~\cite{Brevik:2003wm,Brevik:2004pm,Brevik:2005bj,Brevik:2001ed,Brevik:2010jv}, and
additionally the
related Ref.~\cite{Brevik:2004sd}.

 We start with the Cardy entropy formula for an (1+1) dimensional CFT:
 \begin{equation}
 S=2\pi \sqrt{\frac{c}{6}\left( L_0-\frac{c}{24}\right)}, \label{3.48}
 \end{equation}
where $c$ is the central charge and $L_0$ the lowest Virasoro generator
\cite{Cardy:1986ie,Cardy:1986gw}.
Comparing with the first Friedmann equation for a closed universe ($k=+1$) when
$\Lambda=0$, namely
\begin{equation}
H^2=\frac{8\pi G}{3}\rho-\frac{1}{a^2}, \label{3.49}
\end{equation}
we deduce (as pointed out by Verlinde \cite{Verlinde:2000wg}) that formal agreement is
achieved if we choose
\begin{equation}
L_0 \rightarrow \frac{1}{3}Ea, \quad c\rightarrow \frac{3}{\pi}\frac{V}{Ga}, \quad
S\rightarrow  \frac{HV}{2G}, \label{3.50}
\end{equation}
where $E=\rho V$ is the energy in the volume $V$. One noteworthy fact is evident already
at this stage: the correspondence is valid also if the fluid possesses viscosity, since
there is no explicit appearance of viscosity in the first Friedmann equation. Moreover,
the equation of state for the fluid is so far not involved.

In order to highlight the physical importance of the formal substitutions (\ref{3.50}),
let us consider the thermodynamic entropy of the fluid. As is known, there exist several
definitions, the Bekenstein entropy, the Bekenstein-Hawking entropy, and the Hubble
entropy. We will consider only the last quantity here, called $S_H$. Its order of
magnitude can be easily estimated by observing that the holographic entropy $A/4G$  ($A$
is the
area) of
a black hole with the same size as the universe may be written in
the form
 \begin{equation}
 S_H \sim \frac{H^{-2}}{4G} \sim \frac{HV}{4G}, \label{3.51}
 \end{equation}
 since $A \sim H^{-2}$ and hence $V\sim H^{-3}$. Various arguments have been provided to
assume the universe's maximum entropy to be identified with the entropy of a black hole
having the same size as the Hubble radius
\cite{Easther:1999gk,Veneziano:1999ts,Bak:1999hd,Kaloper:1999tt}. Nevertheless, more
precise arguments of Verlinde \cite{Verlinde:2000wg} lead to the
replacement of the factor 4 in the denominator with a factor 2, that is
\begin{equation}
S_H  \sim \frac{HV}{2G}. \label{3.51a}
\end{equation}
Therefore, one can see that this relation coincides with the last relation of
(\ref{3.50}), indicating that the formal substitutions above have a physical basis.

Consider now the Casimir energy $E_C$, defined in this context to be
\begin{equation}
E_C=3(E+pV-TS). \label{3.52}
\end{equation}
We may make use of scaling arguments for the extensive part $E_E$ and the Casimir
part $E_C$ that make up the total energy $E$. These arguments finally give (details
omitted here) $E(S,V)=E_C(S,V)+\frac{1}{2}E_C(S,V)$. An essential point is the property
of conformal invariance, that the products $E_Ea$ and $E_Ca$ are volume independent and
depend only on $S$. Hence, we acquire
\begin{equation}
E_E=\frac{\alpha}{4\pi a}S^{4/3}, \quad E_C=\frac{\beta}{2\pi a}S^{2/3}, \label{3.53}
\end{equation}
where $\alpha, \beta$ are constants. Their product arises from CFT arguments as
$\sqrt{\alpha\beta}=3$ for $n=3$ spatial dimensions. From the formulae above we obtain
\begin{equation}
S=\frac{2\pi a}{3}\sqrt{E_C(2E-E_C)}, \label{3.54}
\end{equation}
which is the Cardy-Verlinde formula. With the substitutions $Ea \rightarrow L_0$ and
$E_C\rightarrow c/12$ it is seen that expressions (\ref{3.53}) and (\ref{3.48}) are in
agreement, apart from a numerical prefactor. This is caused by our assumption about $n=3$
spatial dimensions instead of the $n=1$ assumption in the Cardy formula.

The above arguments were made for a radiation dominated, conformally invariant, universe.
Hence, the question that arises naturally is whether the same arguments apply to a
viscous universe too. The subtle point here is the earlier
pure entropy dependence of the product $Ea$, which is now lost. To analyze this question
we may consider the following equation, holding for a $k=1, \Lambda=0$ universe with EoS
$p=\rho/3$, namely
\begin{equation}
\frac{d}{dt}(\rho a^4)=9\zeta H^2 a^4. \label{3.55}
\end{equation}
This is essentially an equation for the rate of change of the quantity $Ea$. Let us
compare this relation with the entropy production formula
\begin{equation}
n\dot{\sigma}=\frac{9H^2}{T}\zeta, \label{3.56mdotsi}
\end{equation}
where  $n$ is the particle number density and $\sigma$  the entropy per particle. As
we observe, both time derivatives in  (\ref{3.56mdotsi}) and (\ref{3.55}) are
proportional
to
$\zeta$. If $\zeta$ is small we can insert the usual solution for the scale factor of
the nonviscous case,  namely
$a(t)=\sqrt{(8\pi G/3)\rho_{\rm in}a_{\rm in}^4}\, \sin \eta$, with $\eta$ the
conformal time (``in'' denotes the initial time). As the densities $\zeta^{-1}\rho
a^4$ and $\zeta^{-1}n\sigma$ can then be regarded as functions of $t$ (recall that
$\zeta=$constant), we conclude that $\rho a^4$ can be regarded as a function of
$n\sigma$. This implies in turn that $Ea$ can be regarded as a function of $S$. This
property, originally based upon CFT, can thus be carried over to the viscous case too,
assuming that the viscosity is small.

At this stage we should pay attention to the following conceptual point. The specific
entropy $\sigma$ in (\ref{3.56mdotsi}) is a conventional thermodynamic quantity,
whereas the identification $S\rightarrow HV/(2G)$ in (\ref{3.50}) is based on the
holographic principle. The latter entropy is identified with the Hubble entropy $S_H$,
and thus we can set $n\sigma_H=H/(2G)$, with $\sigma_H$ the specific Hubble entropy. The
quantity $\sigma_H$ is holography-based, whereas the quantity $\sigma$ is
not.

Finally, note that the same kind of arguments can be also applied in the
more general situation where the EoS has the form
\begin{equation}
p=(\gamma -1)\rho,
\end{equation}
with $\gamma$ a constant. For the non-viscous case this analysis was performed by Youm
\cite{Youm:2002ea}, with the result
\begin{equation}
S=\left[ \frac{2\pi a^{3(\gamma-1)}}{\sqrt{\alpha
\beta}}\sqrt{E_C(2E-E_C)}\right]^{\frac{3}{3\gamma-1}}. \label{3.57}
\end{equation}
Lastly, in this case the application to  weak viscosity can also be performed as
in \cite{Brevik:2003wm,Brevik:2004pm,Brevik:2005bj}, and when $\gamma=4/3$ the
radiation dominated result is recovered.

\section{Conclusions}
\label{Section5}

From a hydrodynamicist's point of view the inclusion of viscosity concepts in the
macroscopic theory of the cosmic fluid seems  most natural, as an ideal fluid is after all
an abstraction (unless the fluid is superconducting). Modern astronomical
and cosmological observations permit us to look back in history, evaluating the Hubble
parameter up to a redshift $z$ of about 2. Armed with such observational data, and having
at one's disposal the  formalism of FRW cosmology with bulk viscosity included, one would
like to extrapolate the description of the universe back in time up to the
inflationary era, or go to the  opposite extreme and analyze the probable  ultimate
fate of the universe, which  might well be in the form of a Big Rip singularity. In the
present review we have undertaken this quite extensive program.

After fixing the notation
in subection  \ref{subection1}, we began in Section  \ref{Section2} with a presentation of
the theory of the inflationary epoch, covering cold as well as warm
inflation in the presence of bulk viscosity. We investigated in detail the viscosity
effects on the various inflationary observables, showing that they can be significant. A
point to be noted  in this context is that viscous effects may be represented  by a
generalized and inhomogeneous equation of state.

In Section  \ref{Section3}  we turned to viscous theory in the late universe. We
considered the phantom era with its characteristic singularities. Additionally, we
discussed how one can describe in a unified way the inflationary and late-time
acceleration in the framework of viscous cosmology. The simplest way to
achieve this task is to introduce scalar fields. Moreover, we investigated the
cosmological scenario of holographic dark energy in the presence of a viscous fluid, a
subject which is related to black hole thermodynamics.

In the  final Section  \ref{Section4}  of our review we dealt with specific topics. We
classified various options for the ultimate fate of the universe. We gave an analysis of
whether the magnitude of bulk viscosity derived from observations is sufficient to
drive the cosmic fluid from the quintessence into the phantom region. Numerical estimates
indicated that such a transition might well be possible. Furthermore, we investigated
viscous bounce cosmology, and we made use of isotropic turbulence theory from
hydrodynamics to describe the late cosmic fluid. Moreover, we discussed the Little Rip
occurrence in the presence of viscosity. Finally, we examined how viscosity influences
the Cardy-Verlinde formula, which is a topic that relates cosmology with thermodynamics,
and falls within the emergent gravity program.

We close this work mentioning that it would be both interesting and necessary to apply 
the cosmography formalism \cite{bamba12e,Capozziello:2008qc,Aviles:2014rma,Harko:2011kv} 
in order to impose constraints on the viscosity parameters.  Contrary to standard 
observational constraints, the advantage of cosmography is that it is model-independent, 
since one expands the scale factor independently of the solution of the cosmological 
equations. In particular, one introduces $H = \frac{1}{a} \frac{da}{dt}$, $q = - 
\frac{1}{a} 
\frac{d^2a}{
dt^2} \ H^{-2}$, $j = \frac{1}{a} \frac{d^3a}{dt^3} \ H^{-3}$, $s = \frac{1}{a} 
\frac{d^4a}{dt^4} \ H^{-4}$, $l = \frac{1}{a} \frac{d^5a}{dt^5} \ H^{-5}$, known 
respectively as    Hubble, deceleration, jerk, snap and  lerk  parameters \cite{bamba12e}.
One can show that these parameters are related through 
\cite{Capozziello:2008qc,Aviles:2014rma,Harko:2011kv},
\begin{eqnarray}
&&\dot{H} = -H^2 (1 + q) \nonumber\\
&&\ddot{H} = H^3 (j + 3q + 2) \ ,
 \nonumber\\
&&\dddot{H} = H^4 \left [ s - 4j - 3q (q + 4) - 6 \right ] \ ,
 \nonumber\\
&&d^4H/dt^4 = H^5 \left [ l - 5s + 10 (q + 2) j + 30 (q + 2) q + 24 \right ],\nonumber
\end{eqnarray}
which can easily lead to the distance\,-\,redshift relation \cite{bamba12e}. Hence, using 
the Friedmann 
equation in the case of viscous cosmology, we can relate the above cosmographic 
quantities with the present value of viscosity parameter. This could be a significant 
advantage, since the obtained constraint would be model-independent, and thus 
more robust that the ones discussed in subsection \ref{estimatess}. Nevertheless, the 
detailed investigation of this subject  lies beyond the scope of this review, and it is 
left for a future project.

Mostly, this review is based on a theoretical approach. We have however provided
information concerning quantities related to observations, giving estimations on the
inflationary observables, as well as on the magnitude of the
current bulk viscosity itself.

In summary, from the above analysis one can see the important implications and the
capabilities of the incorporation of viscosity, which make viscous cosmology a good
candidate for the description of Nature.

\begin{acknowledgments}

This work is supported by MINECO (Spain), Project FIS2013-44881, FIS2016-76363-P and by
CSIC I-
LINK1019 Project (S.~D.~O. and I.~B.) and by Ministry of Education and Science of Russia,
Project N.
3.1386.2017.

\end{acknowledgments}

\end{document}